\documentclass[structabstract]{aa}  
\usepackage{natbib}
\usepackage{graphicx}
\usepackage{float}
\usepackage{txfonts}

\newcommand{\comment}[1]{}
\newcommand{\beq}{\begin{equation}}
\newcommand{\eeq}{\end{equation}}
\newcommand{\beqa}{\begin{eqnarray}}
\newcommand{\eeqa}{\end{eqnarray}}

\begin{document}
\title{Self-consistent physical parameters for five intermediate-age SMC stellar clusters from CMD modelling\thanks{Based on observations obtained at the Southern Astrophysical Research (SOAR) telescope, which is a joint project of the Minist\'{e}rio da Ci\^{e}ncia, Tecnologia, e Inova\c{c}\~{a}o (MCTI) da Rep\'{u}blica Federativa do Brasil, the U.S. National Optical Astronomy Observatory (NOAO), the University of North Carolina at Chapel Hill (UNC), and Michigan State University (MSU).}}


   \author{B. Dias\inst{1,2} 
	\and L. O. Kerber\inst{1,3} 
	\and B. Barbuy\inst{1} 
	\and B. Santiago\inst{4}
	\and S. Ortolani\inst{5} 
	\and E. Balbinot\inst{4}
}

   \institute{Instituto de Astronomia, Geof\'\i sica e Ci\^encias Atmosf\'ericas,
	Universidade de S\~ao Paulo, Rua do Mat\~ao 1226, 
	Cidade Universit\'aria, S\~ao Paulo, 05508-900, SP, Brazil
	\and 
        European Southern Observatory, Alonso de Cordova 3107,
        Santiago, Chile\\
	\email{bdias@astro.iag.usp.br}
        \and
	LATO-DCET-UESC, Rodovia Ilh\'eus-Itabuna km 16, Ilh\'eus, Bahia, 
	45662-000, Brazil
	\and 
	Universidade Federal do Rio Grande do Sul, IF, CP 15051, Porto Alegre,
	91501-970, RS, Brazil
	\and
	Dipartimento di Fisica e Astronomia Galileo Galilei,
        University of Padova, vicolo dell'Osservatorio 3, 35122,
        Padova, Italy.}
   \date{Received ; accepted }

 
  \abstract
   {Stellar clusters in the Small Magellanic Cloud (SMC) are useful probes
     for studying the chemical and dynamical evolution of this 
     neighbouring dwarf galaxy, enabling inspection of
     a large period covering over 10 Gyr.}
    {The main goals of this work are the derivation of age,
     metallicity, distance modulus, reddening, core radius and central
     density profiles for six sample clusters, in order to place them in the context
     of the Small Cloud evolution. The studied clusters
     are AM~3, HW~1, HW~34, HW~40, Lindsay~2, and Lindsay~3;
     HW~1, HW~34, and Lindsay~2 are studied for the first time.} 
   {Optical Colour-Magnitude Diagrams (V, B-V CMDs) and radial density
     profiles were built from images obtained with the 4.1m Southern Astrophysical 
     Research (SOAR) telescope, reaching V$\sim$23. The determination of
     structural parameters were carried out by applying King profile fitting. 
     The other parameters were derived in a self-consistent way by means
     of isochrone fitting, which uses likelihood statistics to
     identify the synthetic CMDs that best reproduce the observed
     ones. Membership probabilities were determined comparing the
     cluster and control field CMDs. Completeness and photometric
     uncertainties were obtained by performing artificial star tests.} 
   {The results confirm that these clusters (except HW~34, identified
     as a field fluctuation) are intermediate-age clusters, with ages
     between 1.2 Gyr (Lindsay\,3) and $\sim$ 5.0 Gyr (HW\,1). 
     In particular HW~1, Lindsay~2 and Lindsay~3 are located in a
     region that we called West Halo, where studies of ages and
     metallicity gradients are still lacking.
     Moreover Lindsay~2 was identified as a moderately metal-poor
     cluster with [Fe/H] = -1.4$\pm$0.2 dex, lower than
     expected from the age-metallicity relation by Pagel \&
     Tautvaisiene (1998). We also found distances varying from $\sim
     $53~kpc to 66~kpc, compatible with the large depth of the SMC.}
   {}
\titlerunning{Self-consistent physical parameters for 
5 intermediate-age SMC clusters from CMD modelling}
\authorrunning{Dias et al.}

   \maketitle
%

\section{Introduction}
\label{Intro}

Star clusters (SCs) are useful objects for studying the complex stellar 
content observed in nearby galaxies, as most of them may be modelled 
as simple stellar populations (SSP) of a fixed age and metallicity.
The Magellanic Clouds (SMC for the Small Cloud, 
LMC for the Large Cloud, and MCs
for both) form a rich system with over $>$~3700 
stellar systems \citep{bica+08a},
with combinations of age and metallicity that are not found in the
Milky Way \citep{santos+04}. This information can be used to probe the
dynamical and chemical evolution of these neighbouring and interacting 
dwarf irregular galaxies, and in particular the age-metallicity relation
(AMR) of the Magellanic Clouds.

According to \cite{holtzman+99}, the age distribution based on 
clusters is probably distinct from the star formation history (SFH)
as inferred from field stars in the LMC.
For the SMC instead, \cite{rafelski+05} 
analysed a sample of 195 clusters, 
showing that the populations of field stars and star clusters are similar.
 In particular, \cite{piatti+05b} showed evidence
of two peaks in the ages of SMC star clusters at 6.5 Gyr and 2.5
Gyr, and \cite{piatti12a} indicates peaks of star formation for
field stars at 2 Gyr and 7.5 Gyr.
The more recent peak could be due to an encounter with the
 LMC. This leads to a model of bursts of star formation,
in contrast with suggestions of continuous star formation
\citep{DCH98}.
The large period of quiescent star formation in the MCs 
between $\sim$~3~Gyr and 10~Gyr \citep{harris+01,harris+04} is
indicated by the low number of populous SCs 
with these ages \citep{rich+00,rich+01}, and
almost all of them are in the SMC 
\citep{MSF98,piatti+05a,piatti+05b,piatti+07a,piatti+07b,piatti+07c}.

\cite{Cignoni+12} studied the SFH of two fields in the
SMC, and compared their results to the SFH behaviour suggested
by \cite{harris+04}. They concluded that stars older than 8.4 Gyr do not
dominate the SMC stellar population, and the period 
between 2~Gyr and 8.4~Gyr agrees with \cite{harris+04} in one region,
but has a much higher star formation rate (SFR) in another region.
Therefore, further studies of the SMC cluster and field AMR are needed.

Metallicity values for SMC star clusters, as given in
the literature compilation by \cite{parisi+09} for example,
seem to be slightly underestimated when compared with the 
 chemical evolution model predictions
of \cite{PT98},  for ages in the range 3~Gyr to 10~Gyr. 
By adding more data points to this region of the AMR, \cite{piatti11b}
included a few metal-poor intermediate age star clusters
(IACs). 
To further improve the age-metallicity relation of the SMC,
it is important to identify other IACs, and the metal-poor ones
are particularly interesting.

In spite of its great interest,  the SMC cluster system 
has been studied less than that of the LMC.
Literature data on SMC star clusters were reviewed by \cite{dias+10}. 
In their Table 6, ages and metallicities for 33 among the
most well-studied SMC star clusters are reported, giving particular
attention to old/intermediate-age ones. In addition, based on
integrated spectra, the analysis by Dias et al. showed that the clusters
HW\,1 and Lindsay\,3 (as well as NGC\,152) can be added to the list of
intermediate/old clusters. 
Since then, newly revealed intermediate-age star clusters (IACs) were
studied by \citet{piatti11a, piatti11b}, \citet{piatti+11} and \citet{piatti12b}.

The compilation of ages and metallicities for SMC
  clusters by \cite{dias+10} showed that the literature results
  have variations up to 7~Gyr in age and
  0.9~dex in metallicity, for a given cluster. 
In the present work, B and V photometry combined with self-consistent statistical
tools are employed to determine ages and metallicities
for \object{AM\,3}, \object{HW\,1}, \object{HW\,40}, \object{Lindsay\,2} and \object{Lindsay\,3}, in order to provide more precise constraints for the AMR of the
  SMC. 
For these clusters, there are the spectroscopic analysis by
\cite{dias+10} of HW~1 and Lindsay~3, and the Washington photometry
analysis of AM\,3, HW\,40 and Lindsay\,3 \citep{piatti11a, piatti+11}.
The target \object{HW~34} is probably only a field fluctuation. For HW~1 and Lindsay~2 no
previous CMD data were available in the literature. The confirmation of some
of these clusters as intermediate or old age significantly improves
the poor census in the age range corresponding to the age gap for the
LMC clusters ($\sim$3-10 Gyr).

In Sect. 2 the observations, data reduction, and photometry are described. 
In Sect. 3 the modelling of CMDs and statistical comparisons carried
out to find the best fit of synthetic vs. observed CMDs are
detailed. In Sect. 4 the results and a discussion of each cluster are given.
In Sect. 5 comparisons with the literature, and the age-metallicity relation
 for the SMC are presented. Finally, a summary is given in Sect. 6.


\section{The data}
\label{thedata}

\subsection{The SOAR/SOI data}

Using the SOAR Optical Imager (SOI) mounted on the 4.1m Southern Astrophysical 
Research (SOAR) Telescope, B and V images were obtained for SMC SCs,
under projects SO2007B-013 and SO2008B-017. 
We chose filters B and V since they provide the best temperature
  resolution for -0.2$<$B-V$<$1.4, which corresponds
to the colours of the present CMDs; V, I would be better in high extinction
fields, which is not the case of our fields, and
 B, V is less affected by differential reddening than is V, I \citep{solderaphd}.
This imager has a field of view of 5.26$\arcmin \times$
5.26$\arcmin$, and a pixel scale of 0.077$\arcsec$/pixel, which is
converted to 0.154$\arcsec$/pixel because of the 2x2 binned observations
presented here.
The seeing was $\sim$~0.8~arcsec, and magnitudes up to
V~$\sim$~23 were detected. The log of observations is reported in
Table \ref{log}.

\begin{table}[!htb]
\caption{Log of observations. The CCDs were displaced by $\sim
  20\arcsec$ from the cluster centre to avoid the gap
  between the set of two E2V CCDs in SOI, as shown in Figure
  \ref{soar_soi-plots}. The ($\alpha$, $\delta$) coordinates are from
  \cite{bica+08a}.}
\label{log}
\centering
\begin{tabular}{lc@{}c@{}cccc}
\hline
\noalign{\smallskip}
Name(s) & $\alpha$ (2000)& $\delta$ (2000) & Filter  & Exp. & Airmass & seeing  \\
 & h:m:s & $\degr:\arcmin:\arcsec$ &  & sec. &   & $\arcsec$ \\
\noalign{\smallskip}
\hline
\noalign{\smallskip}
\multicolumn{7}{c} {2007-09-07} \\
\noalign{\smallskip}
\hline
\noalign{\smallskip}
AM\,3  & 23:48:59 & -72:56.7 & B & 600 & 1.42 & 1.00  \\
       &          &           & V & 200 & 1.40 & 1.03  \\
Lindsay\,2  & 00:12:55 & -73:29.2 & B & 600 & 1.38 & 1.10   \\
       &          &           & V & 200 & 1.38 & 1.03  \\
\noalign{\smallskip}
\hline
\noalign{\smallskip}
\multicolumn{7}{c} {2008-09-21} \\
\noalign{\smallskip}
\hline
\noalign{\smallskip}
HW\,1  & 00:18:27 & -73:23.7 & B & 600 & 1.48 & 0.86  \\
       &          &           & V & 200 & 1.53 & 0.87 \\
Lindsay\,3  & 00:18:25 & -74:19.1 & B & 600 & 1.42 & 1.13  \\
       &          &           & V & 200 & 1.44 & 0.86  \\
HW\,34  & 00:57:52 &\phantom{--}-73:32.7 & B & 600 & 1.38 &  0.95  \\
       &          &           & V & 200 & 1.39 & 0.82 \\
HW\,40  & 01:00:25 & -71:17.7 & B & 600 & 1.34 & 0.89 \\
       &          &           & V & 200 & 1.33 & 0.91 \\
\noalign{\smallskip}
\hline
\end{tabular}
\end{table} 

Reduction procedures were based on the SOAR/IRAF packages 
and the photometry procedures were based on the DAOPHOT/IRAF 
package \citep{stetson87}. Classical procedures of aperture and then
PSF photometry were performed in the B and V bands. Since PSF fitting
resulted in better quality photometry, all the analyses in this paper
are based on this method.

The standard stars were chosen
from \cite{sharpee+02}, having coordinates close to the SMC direction, 
in order to save time.
Airmass corrections were applied assuming constant values of
0.22$\pm$0.03~mag/airmass and 0.14$\pm$0.03~mag/airmass for the B and V
bands, respectively (as can be found at the CTIO
website\footnote{http://www.ctio.noao.edu/noao/content/13-m-photometric-standards}). Only
standard stars observed with a seeing lower than 1.0$\arcsec$, and
airmasses close to the cluster observations were considered. Moreover,
stars with magnitude variations between an aperture radius of 5 and
8$\times$FWHM larger than 0.010~mag were not considered in
this fit. Then, the following calibration curves were fitted to the standard
stars; the results are presented in Table \ref{calibcoef}. 
The coefficient of determination r$^2$ that indicates how well
  the data fits a line (close to 1 is best) and the low residual
  values show that both nights were photometric, which is confirmed by
  the CTIO
  monitoring\footnote{http://www.ctio.noao.edu/site/phot/sky\_conditions.php}

\begin{center}
\begin{equation}
  M-m=\alpha\cdot (B-V)+ \beta{\rm ,}
  \label{calib-cur}
\end{equation}
\end{center}

\noindent where $M$ corresponds to either B or V, and $m$ corresponds
to the respective instrumental magnitudes 
(given by -2.5~$\times$~log(counts/exptime) already corrected by airmass effects.

\begin{table}[!htb]
\caption{Coefficients of Eq. \ref{calib-cur} from the fits of the
  2007 and 2008 standard stars. }
\label{calibcoef}
\centering
\begin{tabular}{l|cc}
\hline \noalign{\smallskip}
Coef.          & B                      & V                  \\
\noalign{\smallskip}
\hline 
\noalign{\smallskip}
$\alpha$    & 0.30 $\pm$ 0.10   & 0.30 $\pm$ 0.09  \\
$\beta$ (mag) & 27.02 $\pm$ 0.04 & 26.83 $\pm$ 0.07  \\
r$^2$         & 0.83                     & 0.44  \\
$|$residuals$|$ (mag) & $<$0.04                     & $<$0.10  \\
\noalign{\smallskip}
\hline
\end{tabular}
\end{table} 

\begin{figure*}[!htb]
   \centering
   \includegraphics[width=0.29\textwidth]{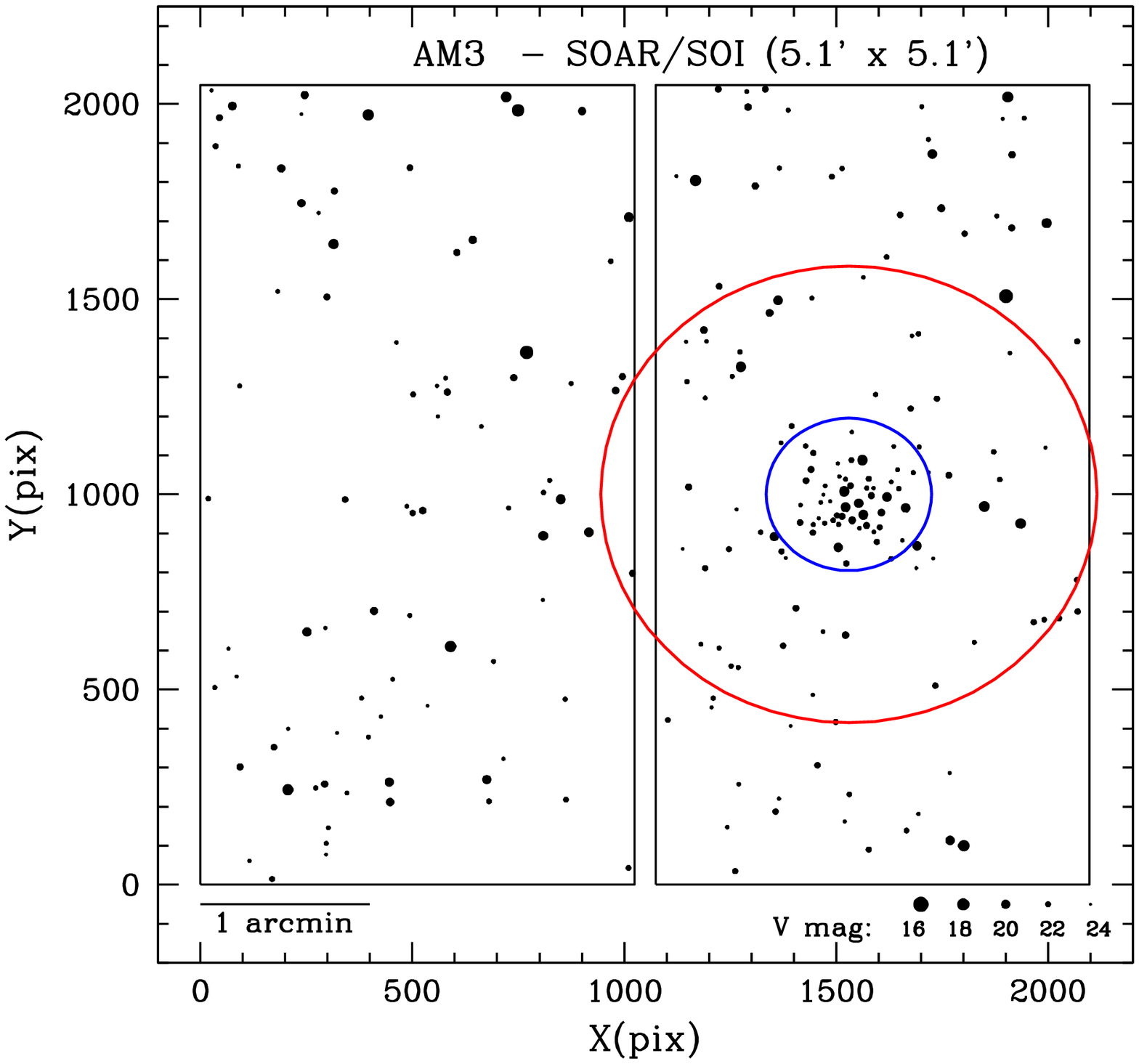}
   \includegraphics[width=0.29\textwidth]{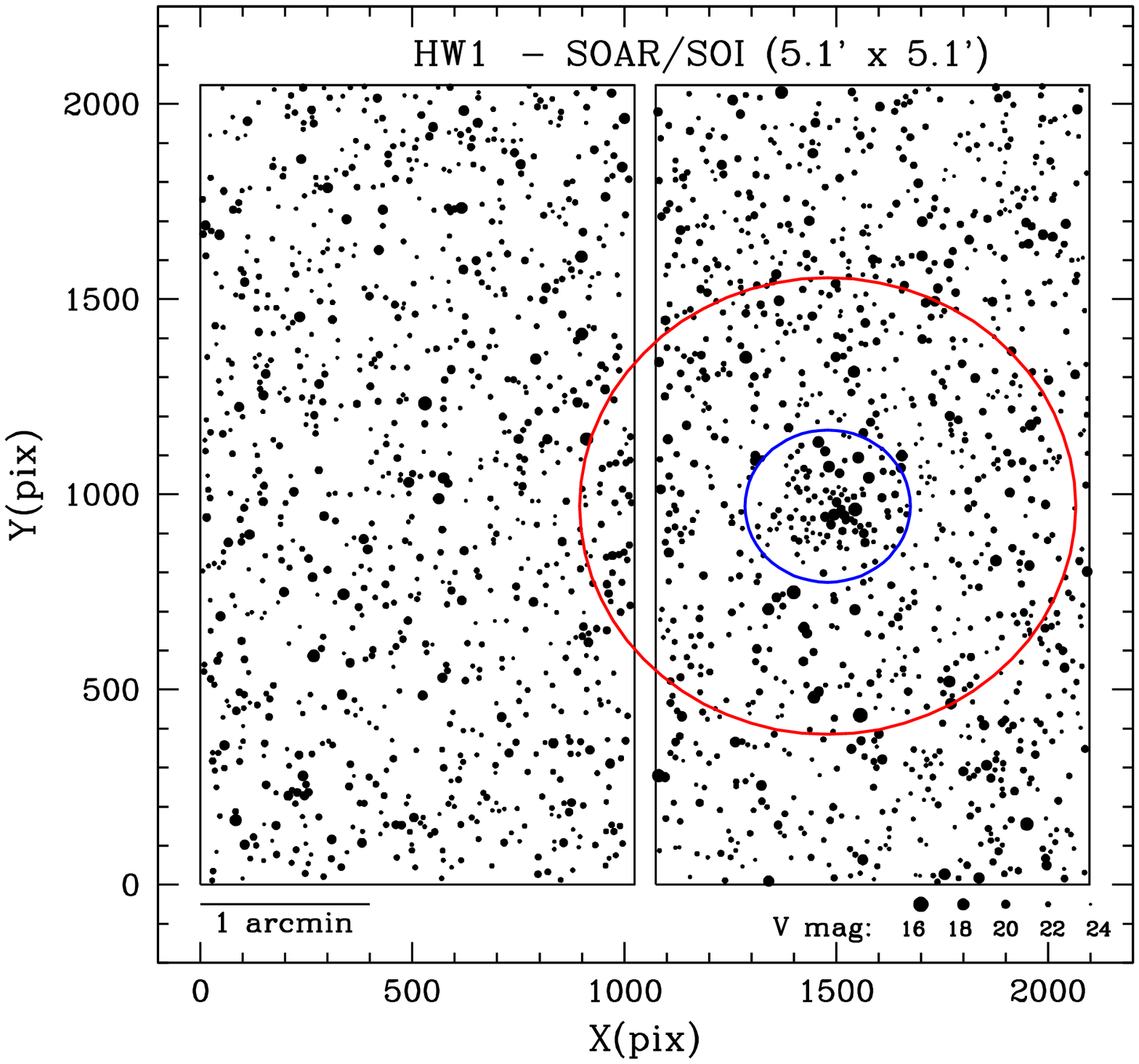}
   \includegraphics[width=0.29\textwidth]{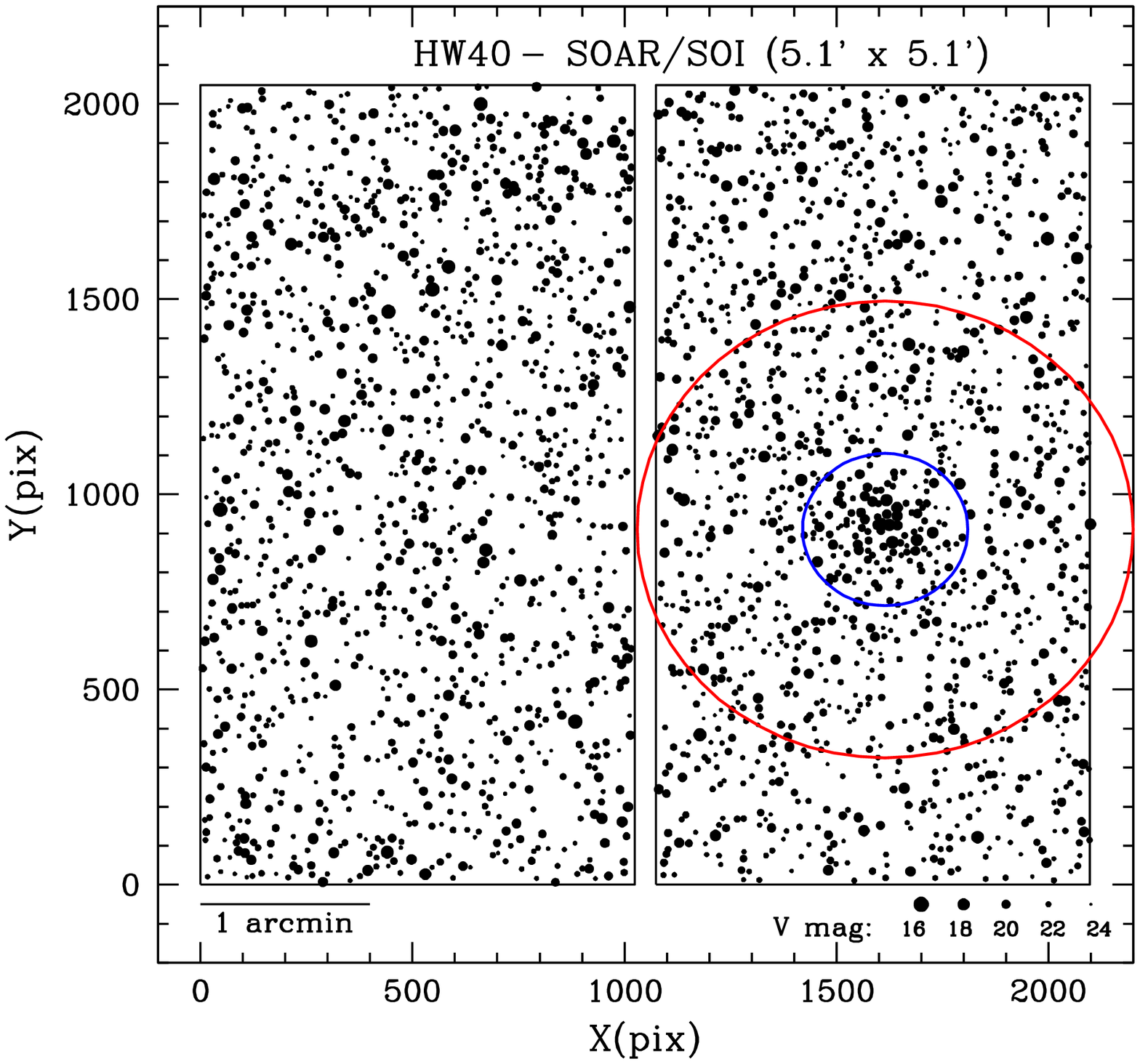}
   \includegraphics[width=0.29\textwidth]{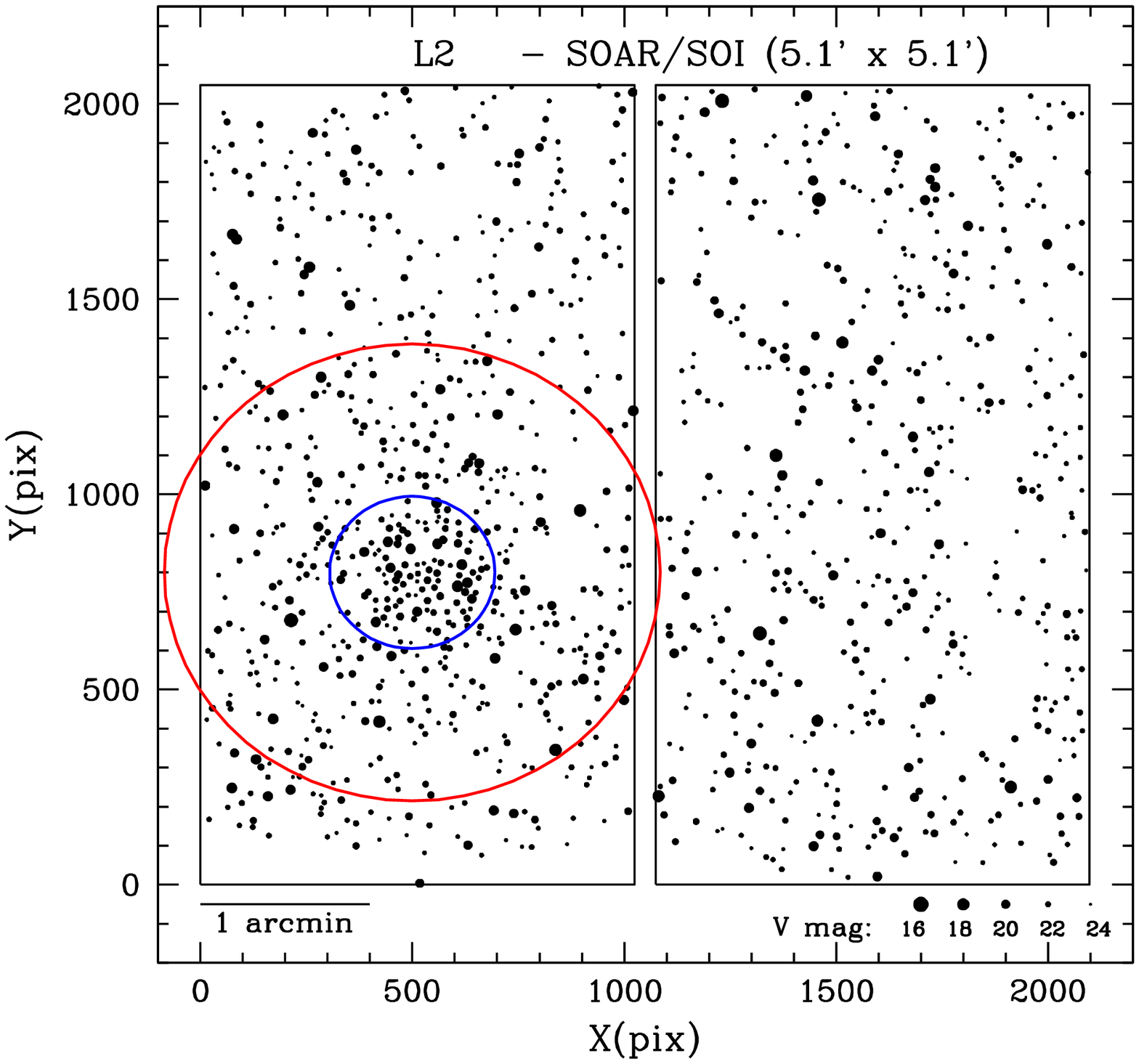}
   \includegraphics[width=0.29\textwidth]{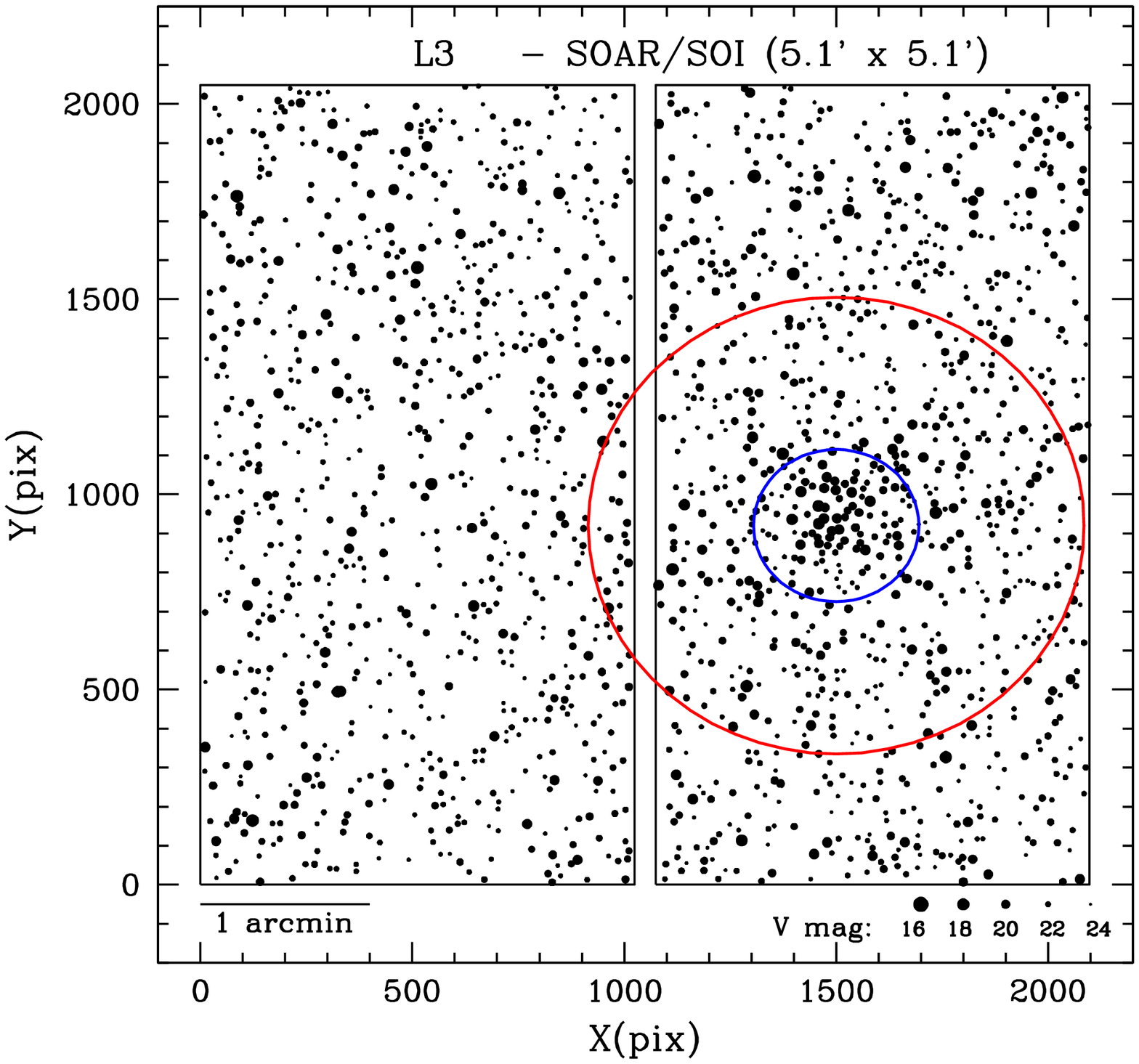}
   \includegraphics[width=0.29\textwidth]{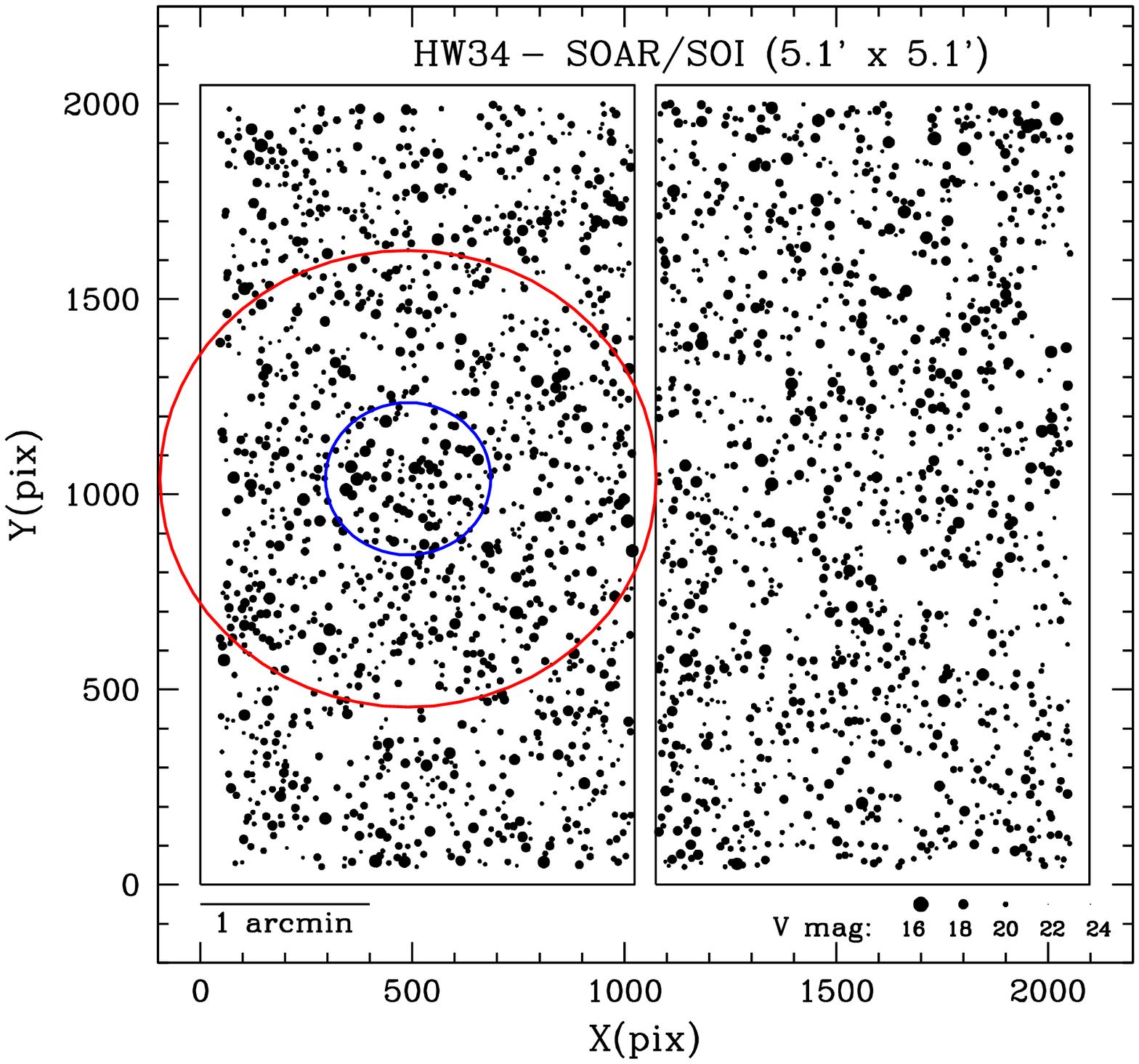}
   \caption{Sky maps showing the stellar distribution for all clusters, 
as imaged by the SOAR/SOI (1 pixel = 0.154 arcsec). The small circle ($R < R_{\rm{clus}}=30$ arcsec)
corresponds to the cluster sample, 
whereas the large circle limits the area for the control field sample
($R > R_{\rm{field}}=90$ arcsec).}
   \label{soar_soi-plots}
   \end{figure*}

Figure \ref{soar_soi-plots} presents the sky map for each cluster,  
showing the stellar spatial distribution of the SOAR/SOI images, and
the selected cluster and field areas.
The adopted cluster ($R_{\rm{clus}}$) and field radii ($R_{\rm{field}}$)
for all clusters are 30 arcsec and 90 arcsec, respectively, covering
solid angles ($\Omega_{\rm{clus}}$ and $\Omega_{\rm{field}}$) 
of about 0.79~arcmin$^2$ and 20.6~arcmin$^2$. These sky maps enable us to
visually identify the overdensity inside the cluster radii
(except for HW~34) and to verify which are the more/less crowded
lines of sight.


%
\subsection{Photometric errors and completeness curves}

\begin{figure*}[!htb]
  \centering
  \includegraphics[width=0.29\textwidth]{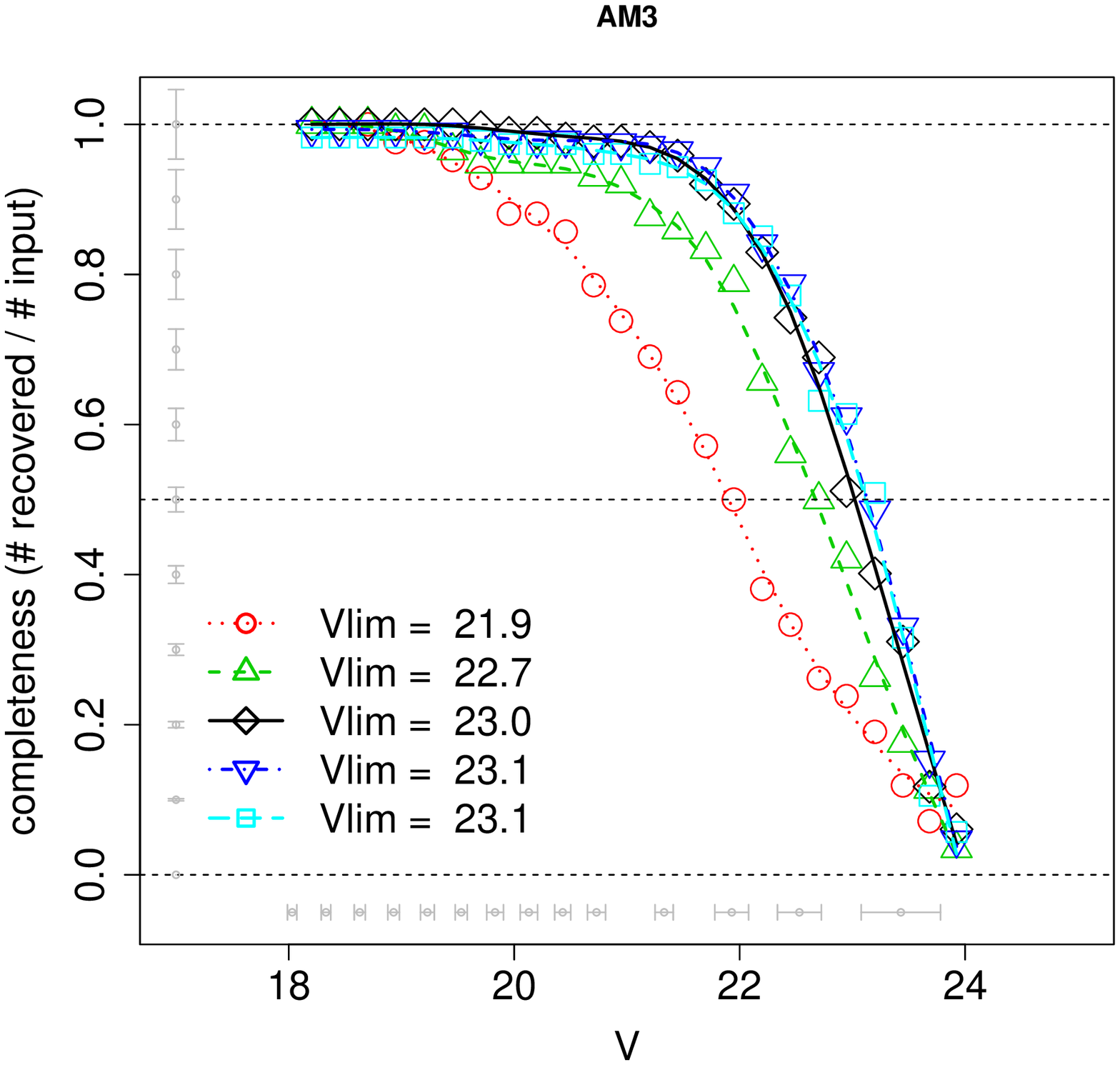}
  \includegraphics[width=0.29\textwidth]{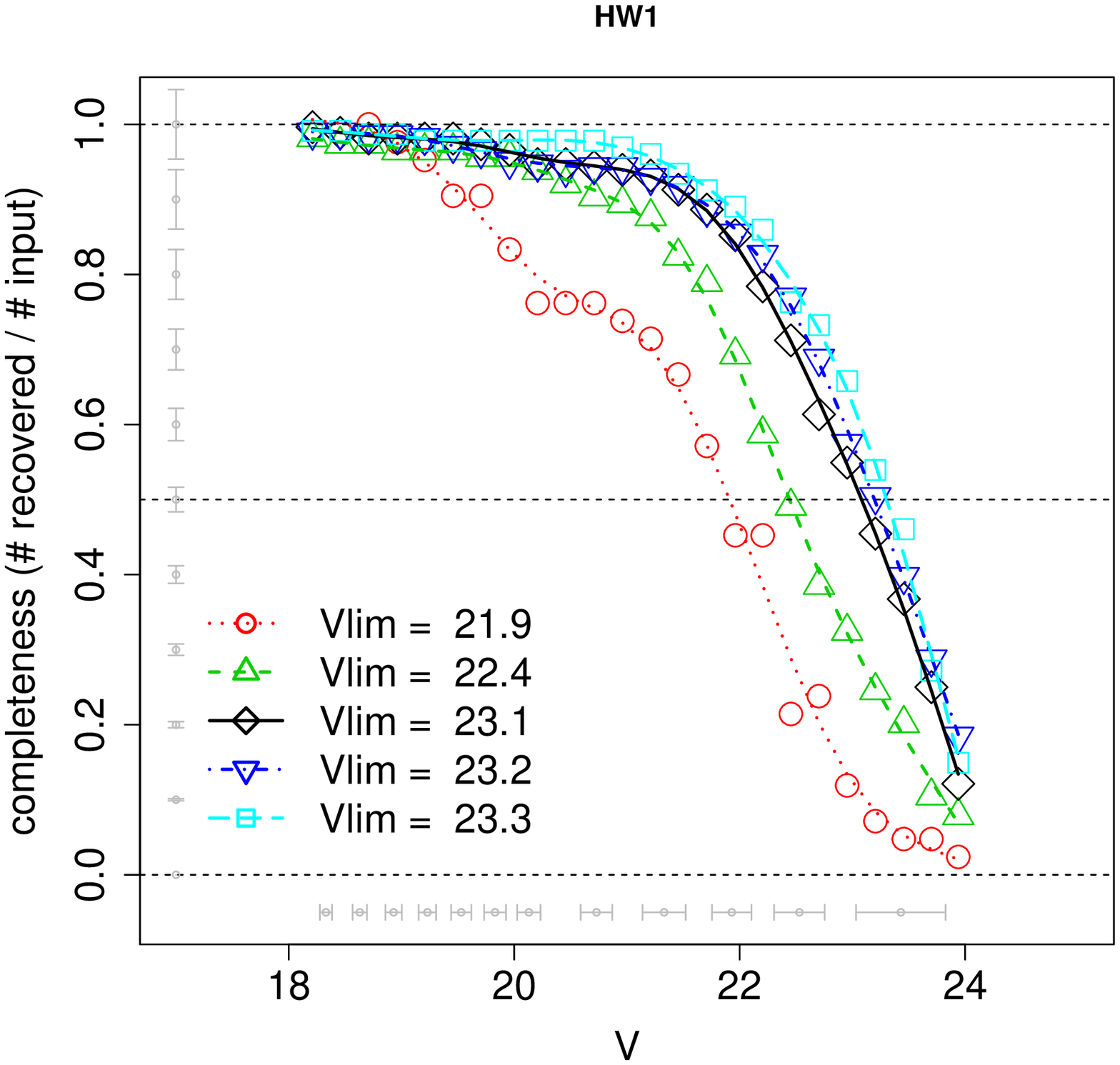}
  \includegraphics[width=0.29\textwidth]{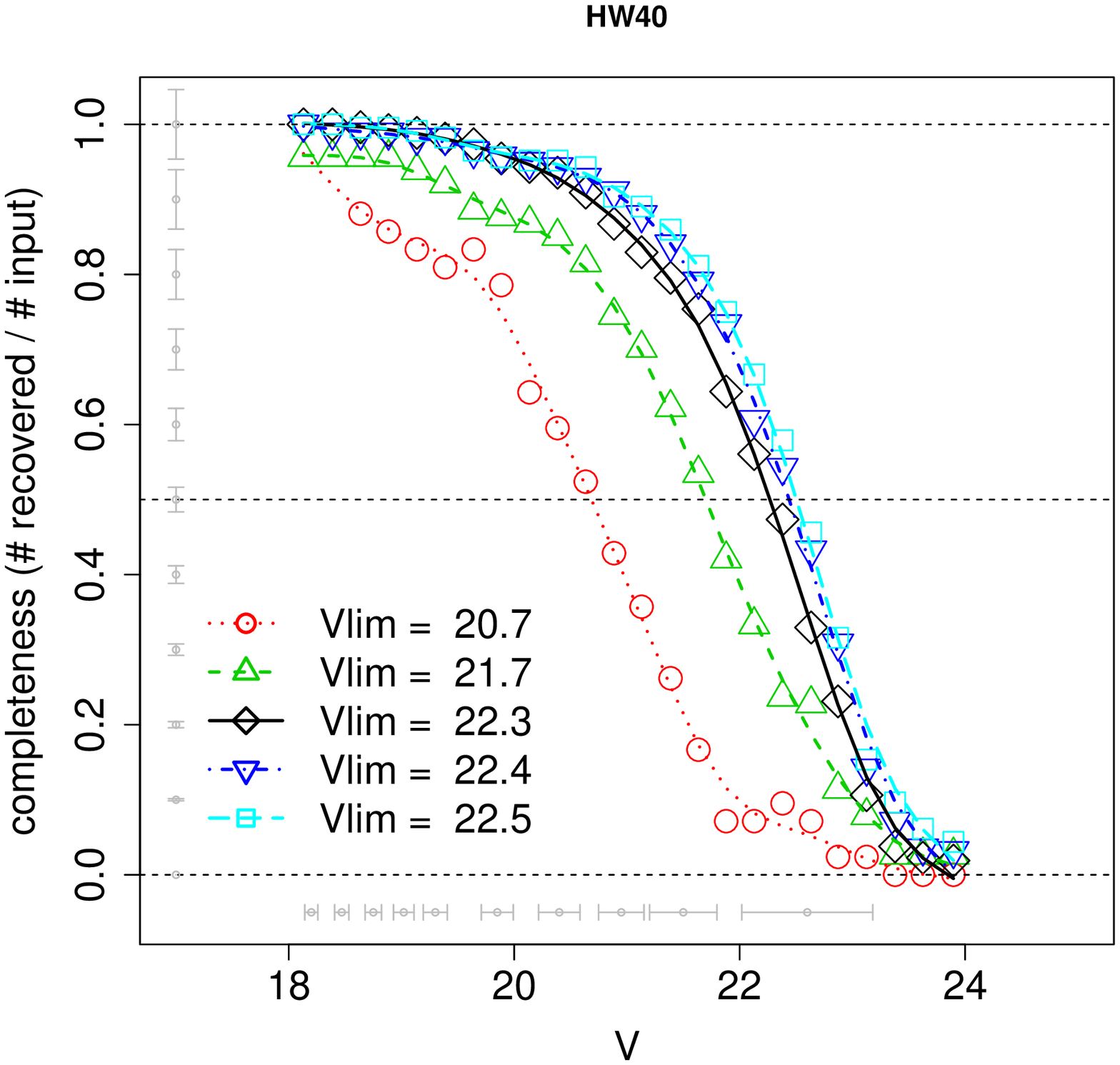}
  \includegraphics[width=0.29\textwidth]{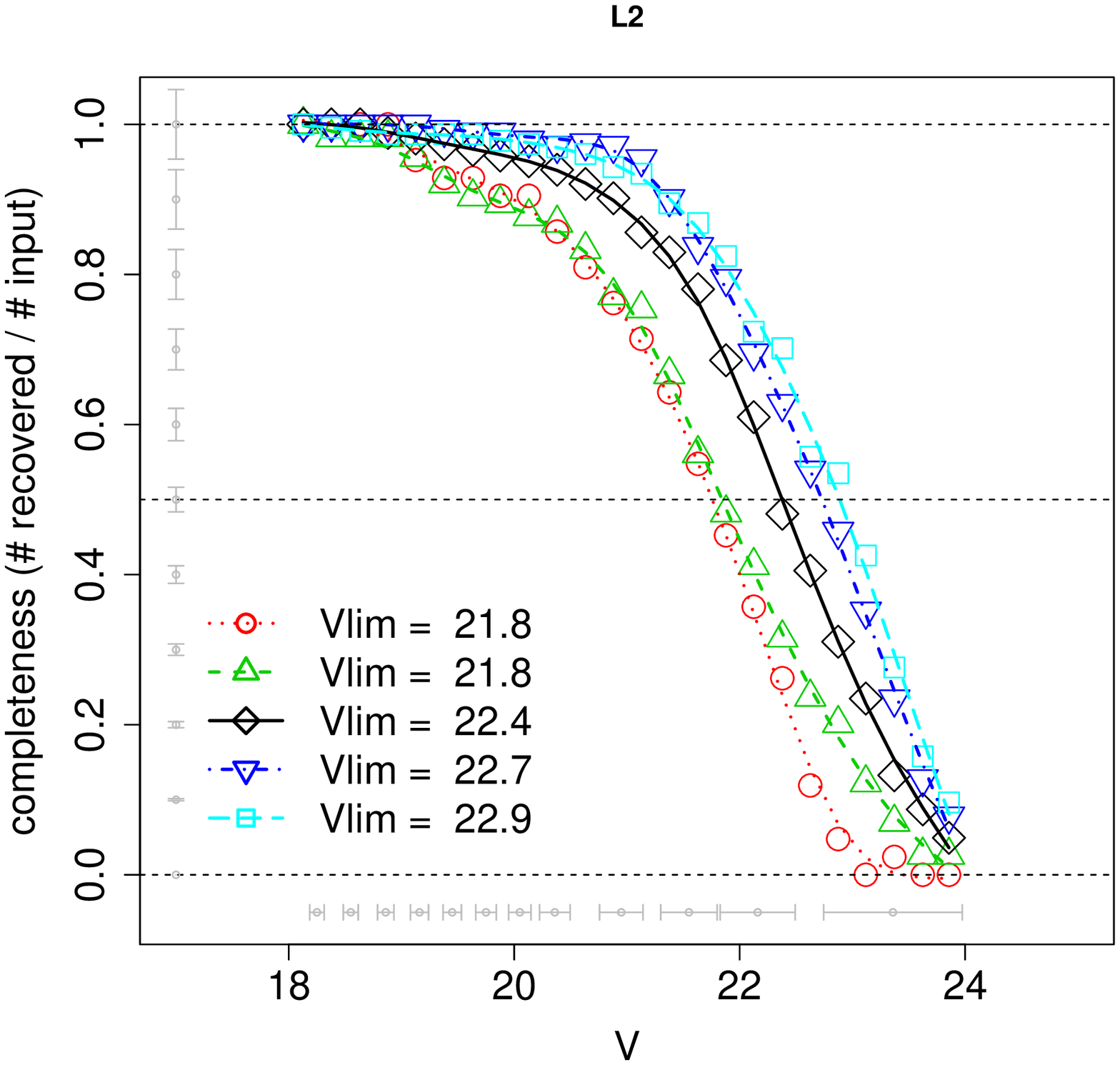}
  \includegraphics[width=0.29\textwidth]{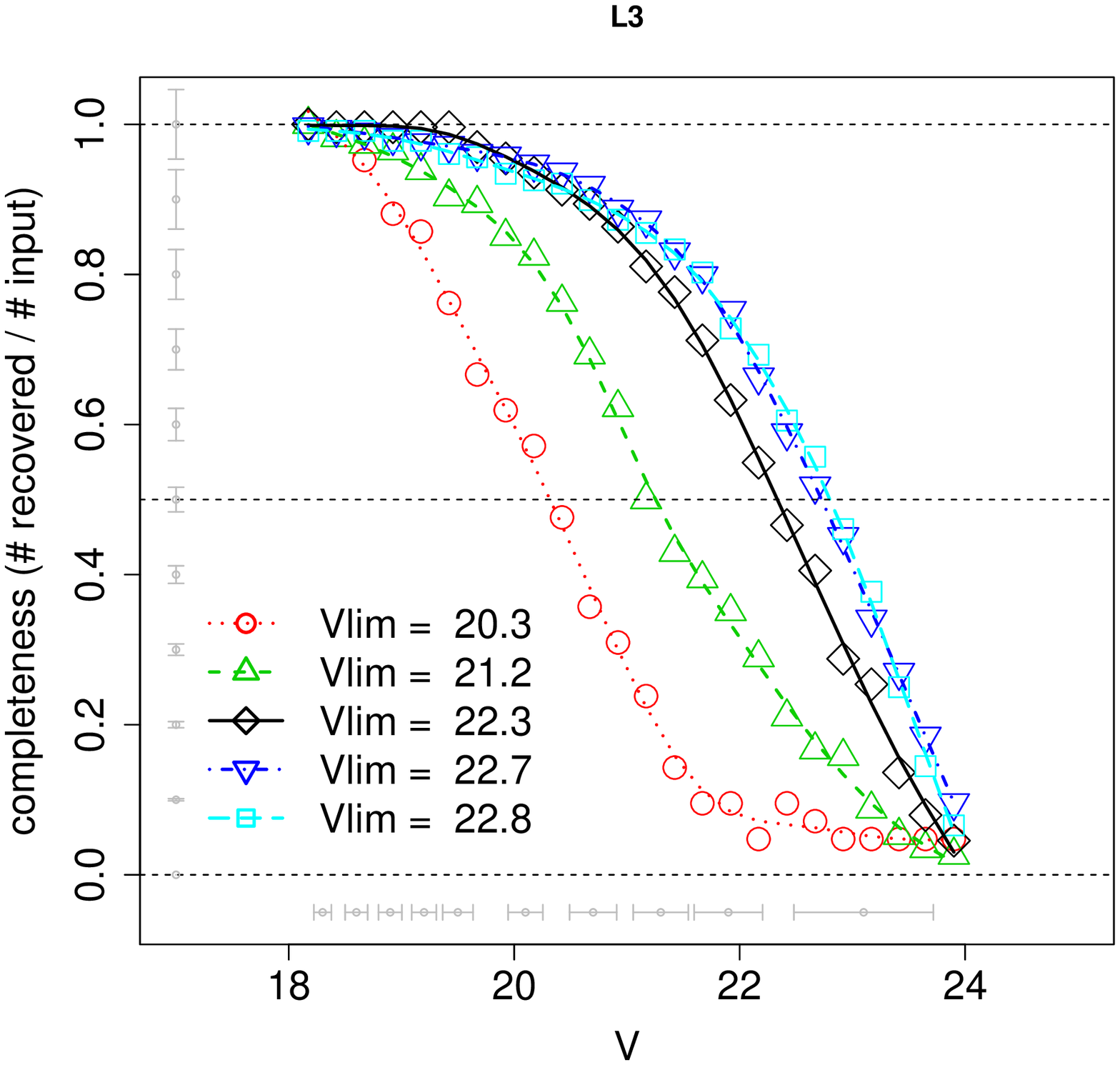}
  \includegraphics[width=0.29\textwidth]{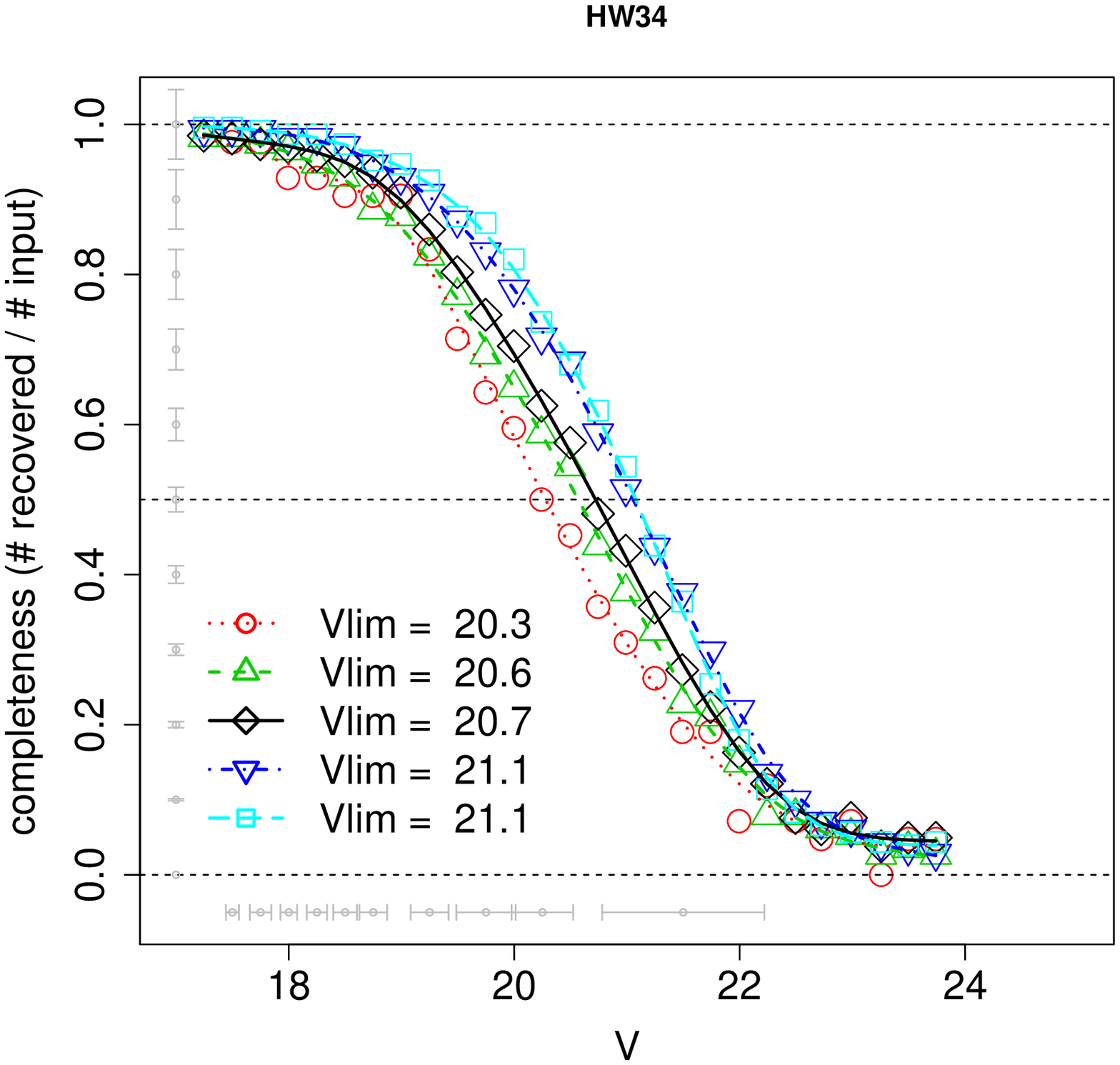}
  \caption{Completeness curves for different distances from
    the cluster centre, based on V-band photometry. These curves
    correspond to a ten-order smoothed spline fitted to the data. Different
    lines are the completeness curves for different annuli, with the
    respective data points:
    $0\leq$~r~$<10\arcsec$ (red dotted/circle);
    $10\leq$~r~$<20\arcsec$ (green
    dashed/triangle); $20\leq$~r~$<30\arcsec$ (black solid/rhombus);
    $30\leq$~r~$<40\arcsec$ (blue dash dotted/inverted triangle);
    $40\leq$~r~$<50\arcsec$ (cyan long dashed/square). The
    magnitude limit corresponding to 50\% of completeness are
    indicated in the plots. Completeness uncertainties are
      indicated in grey vertically (assumed as Poissonian), and
      typical magnitude uncertainties are indicated in grey horizontally.}
  \label{completeness}
\end{figure*}

To properly determine the completeness of the photometry and the
photometric errors we performed artificial star tests 
(ASTs). These consist of adding stars (with known
magnitudes and colours) in random positions to the reduced images, and
then carrying out the photometry exactly as was done with the original data.
The ratio of recovered to input stars is called completeness (see
Figure \ref{completeness}).

The procedure was carried out for both B and V images assuming an
average colour of (B-V)=0.5 for all the stars. The most
important curves are those from the V band, since a V magnitude limit is
applied on Figure \ref{membership-plots}, 
and the curves are used to correct the radial density profiles
(Fig. \ref{profiles-plots}), which are better sampled by V images. In
fact, B completeness curves were used only by internal checks: they
differ by $<$0.5~mag from V curves with the same completeness
values. 
We considered that this is a good
approximation for the observed stars with colours ranging from $\sim$
$0.0 < B-V < 1.0$.
 Therefore, we used only V magnitude limits, as follows.
The total range of V magnitudes covering all observed
stars (from $\sim$17 to $\sim$24~mag)
were divided into 32 intervals. For each interval, 
two sets of 465 stars were simulated inside a radius of 50$\arcsec$ from the
cluster centre, with random magnitudes, separated by at least
3$\arcsec$ ($\gtrsim$ 3.5 $\times <{\rm FWHM}>$) to avoid
introducing additional crowding (e.g. \citealp{rubele+11}). 
The resulting completeness curves as a function of the V magnitude are
presented in Figure \ref{completeness} for diferent annuli around the
cluster centre.
The limiting magnitude is set where the completeness ratio is equal to
50\% (see values in the figure). We adopted the V$_{\rm limit}$ values
of the annulus $20\arcsec < r < R_{\rm{clus}} \equiv 30\arcsec$, 
where R$_{\rm clus}$ is defined as the outer cluster radius used for the
  CMD analysis.

The correspondence between Figures
\ref{soar_soi-plots} and \ref{completeness} is evident. In particular, for the
central regions (red solid curve), completeness for
HW~34, HW~40, and Lindsay~3 are similar, as expected from
 the similar central densities of the stars. Instead, for external
 annuli, HW~40 and Lindsay~3 are 
more complete (2 magnitudes more), while HW~34 presents roughly
similar completeness to the centre. This indicates a crowded and homogeneous field
in the HW~34 direction.

\begin{figure*}[!htb]
  \centering
  \includegraphics[width=0.29\textwidth]{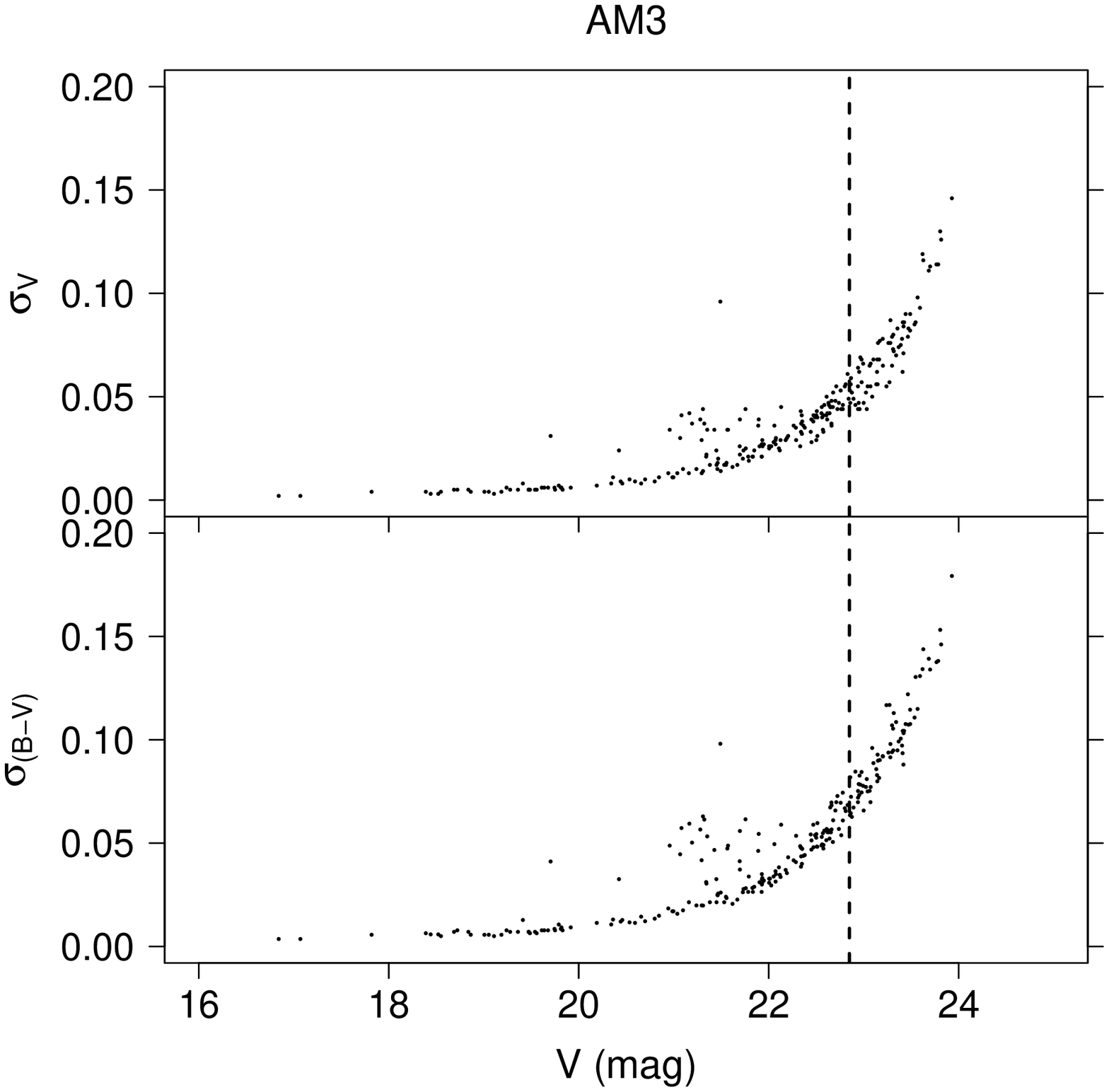}
  \includegraphics[width=0.29\textwidth]{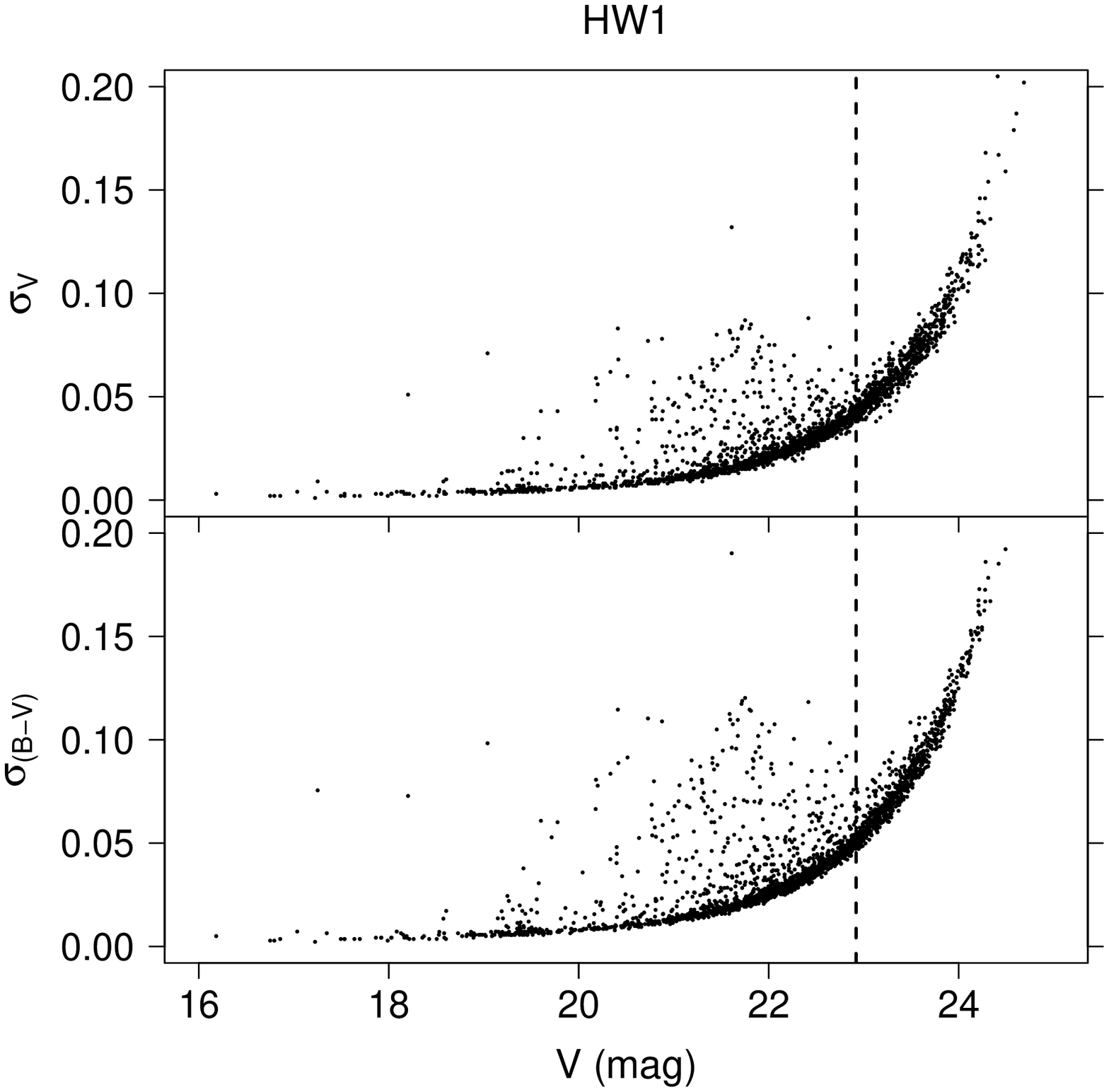}
  \includegraphics[width=0.29\textwidth]{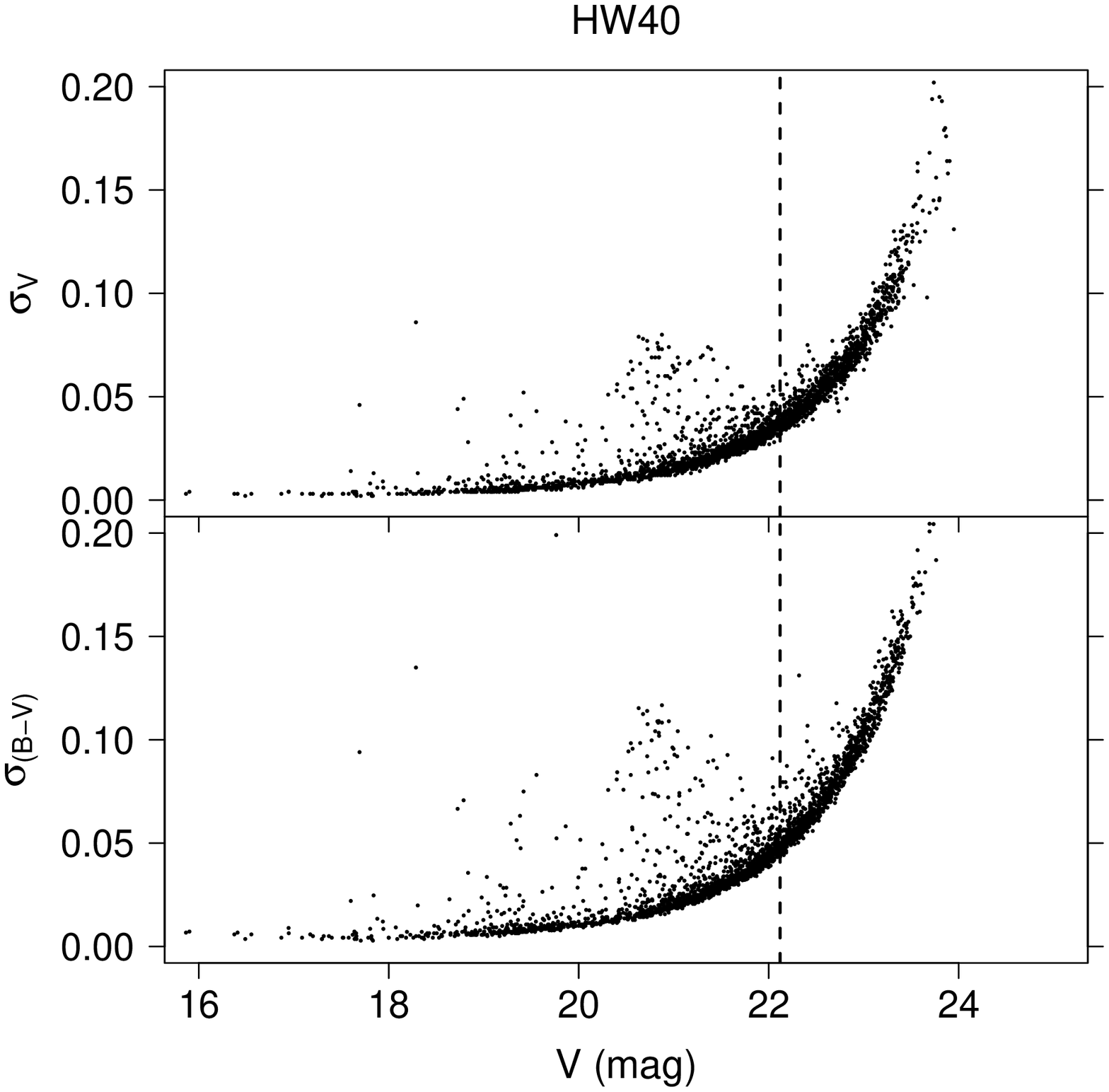}
  \includegraphics[width=0.29\textwidth]{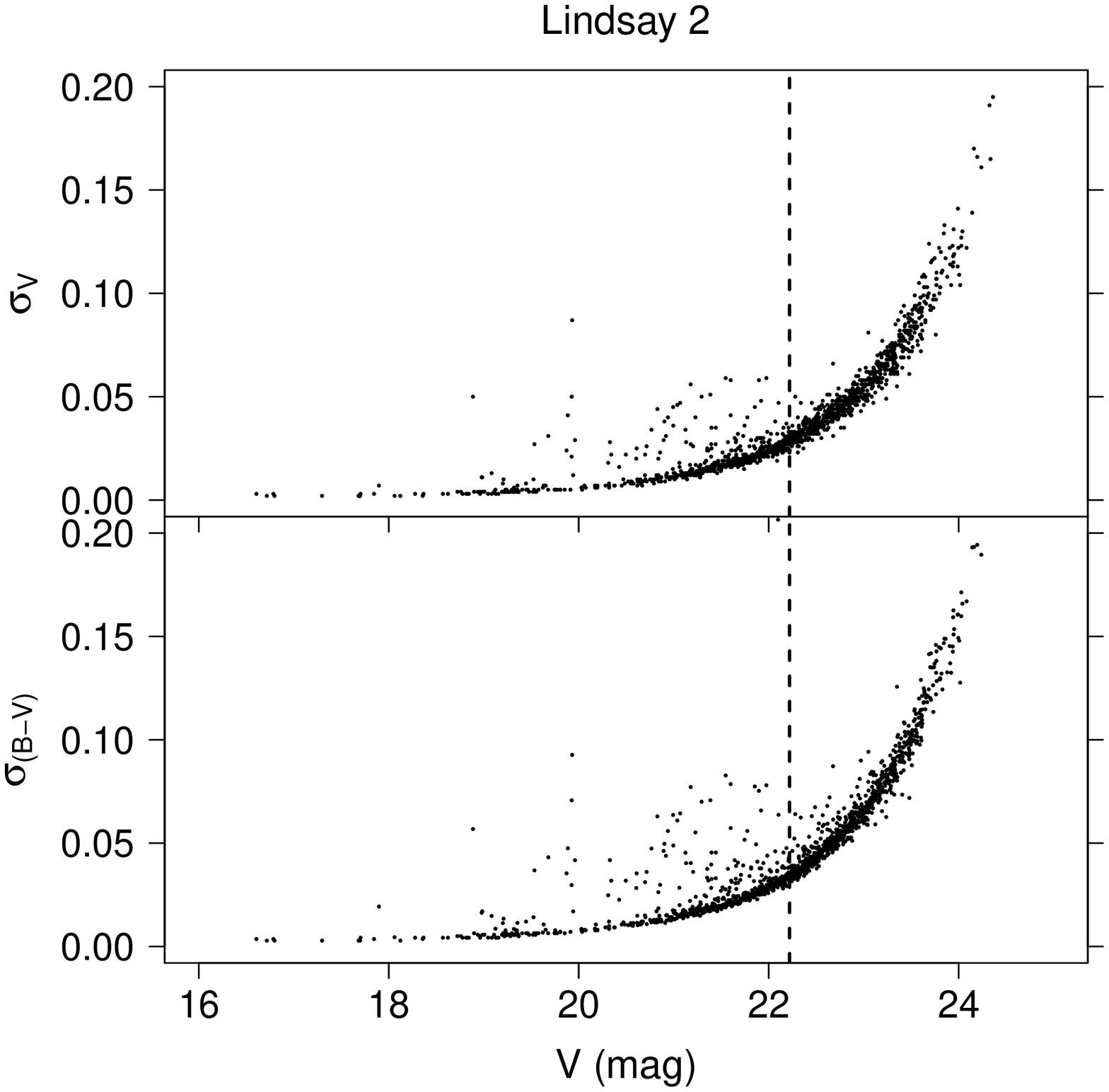}
  \includegraphics[width=0.29\textwidth]{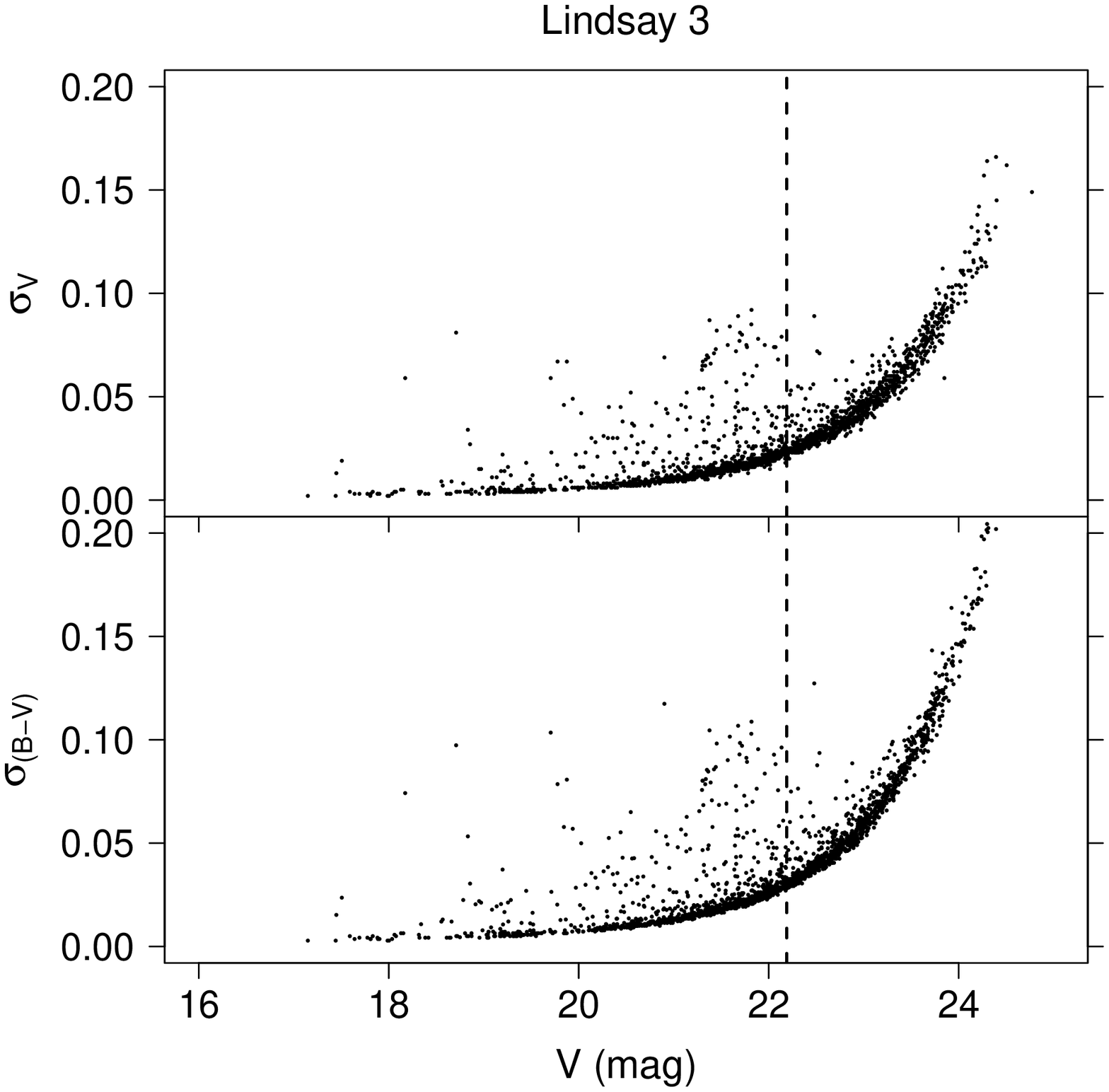}
  \includegraphics[width=0.29\textwidth]{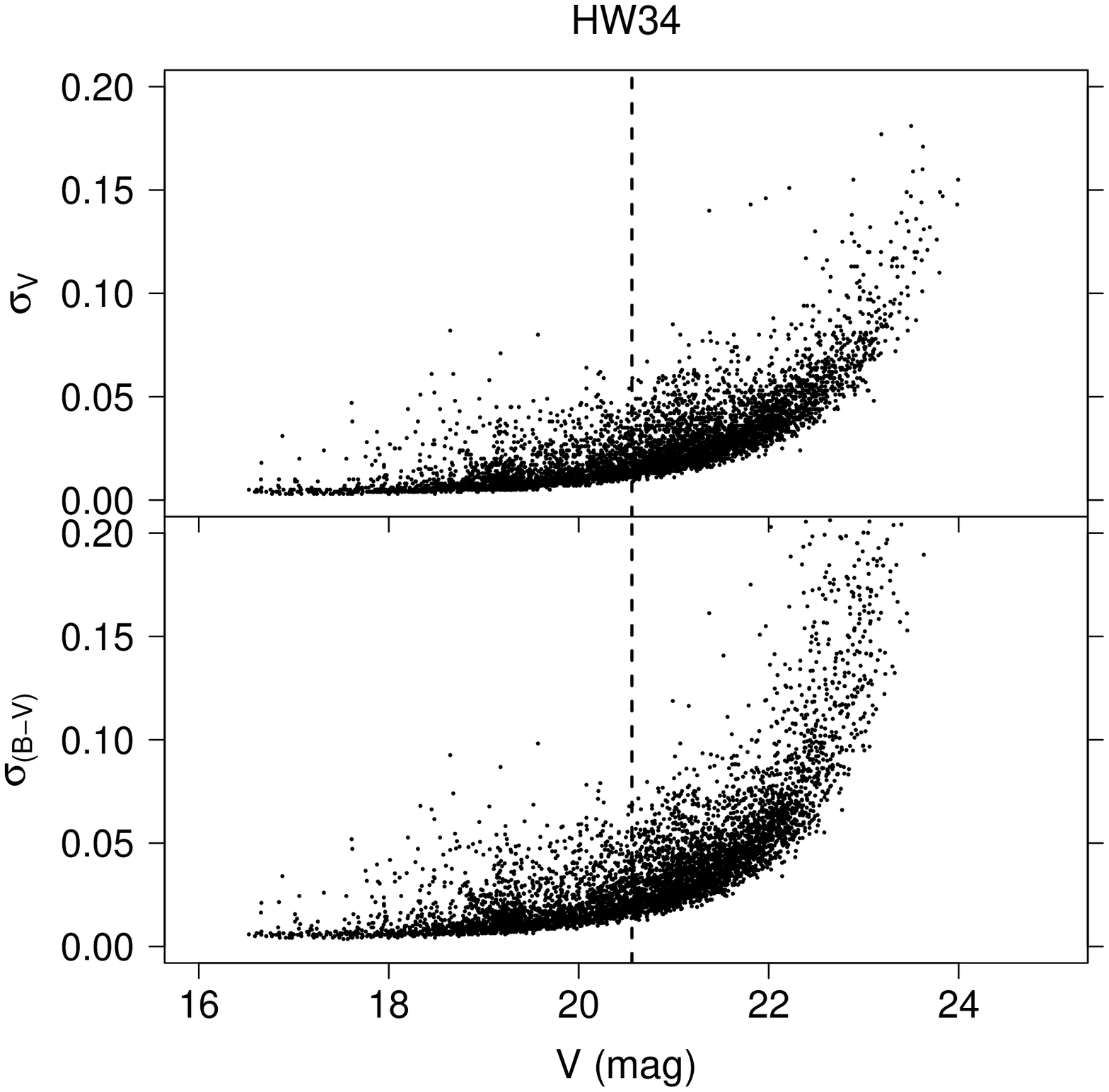}
  \caption{V and (B-V) photometric errors given by DAOPHOT
    outputs. Vertical dashed lines correspond to magnitude limits in a
    radius of 30$\arcsec$, as shown in Figure \ref{completeness}.}
  \label{errors}
\end{figure*}
\begin{figure*}[!htb]
   \centering
   \includegraphics[width=0.29\textwidth]{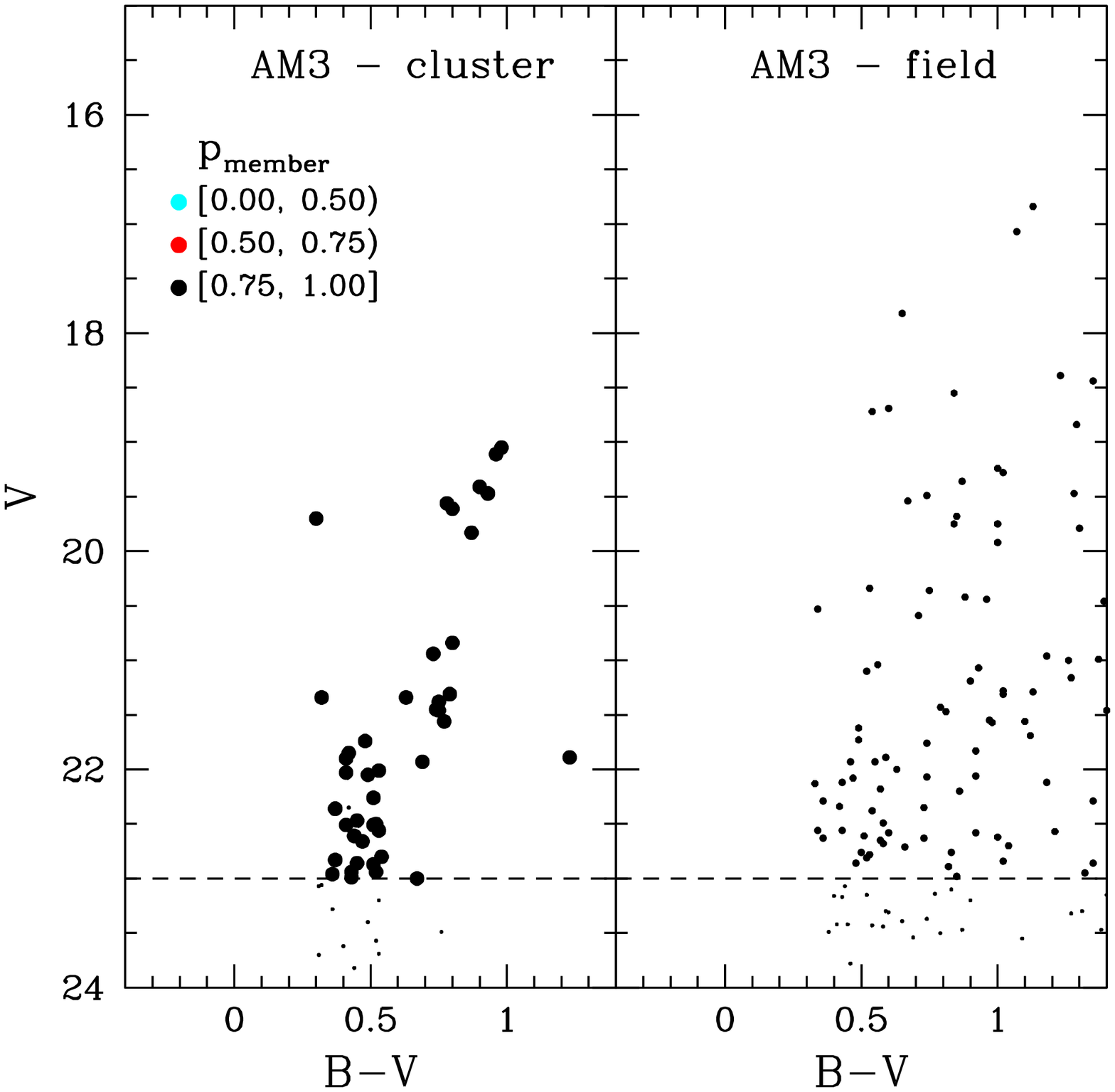}
   \includegraphics[width=0.29\textwidth]{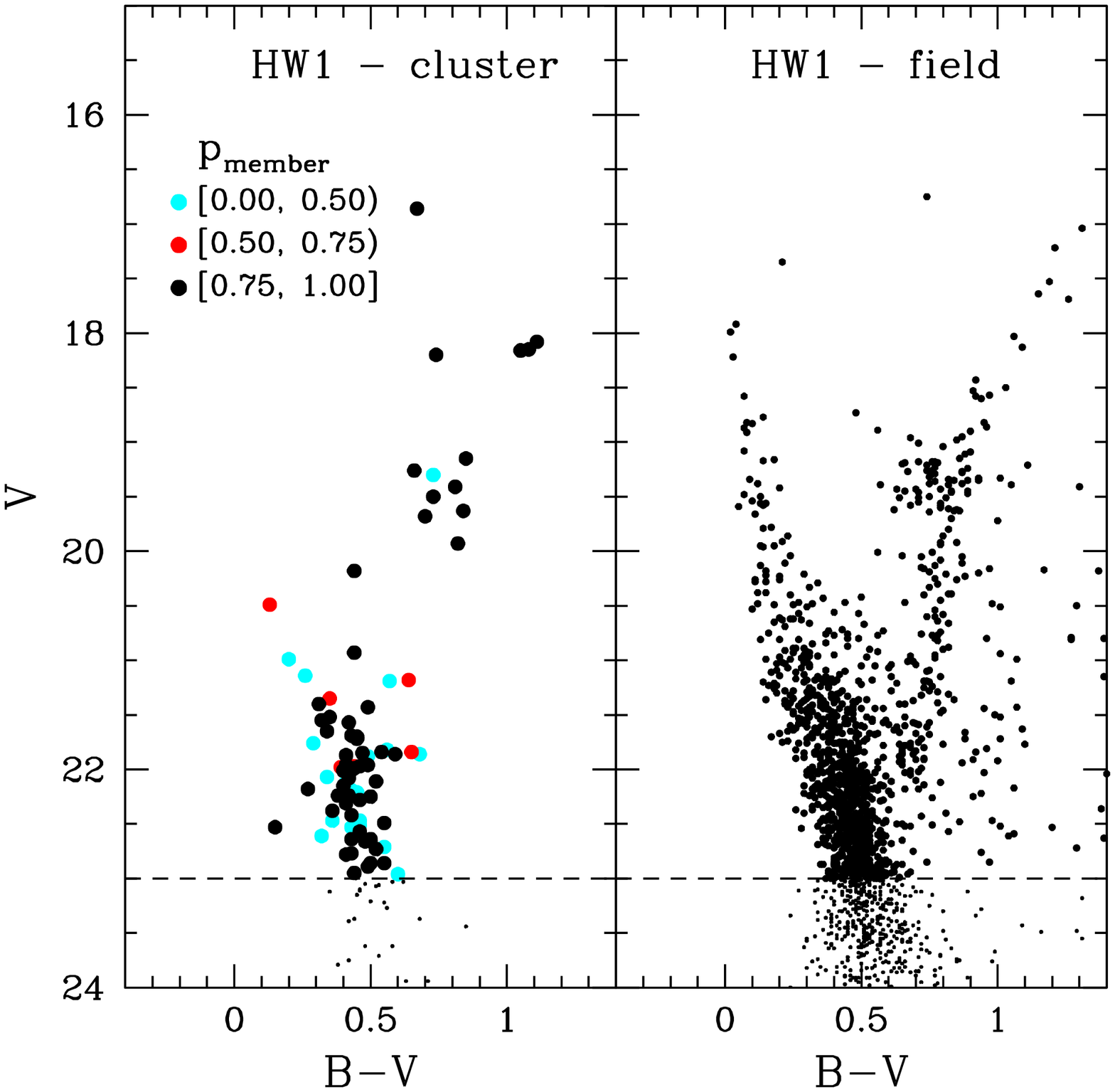}
   \includegraphics[width=0.29\textwidth]{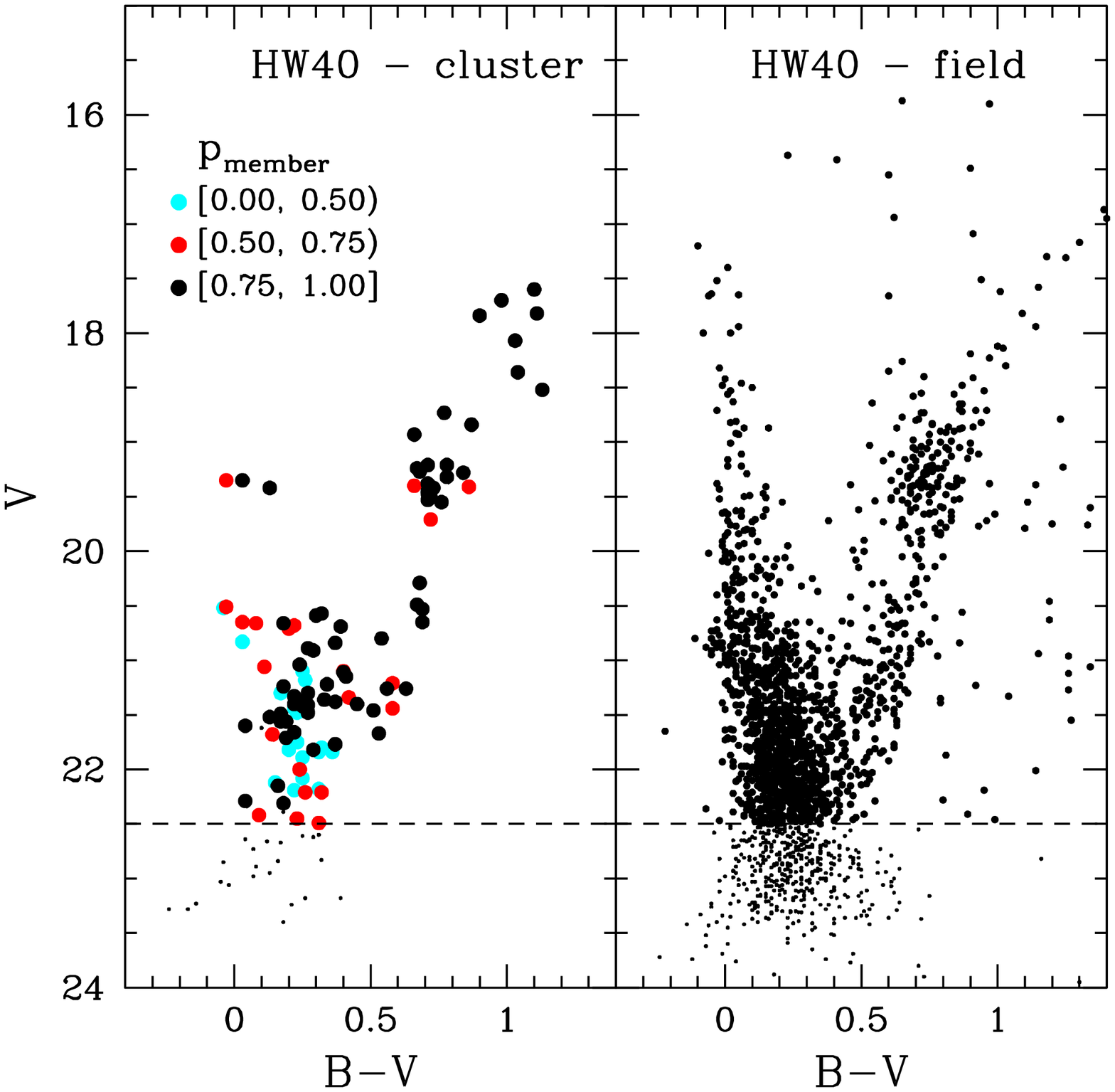}
   \includegraphics[width=0.29\textwidth]{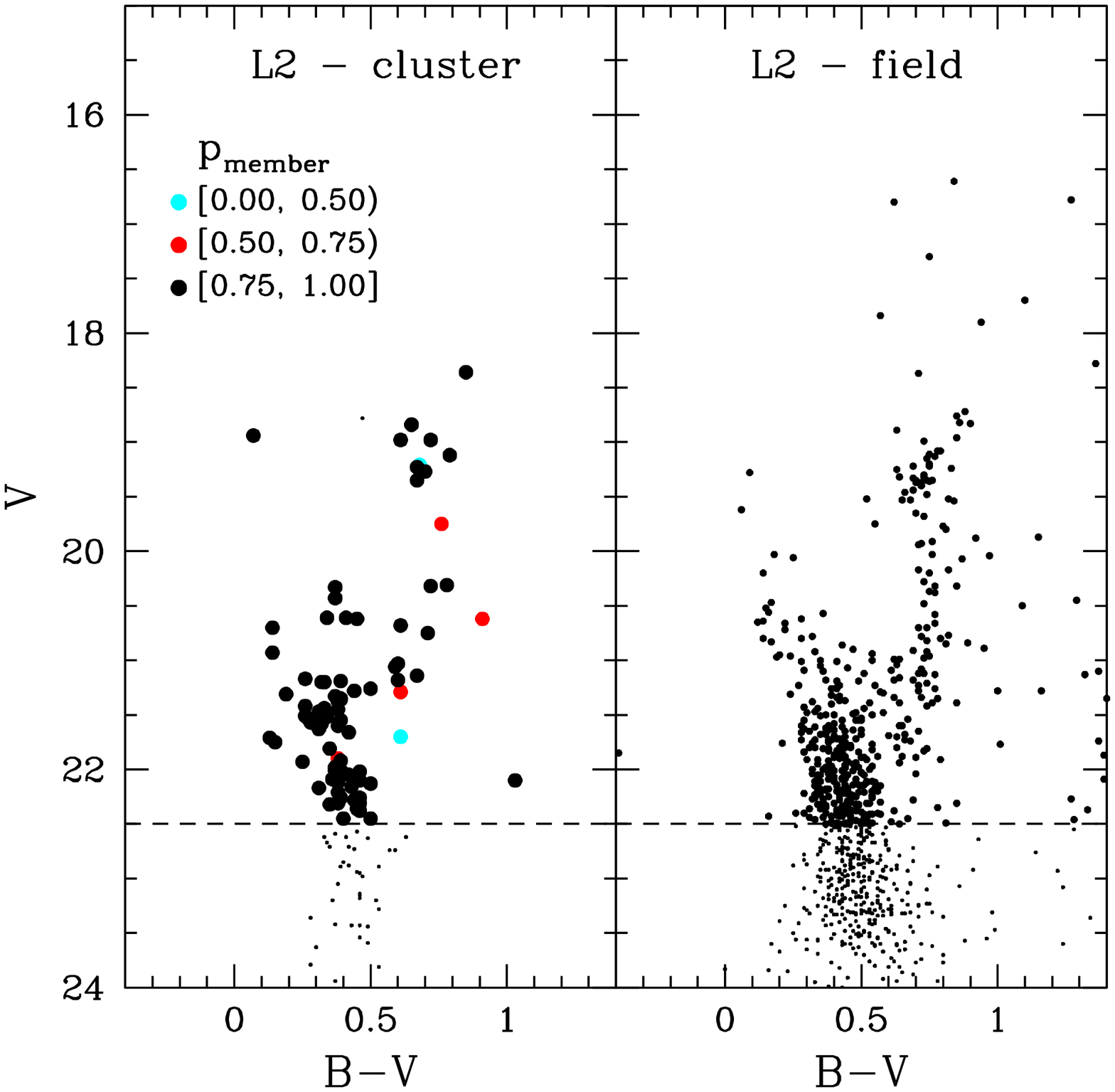}
   \includegraphics[width=0.29\textwidth]{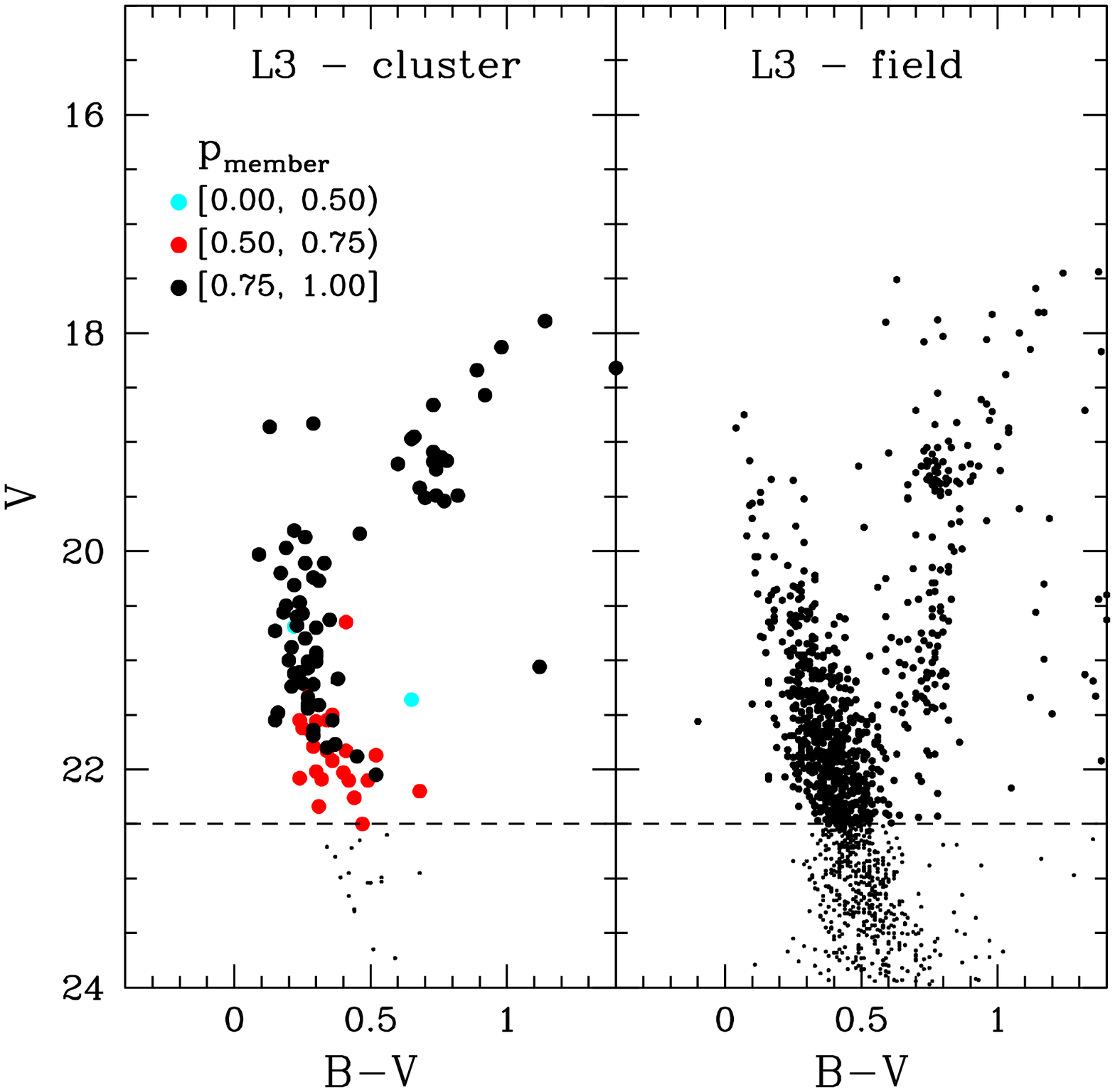}
   \includegraphics[width=0.29\textwidth]{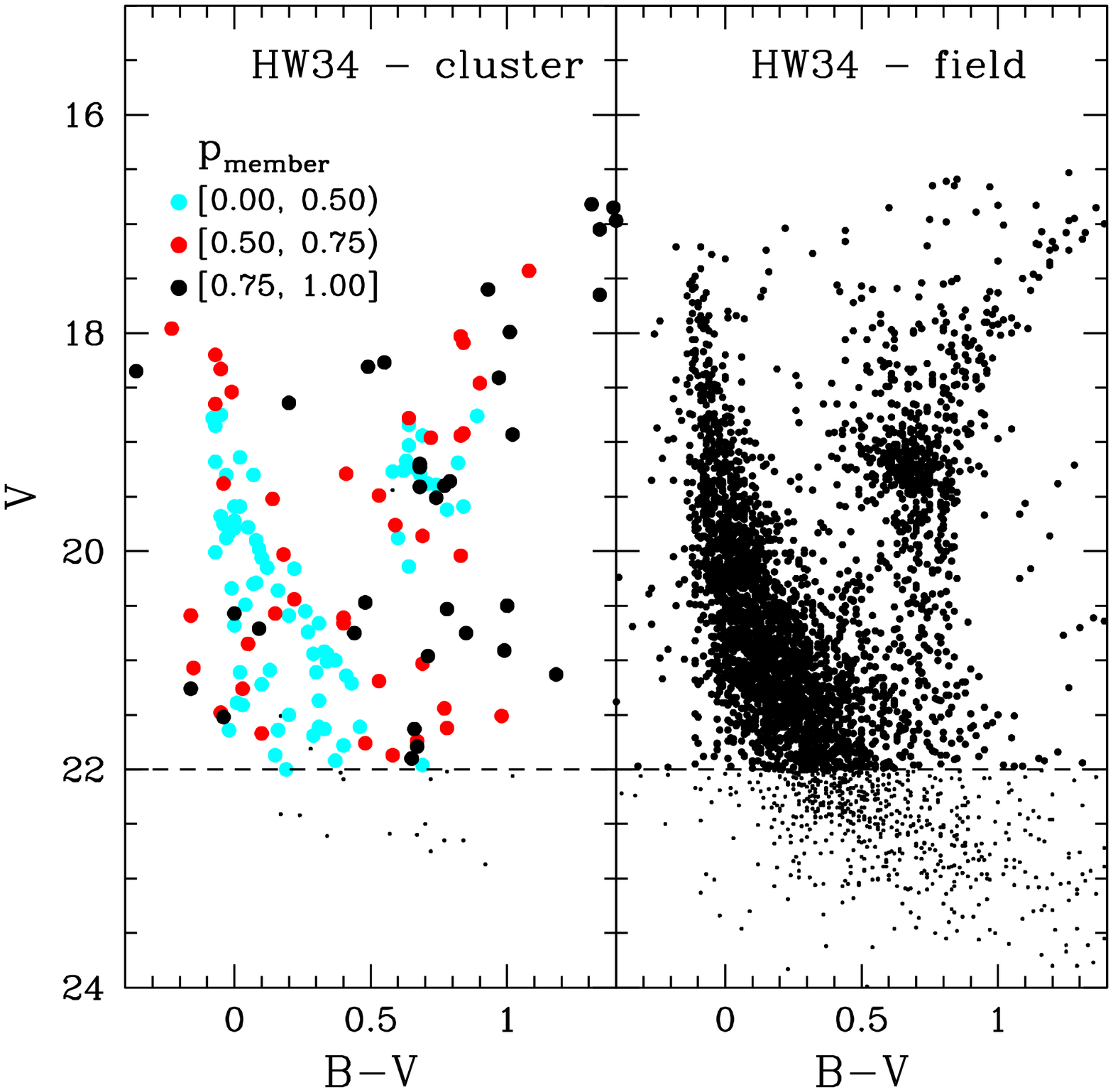}
   \caption{V, (B-V) colour-magnitude diagrams for all clusters.
       Left panels: cluster stars (R $<$ R$_{\rm{clus}}$). Right panels: control field stars (R $>$ R$_{\rm{field}}$).
     The colours depend on the membership probabilities
     (p$_{\rm{member}}$) for each star in the cluster direction (R $<$
     R$_{\rm{clus}}$). The horizontal dashed lines correspond to the
     magnitude limits, derived from completeness curves around R$_{\rm{clus}}$.
   }
   \label{membership-plots}
   \end{figure*}

The observational photometric errors for V and (B-V) as given by DAOPHOT are
presented in Figure \ref{errors}. 
The photometric uncertainties, i.e.,
the standard deviations of the differences between input and output
magnitudes and colours derived from the ASTs, are presented together
with the CMDs (Section \ref{cmds-sec}).

\subsection{The radial density profiles}
\label{profiles-sec}
\begin{figure*}[!htb]
  \centering
  \includegraphics[width=0.29\textwidth]{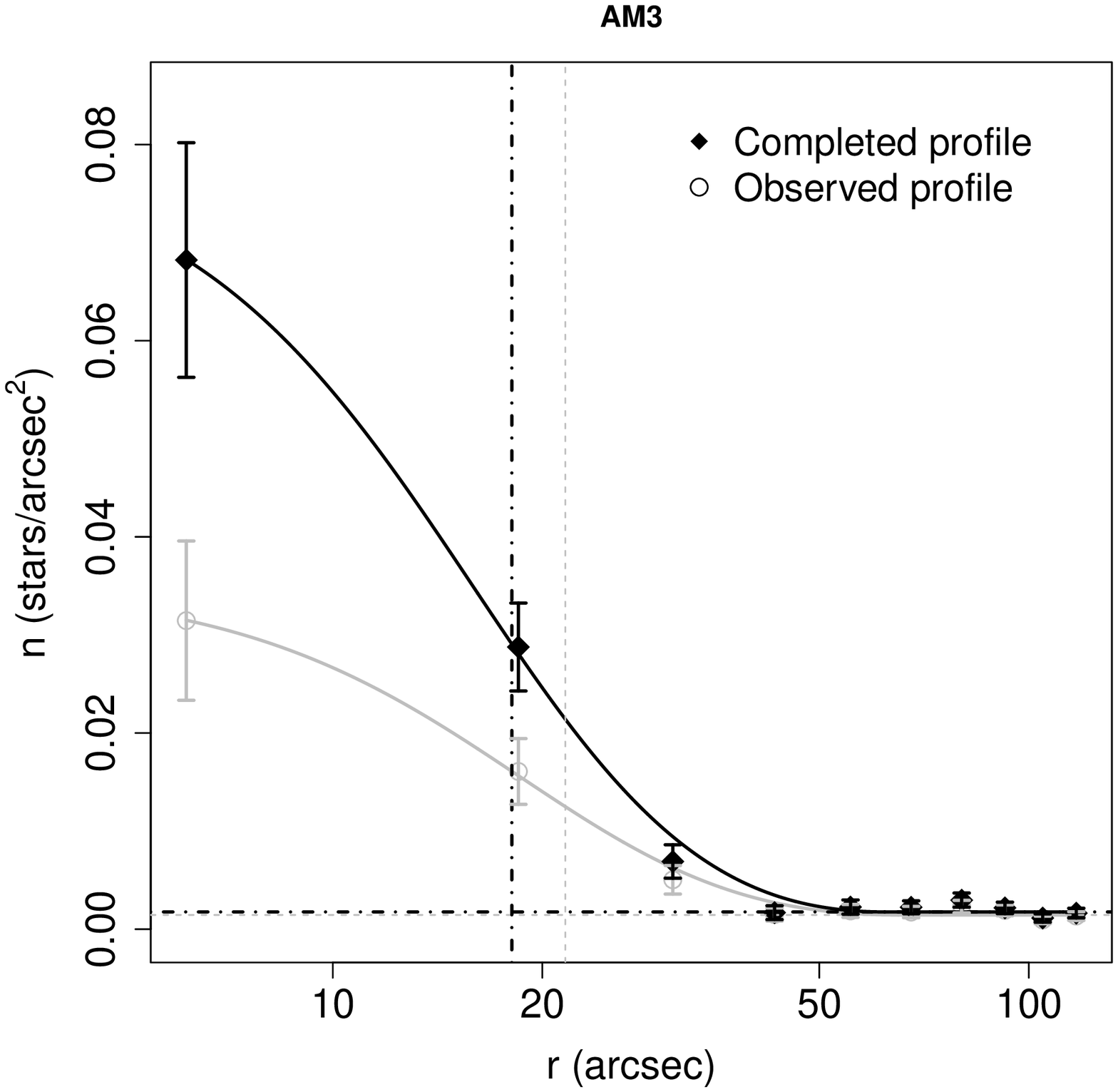}
  \includegraphics[width=0.29\textwidth]{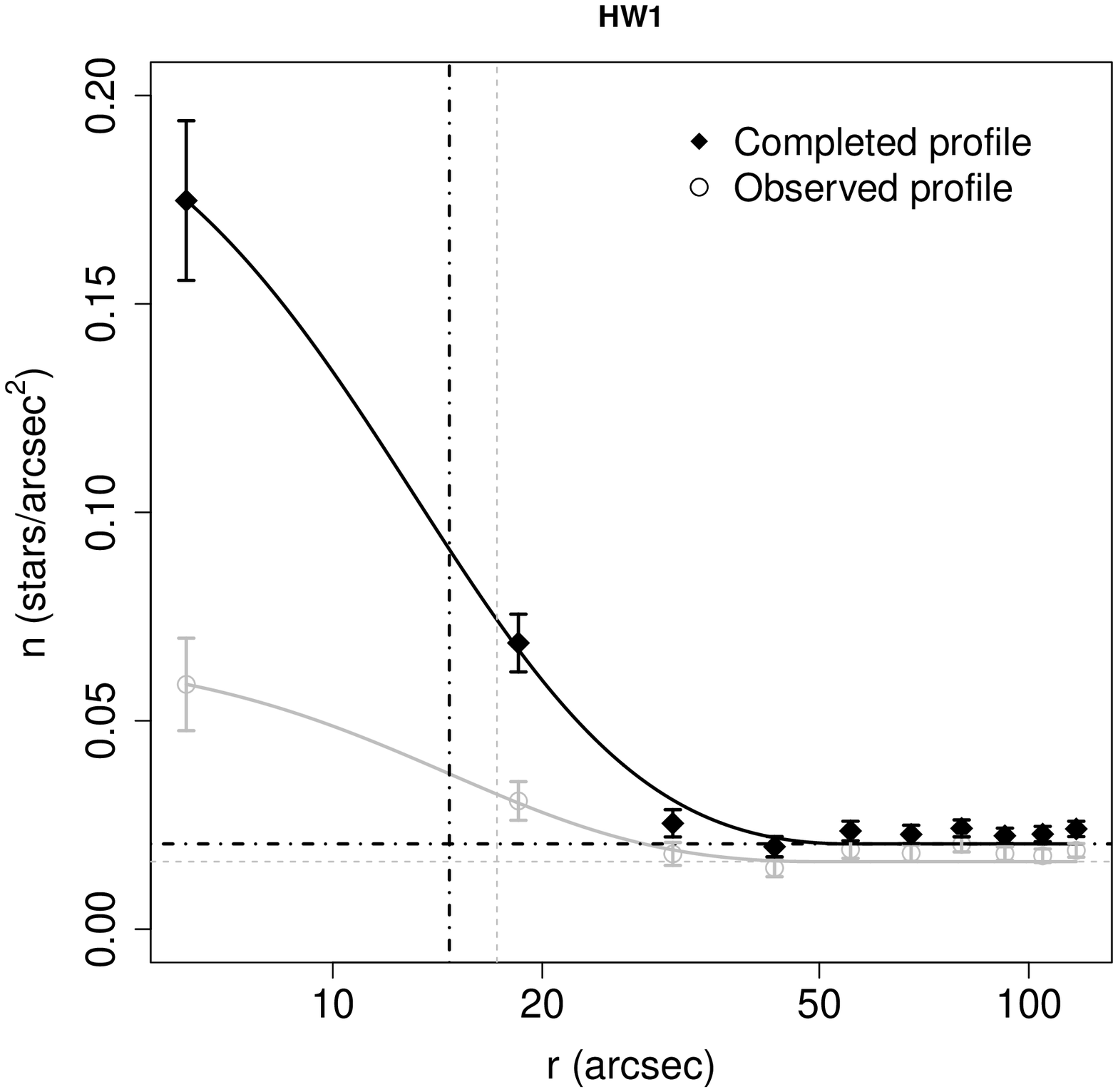}
  \includegraphics[width=0.29\textwidth]{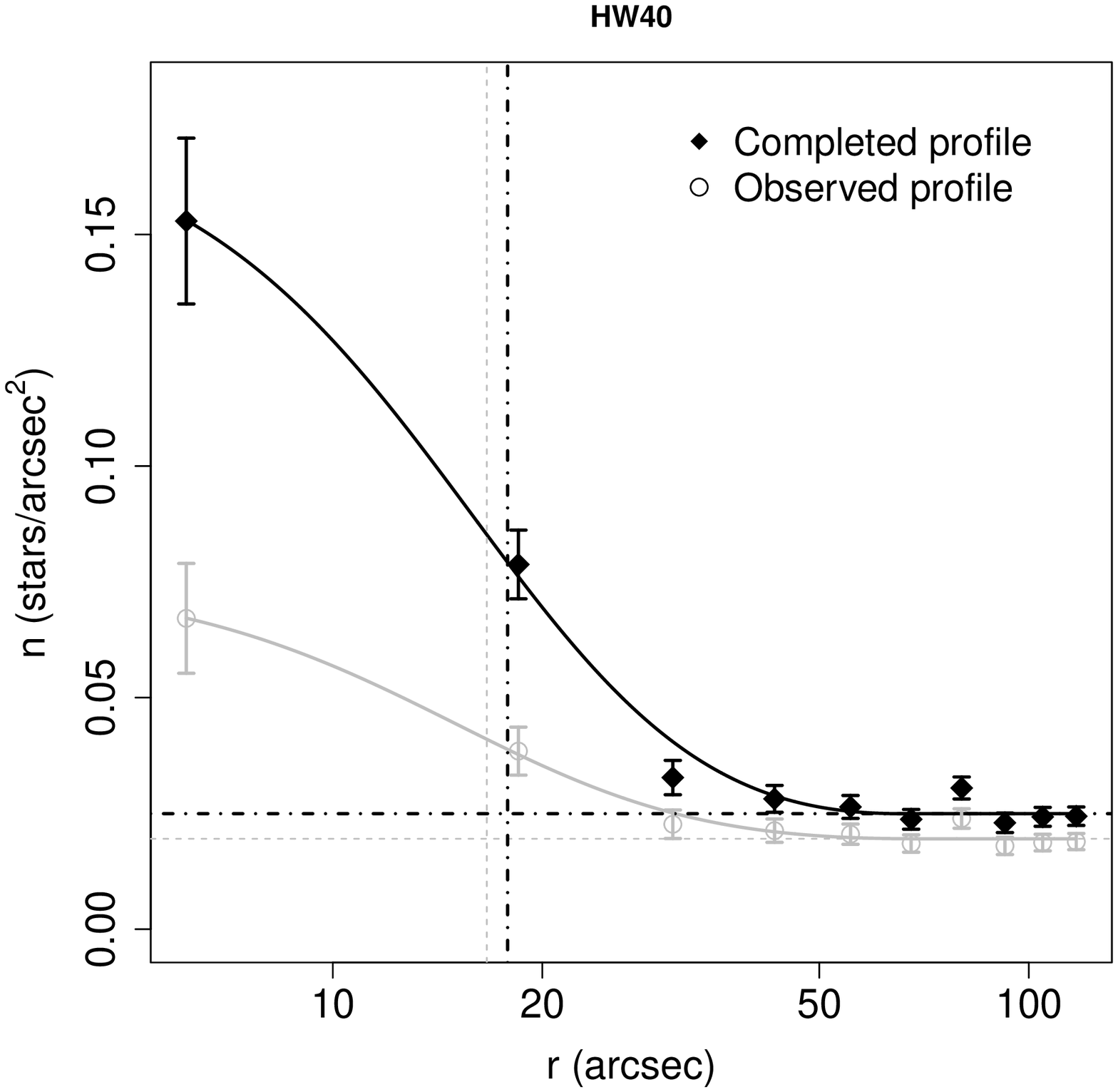}
  \includegraphics[width=0.29\textwidth]{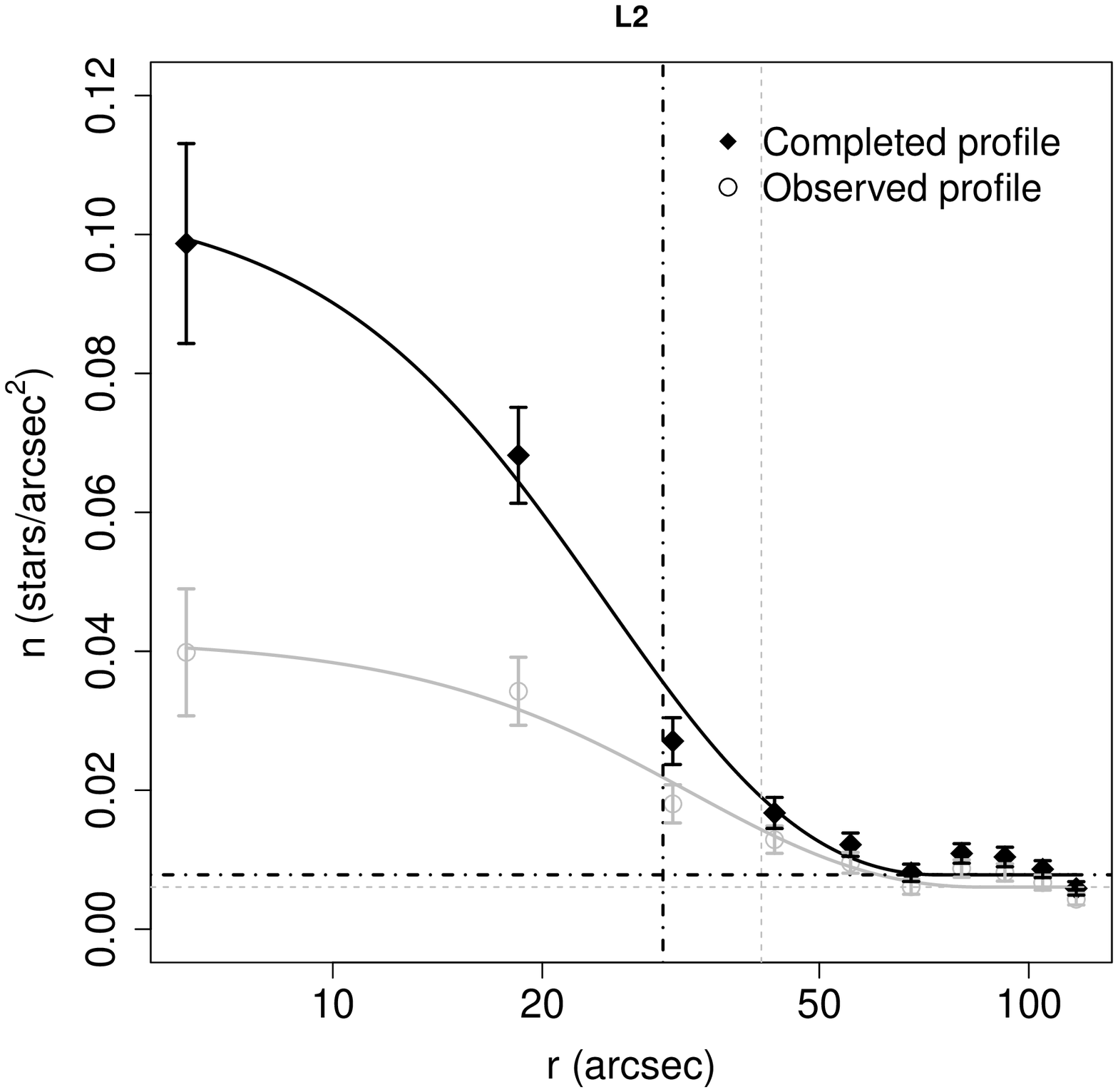}
  \includegraphics[width=0.29\textwidth]{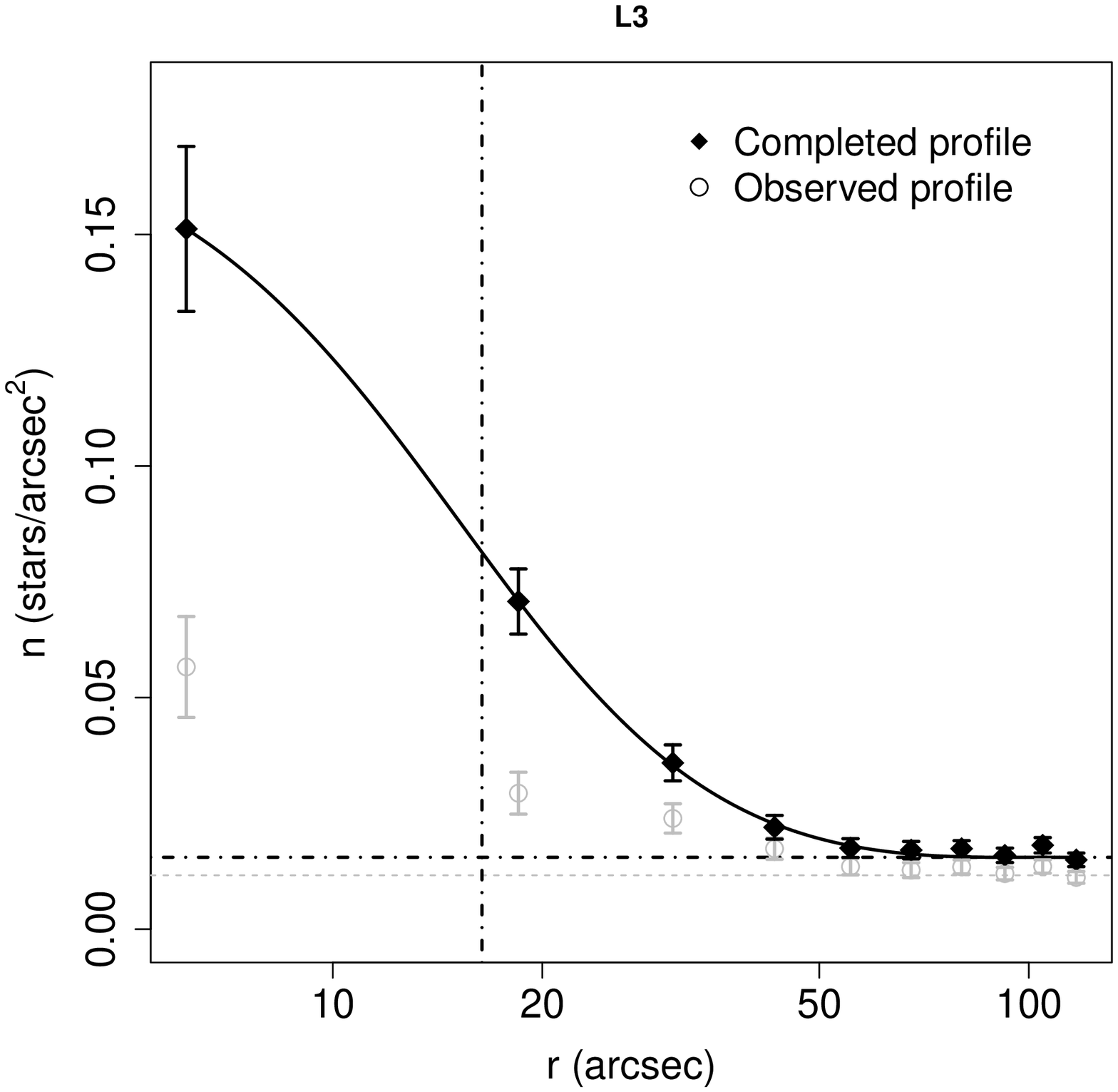}
  \includegraphics[width=0.29\textwidth]{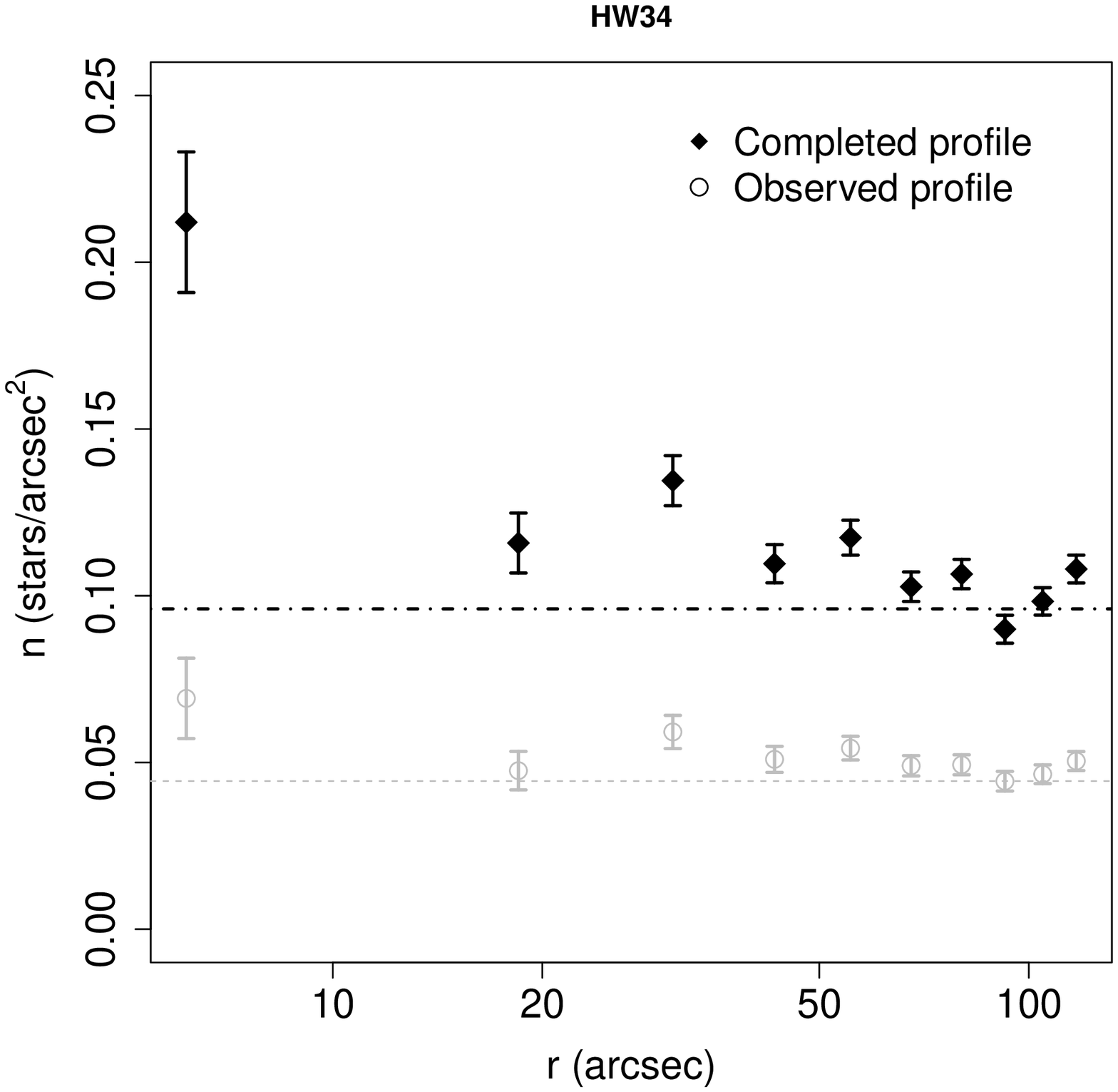}
  \caption{Radial numerical density profiles fitted by King
    profiles. The empty grey circles and filled black diamonds are the
    observed profiles, original and corrected by completeness (Figure
    \ref{completeness}), respectively. Correspondent lines are King
    profiles fitted to the data points. The horizontal and vertical
    dashed grey lines are $n_{\rm field}$ and r$_c$, respectively,
    for the original profile, the dash dotted black lines
    for the corrected profile. HW34 does not present a high enough density
    contrast to be fitted by a King profile.}
  \label{profiles-plots}
\end{figure*}

The radial density profiles for all clusters are shown in 
Fig. \ref{profiles-plots}. Only stars brighter than the magnitude
  limit presented in Fig. \ref{membership-plots} were considered. Two
profiles are presented: a) the lower density, which is observed,
and b) the higher density corresponding to those corrected for
completeness. Each star has a value for the completeness that
  comes from an interpolation in radius and magnitude, based on the curves
  presented in Figure \ref{completeness}.
The error bars are the propagation of the Poissonic errors on the
star counts ($\sqrt{N}$/area), corrected by the incompleteness of the sample.
Each profile
is fitted with the empirical
density law of \cite{king62}, expressed by equation \ref{kinglaw}
below, using a nonlinear least-squares routine

\begin{equation}
n = n_0
\Big\{\frac{1}{[1+(r/r_c)^2]^{1/2}}-\frac{1}{[1+(r_t/r_c)^2]^{1/2}}\Big\}^2
+ n_{\rm field}{\rm ,}
\label{kinglaw}
\end{equation}

\noindent where $n_0$ is the central density of the cluster,
$n_{\rm field}$ is the density of field stars, $r_c$
is the core radius, and $r_t$ is the tidal radius. From these
parameters one can quantify the concentration
parameter $c$, defined as $c \equiv \log(r_t/r_c)$. For each cluster, the
value of $n_{\rm field}$ was assumed to the average density of
stars outside $R_{\rm field}\equiv90\arcsec$, and it was kept constant
in the fits.
The final fits are presented in Fig. \ref{profiles-plots}, whereas all
recovered structural parameters are shown in Table \ref{struct-param}.

The comparison of the panels in Fig. \ref{profiles-plots} and the parameters in 
Table \ref{struct-param} reveal, as would be predicted from a visual analysis
of the sky maps, that the objects present quite different stellar
concentrations and field stellar densities.
In all cases the core radius $r_c < R_{\rm cluster}$ and the tidal
  radius $r_t < R_{\rm field}$, which confirm that the adopted values
  of $R_{\rm cluster}$ and $R_{\rm field}$ are reasonable.
All clusters present roughly the same level of concentration, and
the central density contrast with respect to the field is about 10 to
80 times the field density. We note that the original profile of
Lindsay~2 and both profiles of HW 34 did not converge. The first is due to
incompleteness, and the second does not have enough density contrast
in the centre to be fitted by a King profile,
which can be interpreted as evidence of a non-physical system.


\begin{table*}[!htb]
\caption{Structural parameters derived from King profile fitting, as
  described in Section \ref{profiles-sec}. The overdensity column
  indicates how much bigger the central density is with respect to
  the field density.
{\it Notes}: $a$) 
cases where fit did not converge, and we attribute the
density of the innermost region of the profile as $n_0$ to
estimate a lower limit of the overdensity.}
\label{struct-param}
\centering
\begin{tabular}{lcccccc}
\hline
\noalign{\smallskip}
Cluster & $n_{\rm{field}}$ &  $n_{0}$ & $r_c$ & $r_t$ & $c$ & overdensity \\
 & (stars/$\arcsec^2$) & (stars/$\arcsec^2$) & ($\arcsec$) &  ($\arcsec$) & log($r_t/r_c$) & ($n_0-n_{\rm field}$)/($n_{\rm field}$) \\
\noalign{\smallskip}
\hline
\noalign{\smallskip}
AM\,3 -- original        & 0.0014 & 0.072 $\pm$ 0.009 & 21.6 $\pm$ 1.4 & 65 $\pm$ 6 & 0.48 $\pm$ 0.05  & 50 \\
AM\,3 -- completed    & 0.0018 & 0.150 $\pm$ 0.015 & 18.1 $\pm$ 1.1 & 62 $\pm$ 6 & 0.54 $\pm$ 0.05   & 82 \\
\noalign{\smallskip}
Lindsay\,2 -- original & 0.0061 & 0.11 $\pm$ 0.05 & 41 $\pm$ 8     & 85 $\pm$ 17   & 0.31 $\pm$ 0.12 & 17  \\
Lindsay\,2 -- completed & 0.0078 & 0.25 $\pm$ 0.05 & 29.8 $\pm$ 3.2   & 74 $\pm$ 9   & 0.39 $\pm$ 0.07 & 31  \\
\noalign{\smallskip}
HW\,1 -- original        & 0.0162  & 0.114 $\pm$ 0.009 & 17.2 $\pm$ 1.3       & 49.2 $\pm$ 3.0 & 0.46 $\pm$ 0.04 & 6  \\
HW\,1 -- completed    & 0.0204  & 0.349 $\pm$ 0.021 & 14.7 $\pm$ 0.8      & 55.3 $\pm$ 3.7 & 0.58 $\pm$ 0.04 & 16 \\
\noalign{\smallskip}
Lindsay\,3 -- original$^a$ & 0.0116  & 0.057 $\pm$ 0.011  & ---    & ---  & --- & 3.9 \\
Lindsay\,3 -- completed & 0.0155 & 0.238 $\pm$ 0.010     & 16.4 $\pm$ 0.5     & 89 $\pm$ 8       & 0.73 $\pm$ 0.04 & 14 \\
\noalign{\smallskip}
HW\,40 -- original      & 0.0195   & 0.096 $\pm$ 0.019     & 16.6 $\pm$ 2.3           & 69 $\pm$ 17     & 0.62 $\pm$ 0.12 & 4 \\
HW\,40 -- completed  & 0.0249   & 0.28 $\pm$ 0.04     & 17.8 $\pm$ 1.6           & 65 $\pm$ 9     & 0.56 $\pm$ 0.7 & 10 \\
\noalign{\smallskip}
HW\,34 -- original$^a$      & 0.0444   & 0.069 $\pm$ 0.012 & --- & --- & --- & 0.5 \\
HW\,34 -- completed$^a$  & 0.0960     &  0.212 $\pm$ 0.021 & --- & --- & --- & 1.2 \\
\noalign{\smallskip}
\hline
\end{tabular}
\end{table*}

%
\subsection{The observed CMDs and cluster membership probability}
\label{cmds-sec}

 Figure \ref{membership-plots} shows the V $\times$ B-V CMDs for the
 present sample.  
For each object, the cluster ($R < R_{\rm{clus}}$) and 
field ($R > R_{\rm{field}}$) CMDs are presented,
with stars from V$\sim$17 up to V$\sim$23.
It is possible to identify main sequence (MS) stars as well as 
red giant stars belonging to the subgiant branch (SGB), 
red giant branch (RGB), and to the red clump (RC).  
The comparisons with the field CMDs confirm these stellar identifications.
We note that the field CMDs are much richer in stars than the 
 cluster CMDs, as a natural consequence of the ratio between their
covered areas ($\frac{\Omega_{\rm{field}}}{\Omega_{\rm{clus}}} \sim 26$).

Since the stellar samples in the direction of the clusters are contaminated by 
SMC field stars, a procedure to determine the cluster membership probability
($p_{\rm{member}}$) was applied. 
In summary, the adopted procedure follows the one from \cite{kerber+05}
and \cite{alves+12}, where $p_{\rm{member}}$ for each cluster 
star is determined by comparing the density of stars inside the 
CMD in the cluster direction ($R~<~R_{\rm{clus}}$), 
with a CMD representative of the local SMC field ($R~>~R_{\rm{field}}$)
(see Fig. \ref{membership-plots}). 
The cluster and field CMDs are divided into a grid of small boxes 
in V magnitude and B-V colour, centred on each cluster star.
So, for the i$^{\rm{th}}$ cluster star, the number of cluster 
($N_{\rm{clus},i}$) and field stars ($N_{\rm{field},i}$) contained in
a box in magnitude and colour  
(with a 3-sigma size in photometric errors) are computed, 
taking the inverse of the completeness into account.  
Finally, the cluster membership probability for the i$^{\rm{th}}$ 
cluster star is given by

\begin{equation}
p_{\rm{member},i}= 1 - \frac{N_{\rm{field},i}}{N_{\rm{clus},i}} \times
\frac{\Omega_{\rm{clus}}}{\Omega_{\rm{field}}}{\rm .}
\label{p_member-eq}
\end{equation}

The results of these determinations are shown in
Fig. \ref{membership-plots} in a colour scale.
The analysis of this figure reveals that most probable cluster members
($p_{\rm{member}} > 75\%$) are red giants or stars in the upper MS.
In particular, almost all stars in the AM\,3 cluster sample can be considered
physical cluster members. On the other hand, the MS stars in the HW\,34 
cluster direction have a higher probability of being field stars rather than cluster
stars since they present $p_{\rm{member}} < 50\%$.
This strongly reinforces the argument that this stellar concentration is not
a physical system, but only a field stellar fluctuation 
(see also Sect. \ref{profiles-sec}). 
Table \ref{stellarcounts} summarizes the star counts in the sample CMDs.


\begin{table*}[!htb]
\caption{Stellar counts in the sample CMDs. Columns correspond to: the
  cluster's name, the number of stars without the completeness
  corrections and with them for the cluster (N$_{\rm
    clus}^{\rm (obs)}$ and N$_{\rm clus}^{\rm (comp)}$) and field
  direction (N$_{\rm field}^{\rm (obs)}$ and N$_{\rm field}^{\rm
    (comp)}$), as well as the expected number of field stars  
  in the completed cluster sample, with p$_{\rm member} < 0.5$
  (contamination $C$). The
  last column is the number of cluster stars in
  terms of the field standard deviation (assumed as Poissonian),
  given by the expression $N_{\sigma}=\big(N_{\rm clus}^{\rm (comp)}-C\big)/\sqrt{N_{\rm field}^{\rm (comp)}}$.}
\label{stellarcounts}
\centering
\begin{tabular}{lcccccc}
\hline \noalign{\smallskip}
Target  & N$_{\rm clus}^{\rm (obs)}$ & N$_{\rm clus}^{\rm (comp)}$ & N$_{\rm field}^{\rm (obs)}$ & N$_{\rm field}^{\rm (comp)}$ & $C$ & N$_{\sigma}$ \\
\noalign{\smallskip}
\hline
\noalign{\smallskip}
AM\,3 & 56 & 172 & 135 & 209 & 8 & 11  \\
Lindsay\,2  & 127 & 490 & 847 & 2434 & 93 & 8.1 \\
HW\,1 & 112 & 370 & 1659 & 3686 & 141 & 3.8 \\
Lindsay\,3  & 116 & 342 & 1412 & 3818 & 145 & 3.2\\
HW\,40 & 133 & 393 & 1906 & 3790 & 143 & 4.1 \\
HW\,34  & 171 & 562 & 4014 & 13594 & 515 & 0.4  \\
\noalign{\smallskip}
\hline
\end{tabular}
\end{table*} 

%

\section{Statistical isochrone fitting}
\label{isot_fitting-sec}

The isochrone fitting used in this work follows a numerical and statistical
approach that combines CMD modelling and an objective criterion of
comparing synthetic CMDs with the observed ones. 
It was initially developed to analyse CMDs of rich LMC 
clusters, observed with the Hubble Space Telescope (HST)
 \citep{kerber+02}, and recently it was applied to 
determining physical parameters of Galactic open clusters imaged by
2MASS \citep{alves+12}.
The central idea behind this method is to statistically determine  
the synthetic CMDs that best reproduce the observed CMD, 
recovering as a consequence the physical parameters for the stellar cluster.
In the present work the statistics adopted to select the best models 
is the likelihood, in particular following a Bayesian approach.
This approach has been successfully applied to analysing
CMDs of stellar clusters \citep{NJ06, HVG08, MDC10}
as well as CMDs of composite stellar populations
\citep{HGVG00, HGVG99, vergely+02}.

The generation of synthetic CMDs, as well as the likelihood statistics,
are detailed in the next subsections.

\subsection{CMD modelling}

\begin{figure}[!hb]
\centering 
\includegraphics[width=1.0\columnwidth]{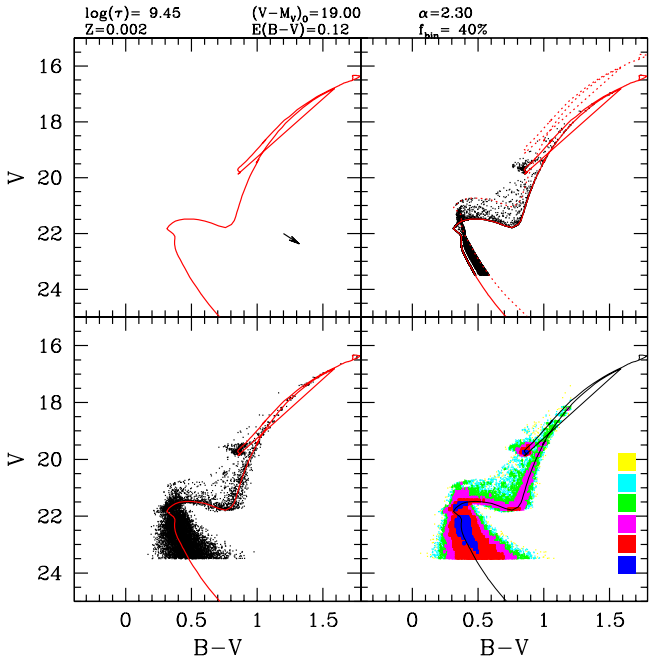}
\caption{The generation of synthetic V vs. B-V 
CMDs. 
{\it Panel a}: the adopted isochrone shifted by distance modulus and reddening.
{\it Panel b}: the distribution of stars in accordance to the IMF and the fraction
of binaries; 
{\it Panel c}: the introduction of photometric errors in magnitude and colour.  
{\it Panel d}: colours following the density of points in the CMD in a logarithmic scale. 
The density of points is related to $p_{\rm{CMD,i}}$ (see
Sect. \#\ref{statistics} ). 
The reddening vector and the values for all physical parameters are shown in the figure.}
\label{modelling-plot}%
\end{figure}

The synthetic CMDs are generated considering that they are 
SSPs, characterized by stars with the same 
age ($\tau$) and metallicity ($Z$). 
The basic steps to generate a specific synthetic CMD
are the following (see Fig. \ref {modelling-plot}):

\smallskip
\noindent
(i) adoption of an evolutionary stellar model. 
The new Padova-Trieste models (PARSEC isochrones, \citealp{bressan+12}) were adopted in this work;
\smallskip

\noindent
(ii) selection of the age and metallicity of the SSP, which is equivalent to
selecting an isochrone from the library;
\smallskip

\noindent
(iii) application of the distance modulus $(m-M)_{0}$
and reddening $E(B-V)$ values to shift the isochrone 
from the absolute magnitudes to the observed ones;
\smallskip

\noindent
(iv) generation of 10$^5$ synthetic stars belonging to stellar systems (single and binary stars) 
following an IMF and a fraction of unresolved binaries ($f_{\rm{bin}}$).
A Salpeter IMF ($dN/dm \propto m^{-2.35}$) and 
$f_{\rm{bin}}=30\%$ were adopted, with a minimum mass ratio 
($q_{\rm{min}}=m_{2}/m_{1}$) of 0.7 and a constant 
mass-ratio distribution ($dN/dq = const.$);
\smallskip

\noindent
(v) introduction of the observational photometric errors 
in magnitude and colour.
\smallskip

To determine the cluster parameters, a wide and regular grid of models 
was build for each cluster, composed typically of $\sim$ 9000 
combinations in log($\tau$/yr), $Z$, $(m-M)_{0}$ and $E(B-V)$ centred
on the parameters found by a visual isochrone fit.
To take into account the line-of-sight depth and the reddening values 
found for the SMC stellar clusters \citep{glatt+08b,crowl+01},
the grids of models span distance modulii
{and reddening values in the range 18.50 $<$ $(m-M)_{0}$ $<$ 19.20
and 0.00 $<$ $E(B-V)$ 0.20}, in steps of 0.05 and 0.01,
respectively. For log($\tau$/yr) the step is 0.05, and for $Z$ all values provided by the isochrones are considered: 0.0001, 0.0004, 0.001, 0.002, 0.004,
0.006, 0.008, 0.010, 0.012 and 0.014.

\subsection{Likelihood statistics}
\label{statistics}

The first step in computing the likelihood for a synthetic CMD 
is to establish the probabilities for each observed star cluster 
that it belong to that choice of SSP ($p_{\rm{CMD}}$).
In general, this is done  using analytical expressions
\cite[e.g.][]{HVG08,MDC10},
assuming that the observed CMD positions would represent stars scattered
from the isochrones following Gaussian distributions for the photometric 
errors. 
However, in the present work these probabilities are computed 
assuming that they are proportional to the density of points generated 
by the synthetic CMD in each of the respective CMD positions ($N[V,(B-V)]$),
which naturally incorporates the effect of
unresolved binaries and photometric uncertainties.
This density map on the CMD is 
commonly called a Hess diagram, even though its application is not 
directly related with the likelihood statistics.
Therefore, the likelihood statistics in this numerical approach is given 
by the product of these probabilities over all the $N_{\rm{clus}}$ 
observed cluster stars as stated by the expression

\begin{equation}
\rm{L} \propto 
\prod_{i=1}^{N_{\rm{clus}}} p_{\rm{CMD},i}
\times p_{\rm{member},i} 
\propto 
\prod_{i=1}^{N_{\rm{clus}}} N[V_i,(B-V)_i] 
\times p_{\rm{member},i}~,
\label{likelihood}
\end{equation}

\noindent where the $i$ index corresponds to the i$^{th}$ observed cluster star.
It can thus be stated that the best model is the one that provides
the synthetic CMD/Hess diagram that maximizes the above expression
($\rm{L_{max}}$).
The final parameters for each cluster, as well as corresponding uncertainties, 
are determined by computing the average and standard deviation over the 
physical parameters of the set of models that have likelihood values
at the 1-sigma level from $\rm{L_{max}}$ (see Appendix A for details).

%
\section{Results}
\label{results}

\begin{figure*}[!htb]
   \centering
   \includegraphics[width=0.3\textwidth]{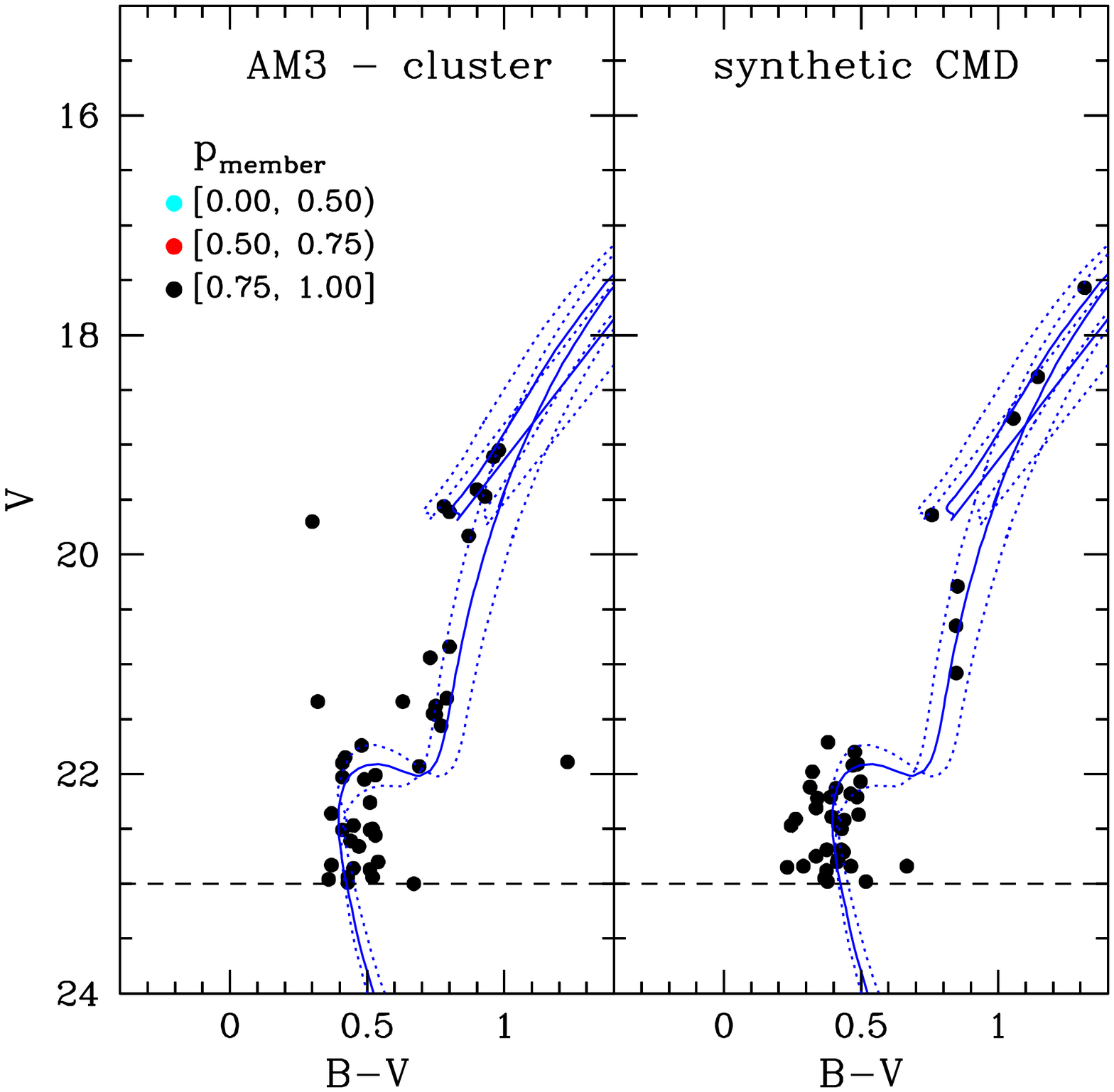}
   \includegraphics[width=0.3\textwidth]{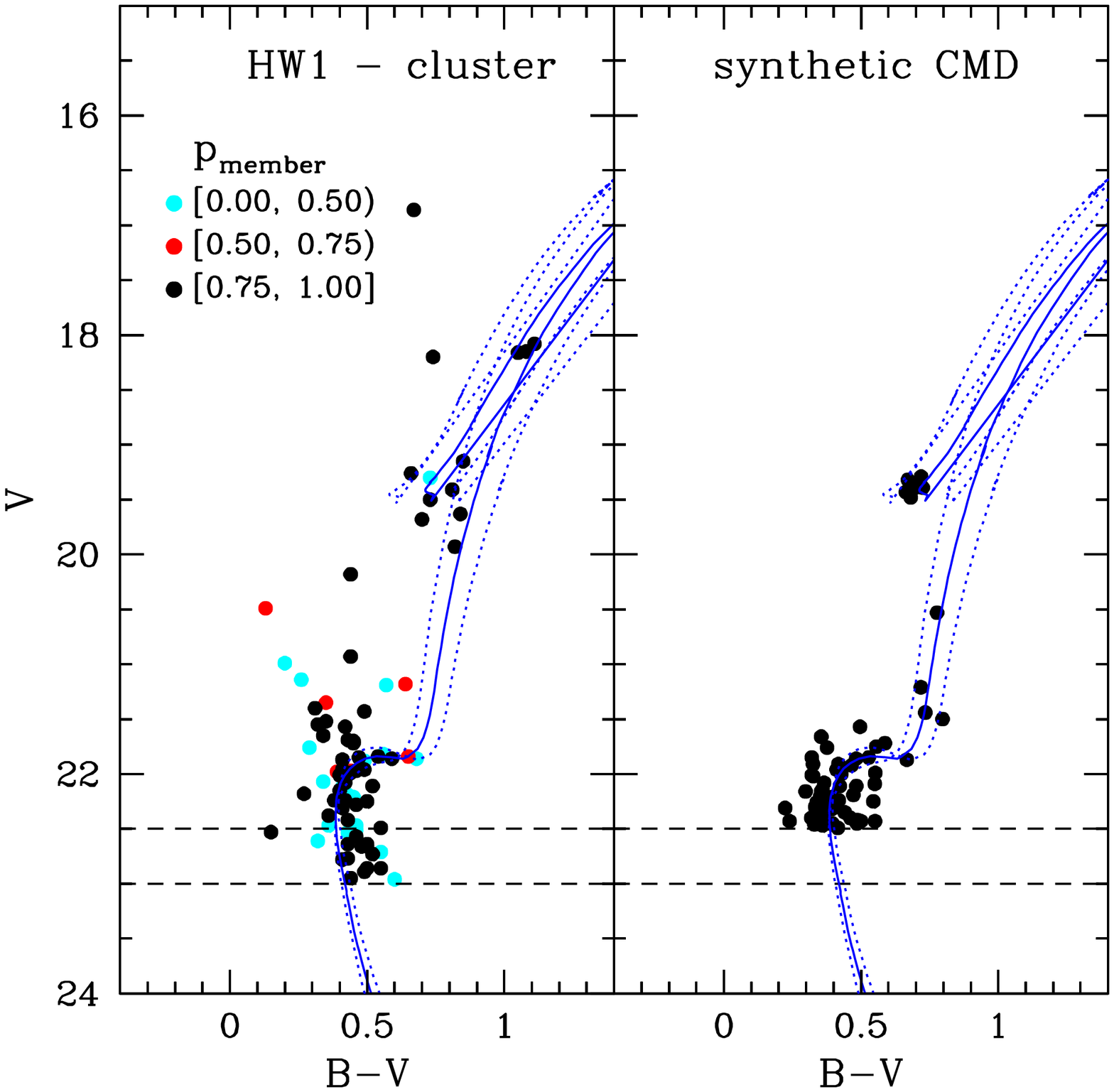}
   \includegraphics[width=0.3\textwidth]{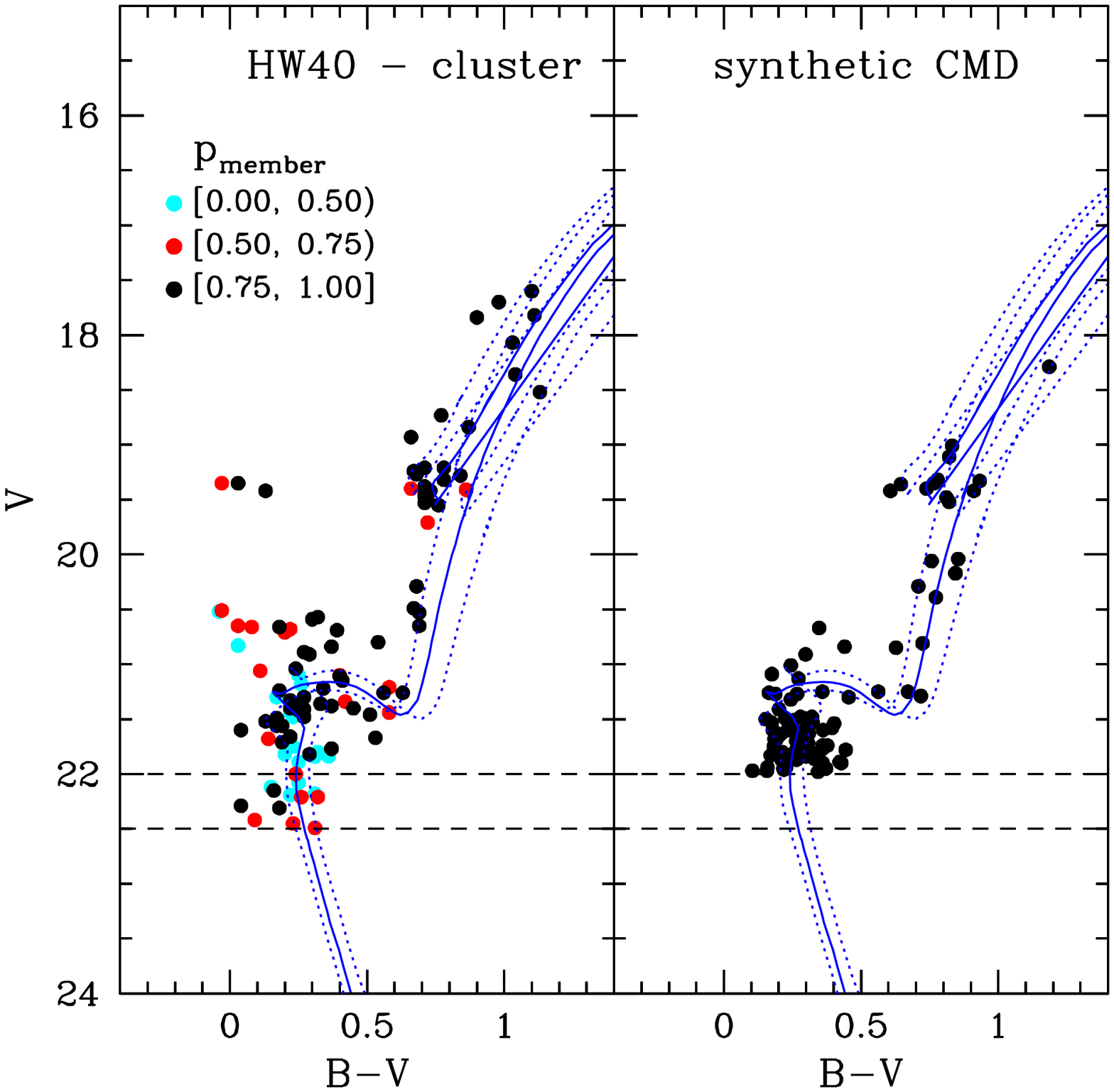}
   \includegraphics[width=0.3\textwidth]{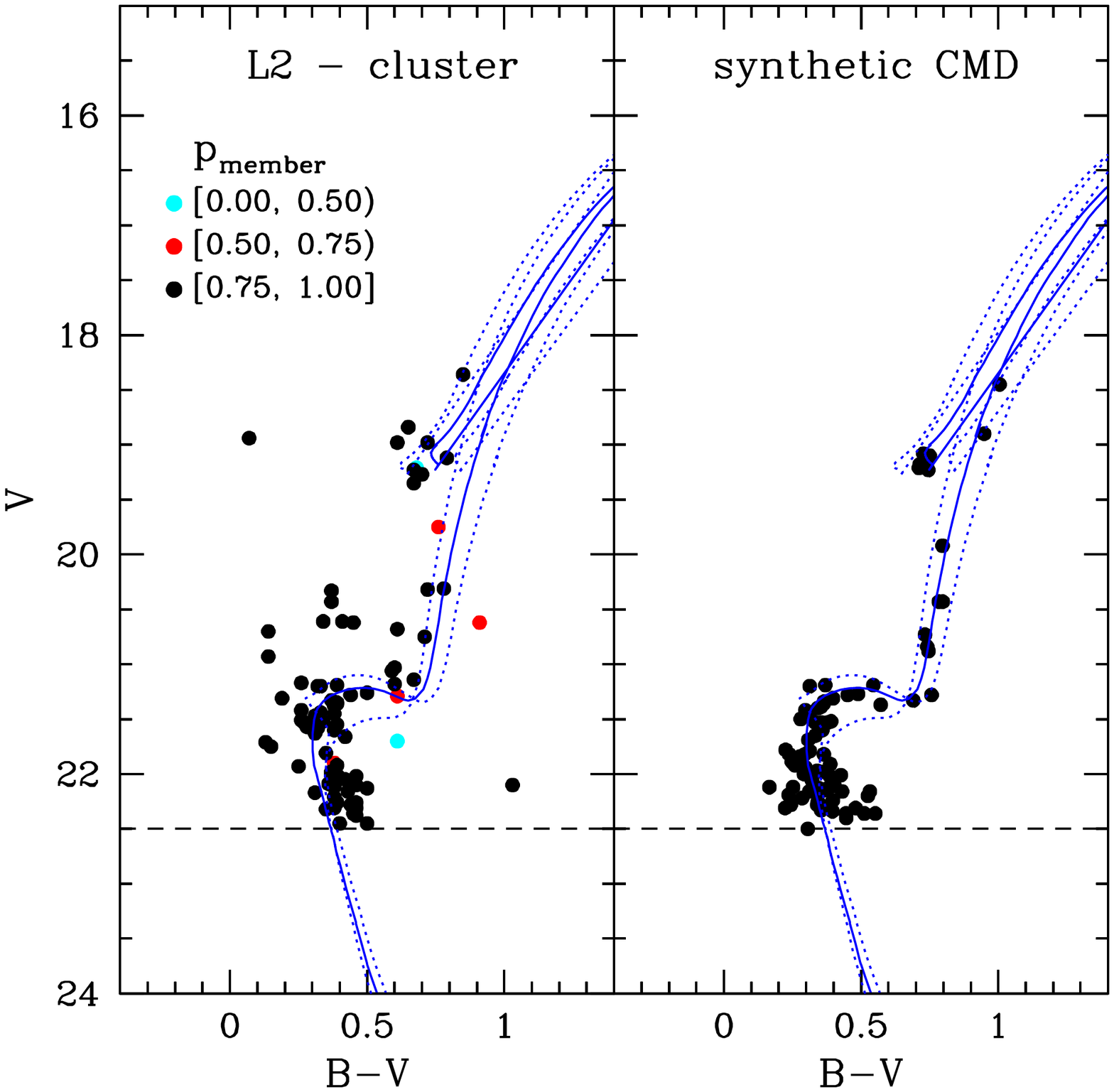}
   \includegraphics[width=0.3\textwidth]{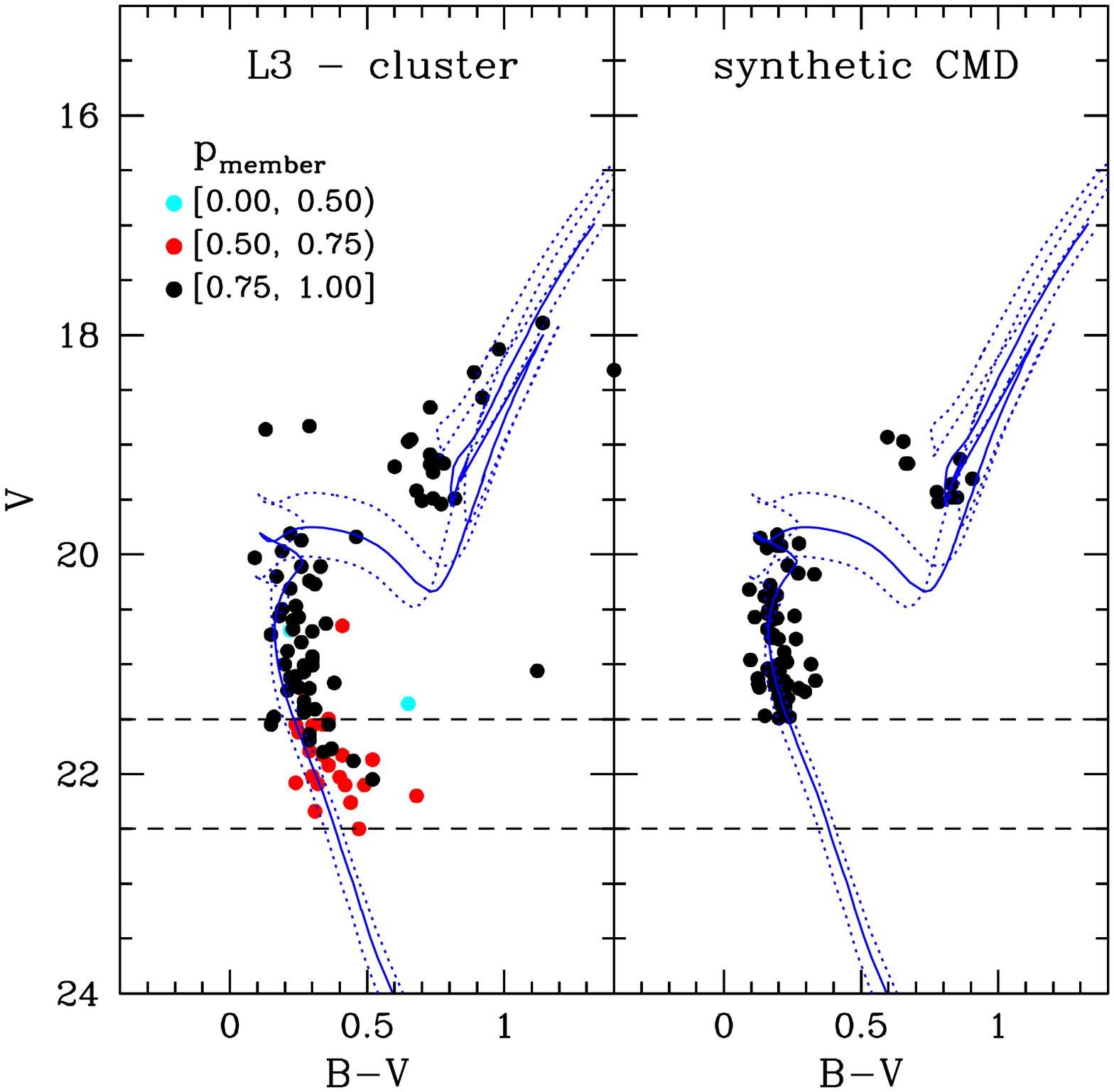}
   \caption{
Best isochrone fittings for all clusters.
Left panels: cluster stars with colour symbols in accordance
to the membership probabilities (p$_{\rm{member}}$).
Right panels: synthetic CMDs generated with the parameters found 
for the best solution. The number of points is equal to the observed
CMDs within the same magnitude limits.
The horizontal dashed black lines correspond to the magnitude limits used to 
compute the likelihood (brighter mag) and p$_{\rm{member}}$ (fainter
mag). The solid and dashed blue lines correspond to the isochrones with the parameters found in
Table \ref{finalparam} and take into account the uncertainties on
ages and metallicities. 
}
   \label{bestfits-plots}
   \end{figure*}

The results of the isochrone fittings for the sample clusters are presented in 
Figure \ref{bestfits-plots} and Table \ref{finalparam}. The panels in
Fig. \ref{bestfits-plots} 
  give isochrones corresponding to the best-fit parameters, together with another
  two isochrones indicating 1-$\sigma$ uncertainties in age and
  metallicity: one younger and more metal-rich, and the other older
  and more metal-poor. The three isochrones cover most of the data
  points, as expected, allowing the reader to visually inspect the
  quality of the fits. 
For the [Fe/H] derivation we adopted a solar metallicity Z=0.0152
\citep{caffau+11} in order to be compatible with the value
assumed in the PARSEC isochrones \citep{bressan+12}. 
These results are discussed cluster by cluster bellow.


\begin{table*}[!htb]
\caption{Physical parameters determined in this work. Columns refer to: cluster name,
  logarithm of the age, age, metallicity Z and [Fe/H] (assuming Z$_{\odot}$=0.0152,
\citealp{caffau+11}), distance modulus, distance, reddening, and
semi-major axis corresponding to the distance of the cluster to the
centre of SMC (as explained in Section \ref{discussion}).}
\label{finalparam}
\centering
\renewcommand{\arraystretch}{2.0}
\begin{tabular}{lcccccccc}
\hline 
Name & log($\tau/yr$) & Age(Gyr) & Z & [Fe/H] & (m-M)$_{0}$ & d(kpc) &
E(B-V)  & $a$($^\circ$) \\
\hline
AM~3  &  9.69  $\pm$  0.15  &  4.9$^{+2.1}_{-1.5}$  &  0.0022  $\pm$  0.0016  &  -0.8$^{+0.2}_{-0.6}$  &  18.99  $\pm$  0.16  &  63.1$^{+1.8}_{-1.7}$  &  0.08  $\pm$  0.05  &  7.3 \\
Lindsay~2  &  9.60  $\pm$  0.09  &  4.0$^{+0.9}_{-0.7}$  &  0.0007  $\pm$  0.0004  &  -1.4$^{+0.2}_{-0.2}$  &  18.68  $\pm$  0.14  &  54.4$^{+1.5}_{-1.5}$  &  0.09  $\pm$  0.04  &  3.9 \\
HW~1  &  9.70  $\pm$  0.12  &  5.0$^{+1.5}_{-1.2}$  &  0.0013  $\pm$  0.0007 & -1.1$^{+0.2}_{-0.3}$  &  18.84  $\pm$  0.16  &  58.7$^{+1.6}_{-1.6}$  &  0.07  $\pm$  0.04  &  3.4 \\
Lindsay~3  &  9.07  $\pm$  0.11  &  1.2$^{+0.3}_{-0.3}$  &  0.0057  $\pm$  0.0020  &  -0.4$^{+0.1}_{-0.2}$  &  18.64  $\pm$  0.14  &  53.4$^{+1.5}_{-1.5}$  &  0.07  $\pm$  0.04  &  2.9 \\
HW~40  &  9.41  $\pm$  0.06  &  2.5$^{+0.4}_{-0.3}$  &  0.0018  $\pm$  0.0005  &  -0.9$^{+0.1}_{-0.2}$  &  19.08  $\pm$  0.14  &  65.6$^{+1.8}_{-1.8}$  &  0.03  $\pm$  0.03  &  2.0 \\
\hline
\end{tabular}
 \end{table*}

\subsection{AM\,3}

Although this is the cluster with the smallest number of stars in our sample,
 it has the advantage of being located in the SMC field with the lowest stellar
density.
The results in the present work indicate an age of $4.9^{+2.1}_{-1.5}$ Gyr.
The recovered metallicity, [Fe/H]$= -0.8^{+0.2}_{-0.6}$, is 
consistent with the value predicted by the SMC chemical evolution model 
from \cite{PT98} for this age.
\citet{piatti11a} also determined physical parameters for this cluster,
 based on the analysis of a CMD built with Washington photometry. 
Taking the uncertainties into account, our results are 
in good agreement with the ones from Piatti et al., although
they derived slightly older ages ($6.0\pm1.0$ Gyr) 
and more metal poor values ([Fe/H]$=-1.25\pm0.25$).
Our results are also close to the parameters derived by \cite{dacosta99}
of [Fe/H]=-1.27 and age of 5.5$\pm$0.5 Gyr.

The differences in age can be explained by the distance modulus. While
we fitted a (m-M)$_0$ = 19.00 for this cluster, \cite{piatti11a}
assumed an average value of 18.90. With a distance
  modulus $\sim$0.1 lower, the isochrone should be $\sim$0.6~Gyr
older to fit the data.

This western SMC cluster was considered a galaxy member by
  \cite{dacosta99} because it is close to the projected plan of the
  SMC, and its red clump magnitude (distance indicator) is comparable
  to other SMC clusters.
It is located in the West Halo region, but it is not as old and
  metal poor as one could expected in the case of gradients in age and
  [Fe/H] for this group (see Sect. \ref{discussion}). Although AM~3 is
  close to the SMC main body in the projected plan, its distance is higher than the
  other West Halo clusters (HW~1, Lindsay~2, and
  Lindsay~3). Therefore its age, metallicity, and distance do not follow
  exactly the general characteristics of the West Halo, which
  indicates a more complex history, possibly a capture by the
  potential well of the SMC.

\subsection{HW\,1 and Lindsay\,2}
The CMD analysis for these clusters which have never been published before,
 identifies these objects as intermediate-age
metal-poor SMC clusters, with similar metallicities ([Fe/H]$\sim -1.1$
and $-1.4$, respectively) and ages (4.98 and 4.0 Gyr).
In particular, the age and metallicity values for HW\,1 confirm the previous results
from the integrated spectral analysis performed by \cite{dias+10}.
We note that these two clusters are located close to each other, in the
West Halo region (Fig. \ref{pos}), at angular distances from the
SMC centre of a = 3.4$^{\circ}$ to 3.9$^{\circ}$,
and distances from the Milky Way of 58.7 and 54.4~kpc, respectively. Their ages
correspond to the epoch of a probable recent tidal interaction between the
Magellanic Clouds, therefore they could be remnants of this period.

\subsection{HW\,40}
The CMD analysis for this cluster indicates that it is 
an intermediate-age SMC cluster with $2.5^{+0.4}_{-0.3}$ Gyr.
The presently derived metallicity ([Fe/H] = -0.9$^{+0.1}_{-0.2}$) is in 
agreement with the results from \citet{piatti11a} based on photometry
([Fe/H]$=-1.10\pm0.25$), and by Parisi et al. (2013, private
  comm.; [Fe/H] = -0.78 $\pm$0.05 dex) using CaT spectra. The age determined
  by Piatti et al. is significantly older ($5.4\pm1.0$ Gyr) than our result.
More accurate CMDs are needed for this cluster to constrain its age.

\subsection{Lindsay\,3}

The age ($1.2\pm0.3$ Gyr) and metallicity -0.4$^{+0.1}_{-0.2}$)
for this cluster is in good agreement with the results 
from \citet{piatti+11}: $1.25\pm0.20$ Gyr and [Fe/H] = $-0.65\pm0.20$ dex. 
A remarkable result from our isochrone fitting is the small 
distance modulus for this cluster ($18.64\pm0.14$),  
suggesting that it is one of the closest SMC clusters. We note that
here too \cite{piatti+11} assumed a distance modulus of 18.90 and also a
reddening of 0.04 instead of our fitted value of 0.07. The
combination of these two values tends to compensate the effects on age 
and metallicity. 
Lindsay~3 is younger 
than the bulk of the surrounding field stellar population in the West
Halo region (HW~1 and Lindsay~2 are $\sim$3~Gyr older, for instance).
If the Magellanic Clouds suffered interactions at about ~4~Gyr ago, 
this could be a resulting cluster in an older field, or else it is possible
that there is an age gradient in the West Halo. More data on the star
clusters of this region are needed to confirm these conclusions (see
Sect. \ref{discussion}).

\subsection{HW~34}

No star cluster was identified, as made evident by the following:

\begin{itemize}
\item{Sky maps given in Fig. \ref{soar_soi-plots} and radial density
    profiles in Fig. \ref{profiles-plots} show no significant
    overdensity in the region centred on HW~34 coordinates;}
\item{the photometric data are of comparable quality to the other clusters (see
    Figs. \ref{completeness} and \ref{errors}), so a cluster would
    be recovered if present;}
\item{Almost all stars in the direction of HW~34 coordinates were
    given the probability of being field stars, as shown in Fig. \ref{membership-plots}
  and Table \ref{stellarcounts};}
\item{Based on the very few RGB stars with p$_{\rm member} > 0.5$, and
    no clear main sequence turnoff, if there was a cluster there, it
    would be older than $\sim 5$ Gyr, which is not expected for the
    central younger regions of the SMC.}
\end{itemize}

%
\section{The age-metallicity relation and spatial distribution}
\label{discussion}

\begin{figure}[!htb]
\centering
\includegraphics[width=\columnwidth]{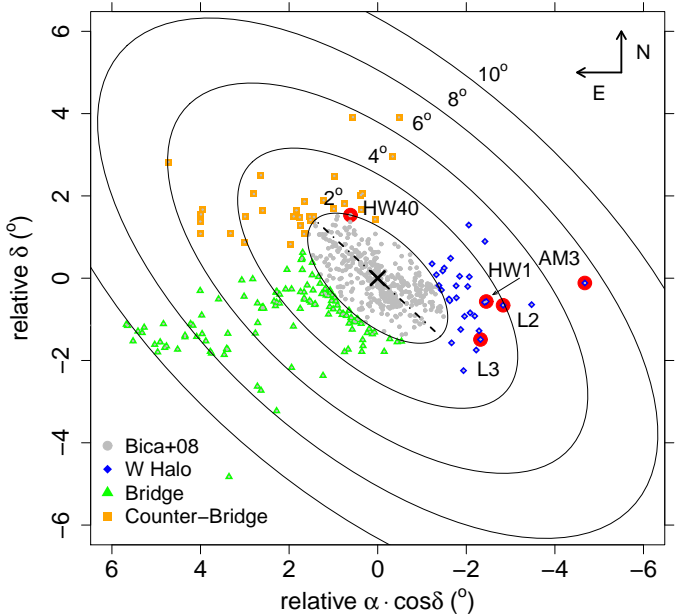}
\caption{On sky distribution of the SMC star clusters, including
  the catalog clusters \citep{bica+08a}, overplotted with the five sample
  clusters. The ellipses are used to illustrate distances from
  the SMC centre (Table \ref{finalparam}). Different colours
indicate subgroups, as indicated in the panel (see text for details). }
\label{pos}
\end{figure}

Figure \ref{pos} shows the 2D positions of the
five clusters in the sky plane, where
clusters from the catalog of \cite{bica+08a} were overplotted.
Ellipses are indicated to find the clusters' distances
from the SMC centre, following the procedure of \cite{piatti+05a}.
We adopt the coordinates of the SMC centre from \cite{crowl+01},
($\alpha$, $\delta$) = ($0^{\rm h} 52^{\rm m} 45^{\rm s}$,
$-72^{\circ} 49\arcmin 43\arcsec$), 
and a minor to major axis ratio of $b/a =1/2$.
Thus, the indicator of distance to the SMC centre would be 
the semi-major axis $a$, which is
more appropriate for a ellipsoidal galaxy like the SMC (see Figure
\ref{pos}). We consider an inclination of 45$^\circ$ for the major
axis of all the ellipses in the projected plane of the figure. The
distances $a$ for all the clusters can be found in Table
\ref{finalparam}.

\begin{figure}[!htb]
\centering
\includegraphics[width=\columnwidth]{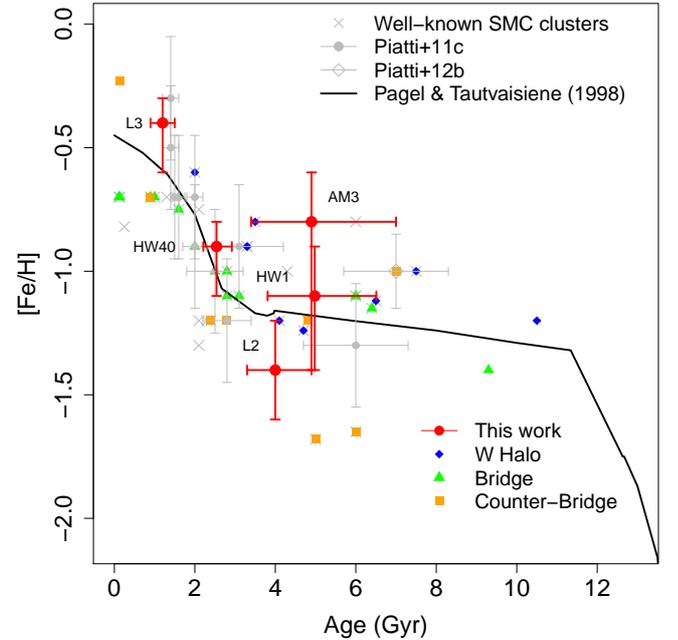}
\caption{Age-metallicity relation for the SMC,
  including the new additions from the present work. The well-known
  clusters are listed in Table 6 of \cite{dias+10}, and 
  other clusters from literature. The model of \citet{PT98} is overplotted.}
\label{amr}
\end{figure}

\begin{figure}[!htb]
\centering
\includegraphics[width=\columnwidth]{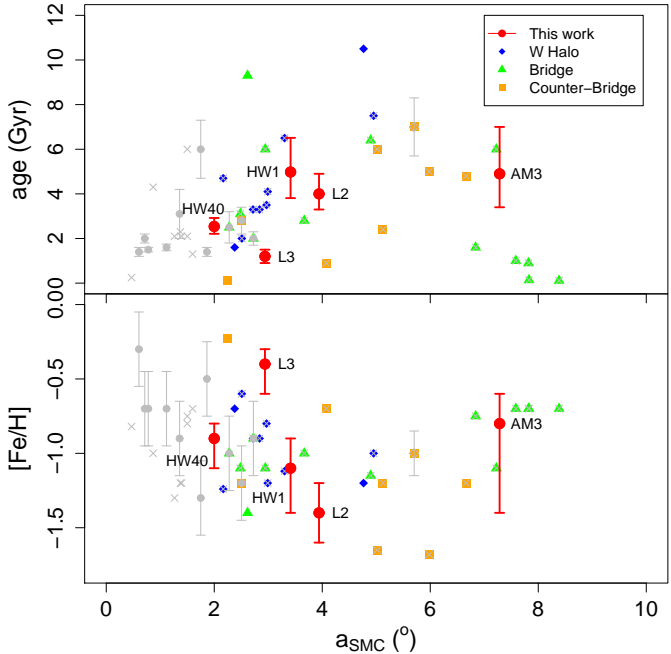}
\caption{Age and [Fe/H] as a function of the distance indicator
  $a$. The dots are the same as described in Figure \ref{amr}.}
\label{agemet-rad}
\end{figure}

Most of the sample clusters are located outside the region sampled by
\cite{harris+04}, but inspired by the different SFR
that they found in
different regions of the SMC, we located our targets in different
groups for a $> 2^{\circ}$ (see Fig. \ref{pos}). These
regions were simply divided into the sky plane based on the HI
structures, as described by \cite{diaz+12}, assuming that these
structures have stellar/star cluster counterparts. They are: Bridge,
Counter-Bridge, and a third region not related with the HI structures
that we have called West Halo. They are indicated by different colours and
symbols in Figs. \ref{pos}, \ref{amr}, and \ref{agemet-rad}.

The 3D structure of the SMC was shown to be complex
\citep{subramanian+12}, with a spread in depth of about 14~kpc, and an
inclination along the main axis with the north-eastern portion towards
the observer. The five clusters studied in this work follow this
dispersion, in particular for West Halo clusters, where we found
distances between $\sim$ 54 and 63~kpc, revealing that more statistics
on this region are needed in order to better describe the 3D structure
of the SMC.

Figure \ref{amr} shows the age-metallicity relation (AMR) 
of the SMC, by plotting data for well-studied star
clusters \citep{dias+10,piatti+11,piatti11a,piatti11b}, overplotted by
the model of \citet{PT98}.
The results for the sample clusters fit well the \cite{PT98} model, and agree with a
dispersion in metallicity for the SMC cluster system in the
metallicity and age ranges of -1.5 $<$ [Fe/H] $<$ -1.0, and 5 $<$ age
$<$ 10~Gyr, as indicated by \cite{piatti11b}, in their Fig. 3. 
The field stars also agree with this statement, as can be seen in
Fig. 8 of \cite{piatti12a}. In this case, the model 
of \cite{PT98} would represent an average of the chemical enrichment
history of the SMC.
Spectroscopic results by \citet{parisi+09} also show a spread in
[Fe/H] among intermediate-age clusters; however, their values are in the range
-1.2 $<$ [Fe/H] $<$ -0.7 dex. Therefore, both photometric and
spectroscopic metallicities show a spread of $\sim$0.5~dex in
metallicity, with limits defined by different metallicity scales.

Another important product of this work concerns AM~3, Lindsay~2, and
HW~1, the oldest clusters of the sample, with
4.9, 4.0, and 4.98 Gyr, respectively.  They are close to the end of the
quiescent star formation as pointed out by \cite{harris+04}, and can
be among the first products of the reactivation of star formation
in the MC system $\sim$ 3-4 Gyr ago.

The cluster distribution in the 2D projected plane of the sky is also
useful for investigating gradients of age and metallicity. This is
presented in Figure \ref{agemet-rad}, where the marked labels correspond
to those of Figure \ref{amr}. 
 The five clusters presented in this work are
located outside the 2$^\circ$ ellipse. The small sample combined with the
error bars do not permit us to establish a strong
gradient. Nevertheless, it is possible to identify that the most
internal regions of the SMC, below $a$ = 1-2$^{\circ}$ might indicate a gradient in
age and metallicity, whereas the outer regions have a spread in these
 parameters. Otherwise, these trends seem to be consistent with
   those of the SMC field population \citep{piatti12a}. 

Even so, if one looks at different
 regions of the projected SMC, they reveal different superimposed
 gradients (which would generate the overall dispersion with no
 gradients), but the dispersion is too high and more data are needed for this to 
be confirmed. 
Lindsay~2, Lindsay~3, and HW~1, which are
 located in the West Halo region, tend to show a possible age gradient in
 this region, as indicated by the blue diamonds in the figure.

%

\section{Summary and conclusions}
\label{conclusions}

We studied intermediate-age stellar
clusters in the SMC. The parameters were derived by statistical
fitting of the observed CMDs with a grid of synthetic CMDs based on the
PARSEC isochrones. Six
targets were observed in the visible filters B and V, and we derived ages
for five of them (AM~3, HW~1, HW~40, Lindsay~2, and Lindsay~3) in the
age range of 1.2~Gyr to 5.0~Gyr, whereas HW~34 was identified as
a field fluctuation. In particular, Lindsay~2 and HW~1 were studied
for the first time in this work, and for the other clusters we derived
self-consistent parameters compatible with the literature, and in
principle with more precision. 

The five clusters essentially follow the chemical enrichment history
modelled by \cite{PT98}.
Figure \ref{amr} shows a strong spread in metallicity for clusters
with ages of  2-5 Gyr, also found by \cite{piatti11b}.
\cite{piatti12a} finds a metallicity spread of $\sim$0.4~dex at ages of
2-4 Gyr, and \cite{parisi+10} finds a dispersion of $\sim$0.32~dex
around the mean value of [Fe/H]=-1.0 from 15 fields.

This suggests that, after an initial period
of rapid enrichment, that brought the metallicity up to around
[Fe/H]$\approx$-1.2, the SMC chemical enrichment is very inhomogeneous,
 since there is a spread in metallicity from $\sim$11 to $\sim$3 Gyr, with a trend for
slow  enrichment. Finally an intense burst occurred at around 2 Gyr
ago, when most of the star clusters started to form.

\cite{piatti+11} suggested the occurrence of
 two recent bursts at $\sim$ 2 Gyr and $\sim$ 6 Gyr ago (preceded
   by a quiescent period lasting $\sim$3 Gyr),
with roughly no metallicity variations.
Our clusters improve the statistics for the two
more recent bursts identifed by \cite{piatti+11}, in particular HW~1
and Lindsay~2, which are located in the West Halo region, at almost
the same distance.

A gradient of age and metallicity can be found in the inner part of
the SMC (a $< 2^{\circ}$), but in the outer regions the high
dispersion in ages and metallicities prevents
the identification of any trends. More data on age, metallicity/abundances,
distance, and kinematics of star clusters are needed in order to prove different
star formation rates and chemical evolution histories in
the  regions due to tidal forces caused by the interactions in the
SMC-LMC-Milky Way system.

%

\begin{acknowledgements}
We acknowledge partial financial support from
CNPq, CAPES and FAPESP. BD acknowledges a CNPq studentship
no. 142047/2010-4, and also an ESO studentship.
B.D., B.B., L.K. thank CAPES / CNPq for their financial support with
the PROCAD project number 552236/2011-0.
\end{acknowledgements}

\bibliographystyle{aa} 
\bibliography{smc} 

%
\begin{appendix}

\section{Likelihood results and degeneracy investigation}

We analysed the likelihood maps 
 to investigate the degeneracy in age, metallicity, reddening,
and distance modulus. The likelihood values were obtained from
a comparison of a given synthetic CMD with the observed one (see Section
3 for details).

To establish the standard deviation of the likelihood statistics,
we proceeded as follows. A synthetic best model CMD was created with the same
numbers of stars as the observed one. We call this a false observation.
We then compared this CMD with another one with the same parameters but
with many more stars (e.g. 10$^5$). From this model vs. model comparison, a
likelihood value is obtained, as previously done for data vs. model comparions.
Then simulating 300 false observations, and comparing them
 with the star rich synthetic CMD, we derived the standard deviation $\sigma$.
This tells how much the likelihood can vary due only to stochastic effects,
i.e., due to the statistical fluctuations on the position of stars in
a CMD with a number of points compatible with that of the observed CMDs.

\begin{figure}[!htb]
   \centering
   \includegraphics[height=0.3\textheight]{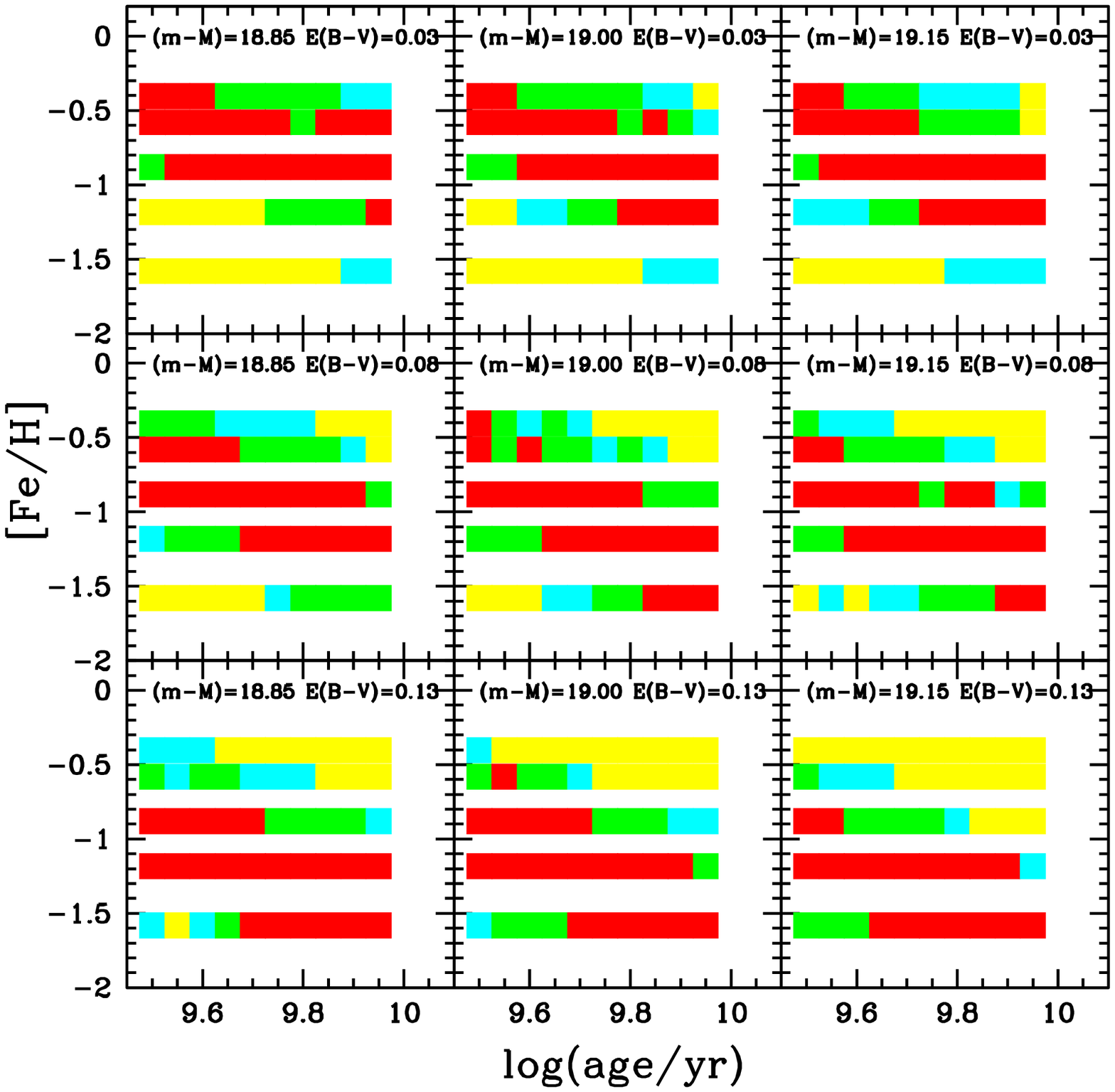}
   \includegraphics[width=0.3\textheight]{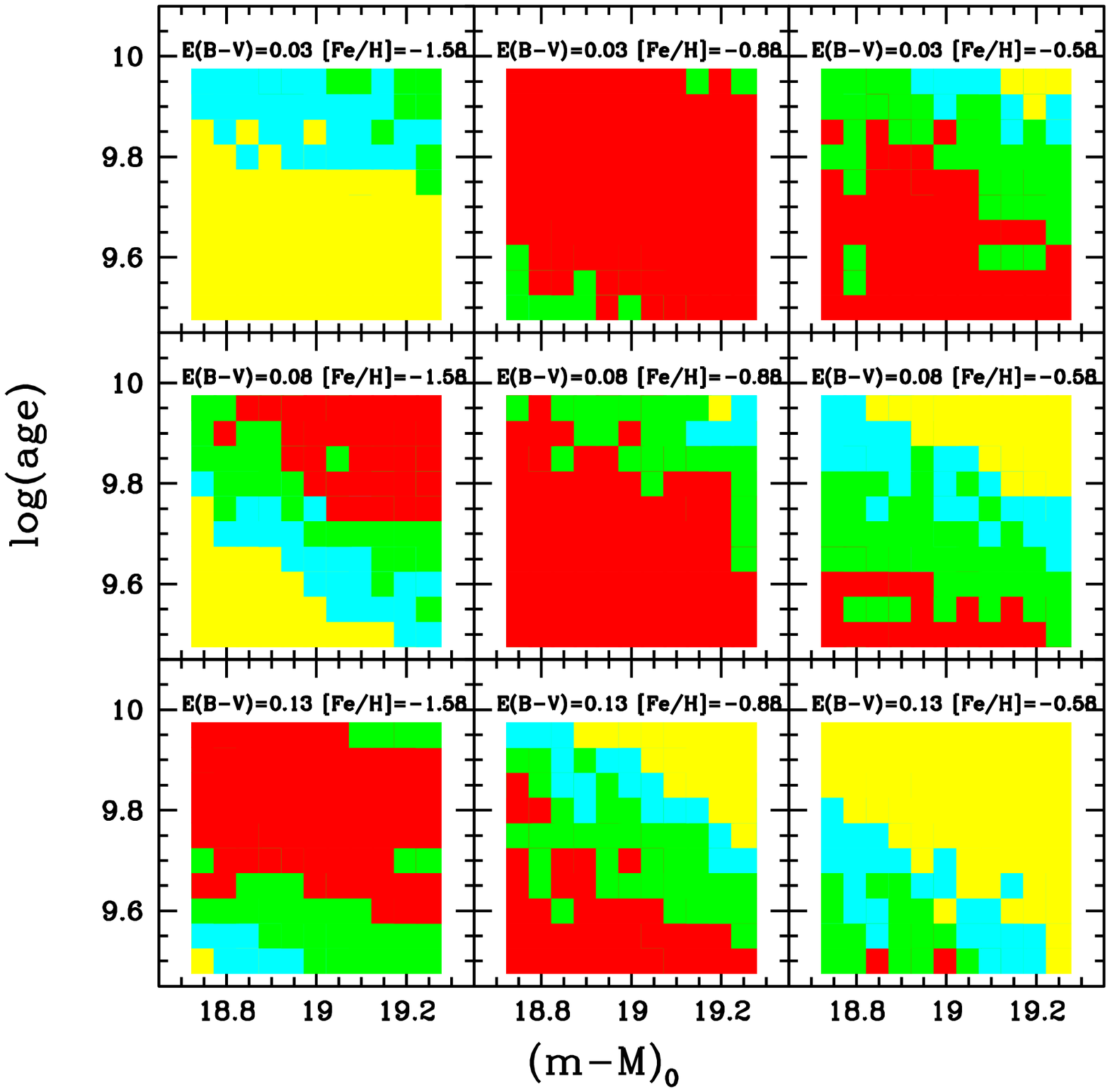}
   \includegraphics[width=0.3\textheight]{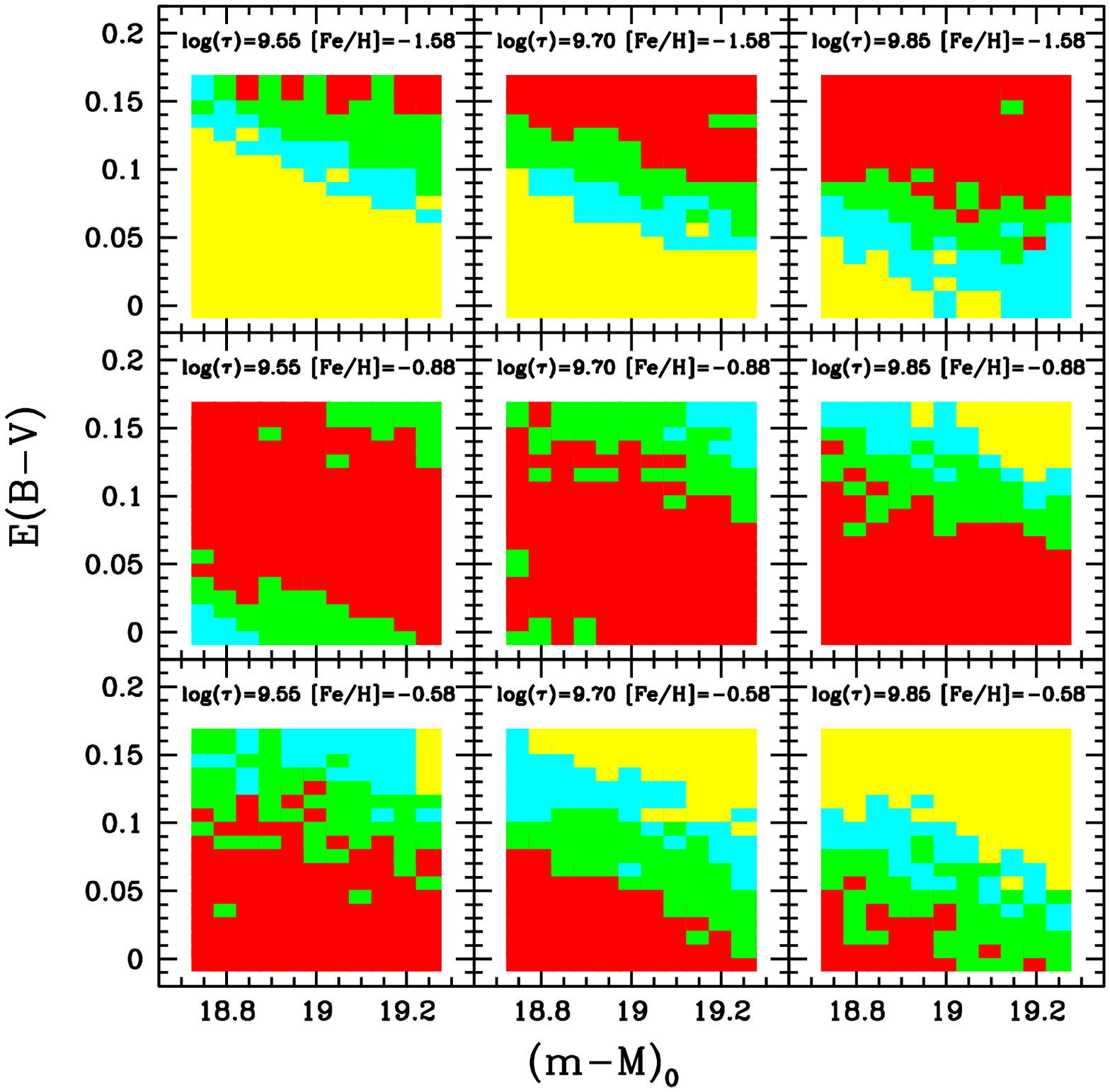}
   \caption{Likelihood from the fits for different combinations of
     age, metallicity, reddening, and distance modulus for AM~3. Upper
     panel: metallicity vs. age; middle panel: age vs. distance
     modulus; bottom panel: reddening vs. distance modulus.
 The colour scale is red, green, cyan or yellow,
corresponding to a difference from the best-fit
   solution up to 1-sigma, between 1- and 2-sigma,
    between 2- and 3-sigma, and larger than
   3-sigma of the best fit, respectively.}
   \label{likelihood-am3}
   \end{figure}

\begin{figure}[!htb]
   \centering
   \includegraphics[height=0.3\textheight]{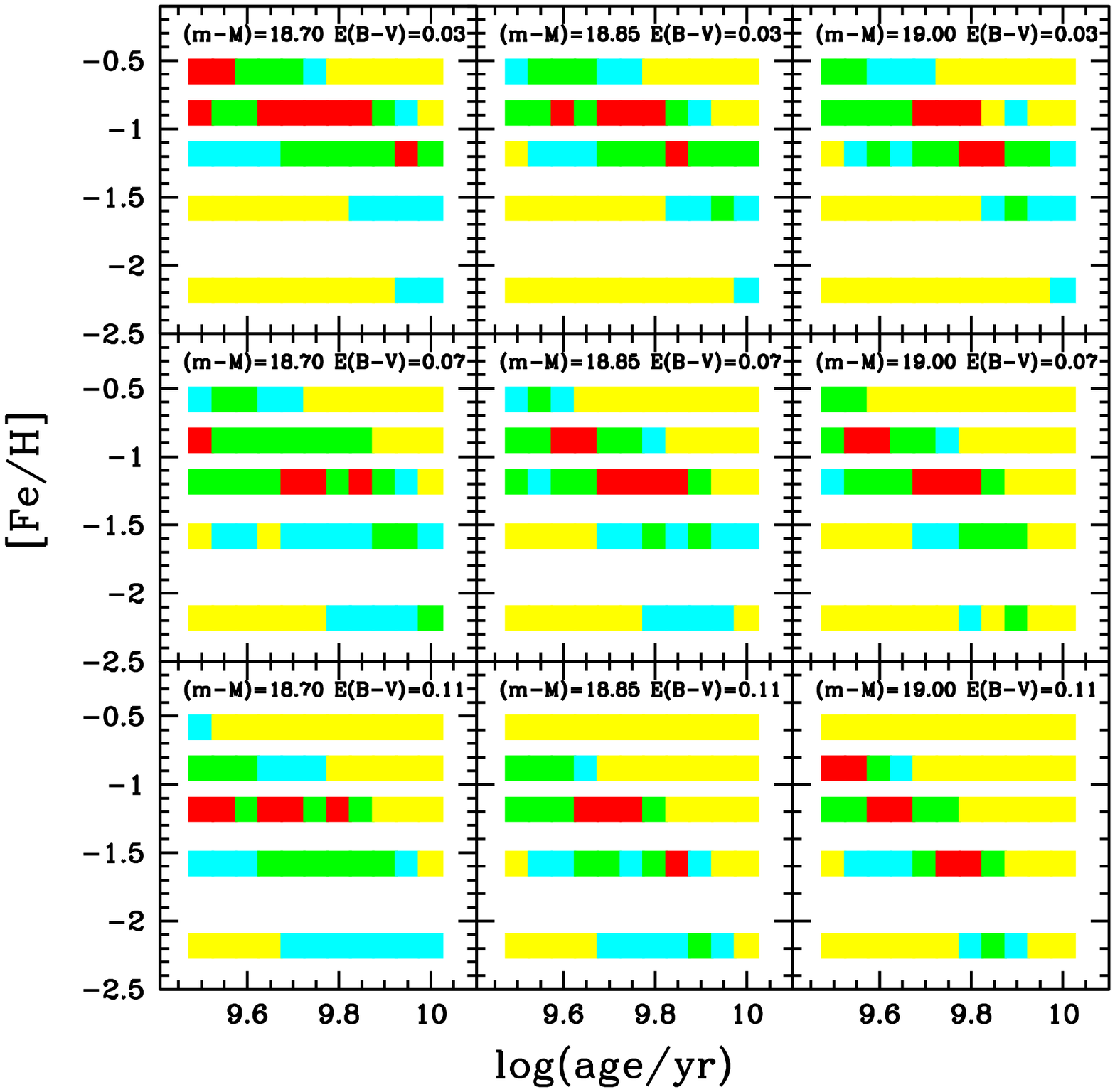}
   \includegraphics[width=0.3\textheight]{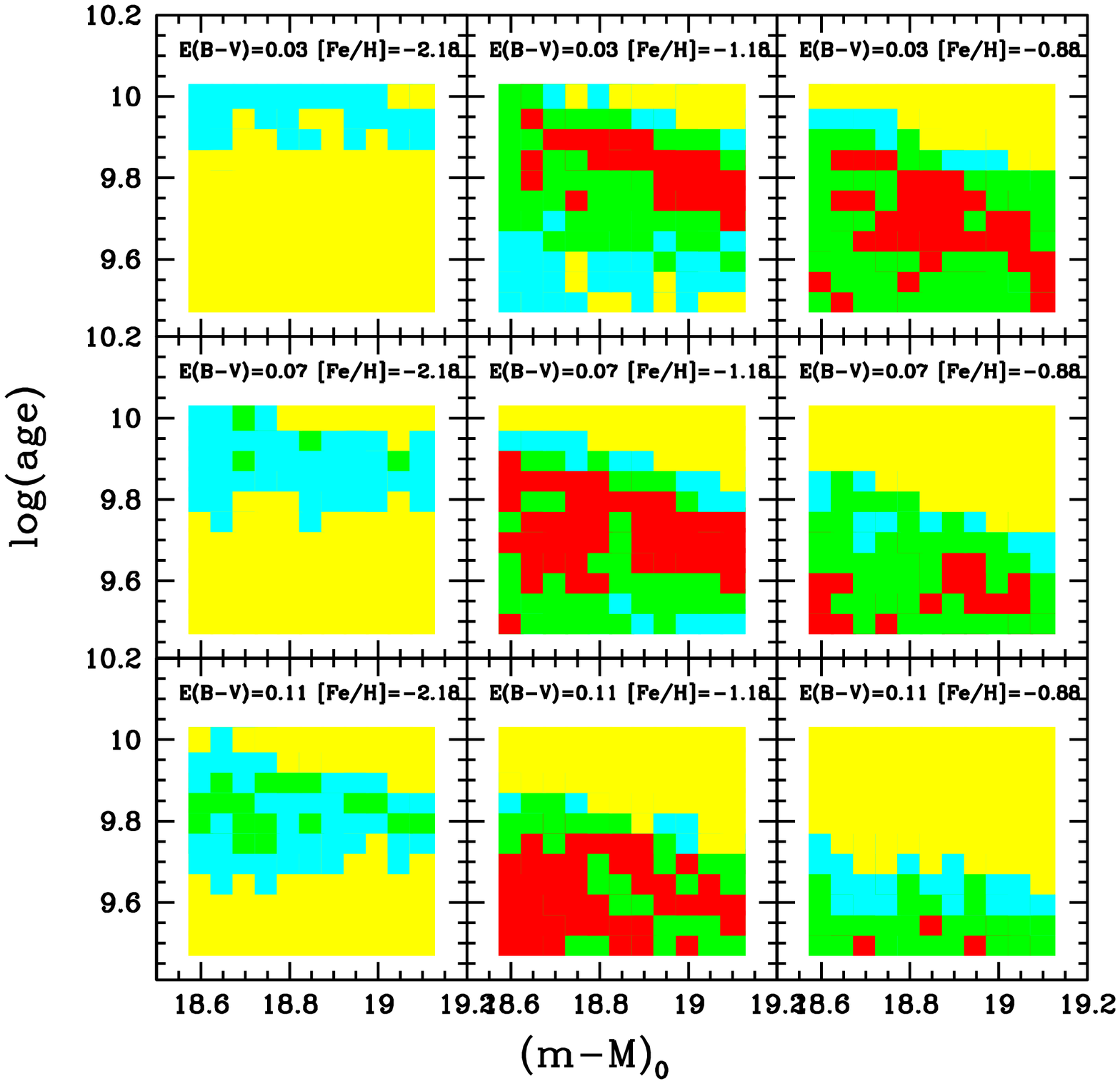}
   \includegraphics[width=0.3\textheight]{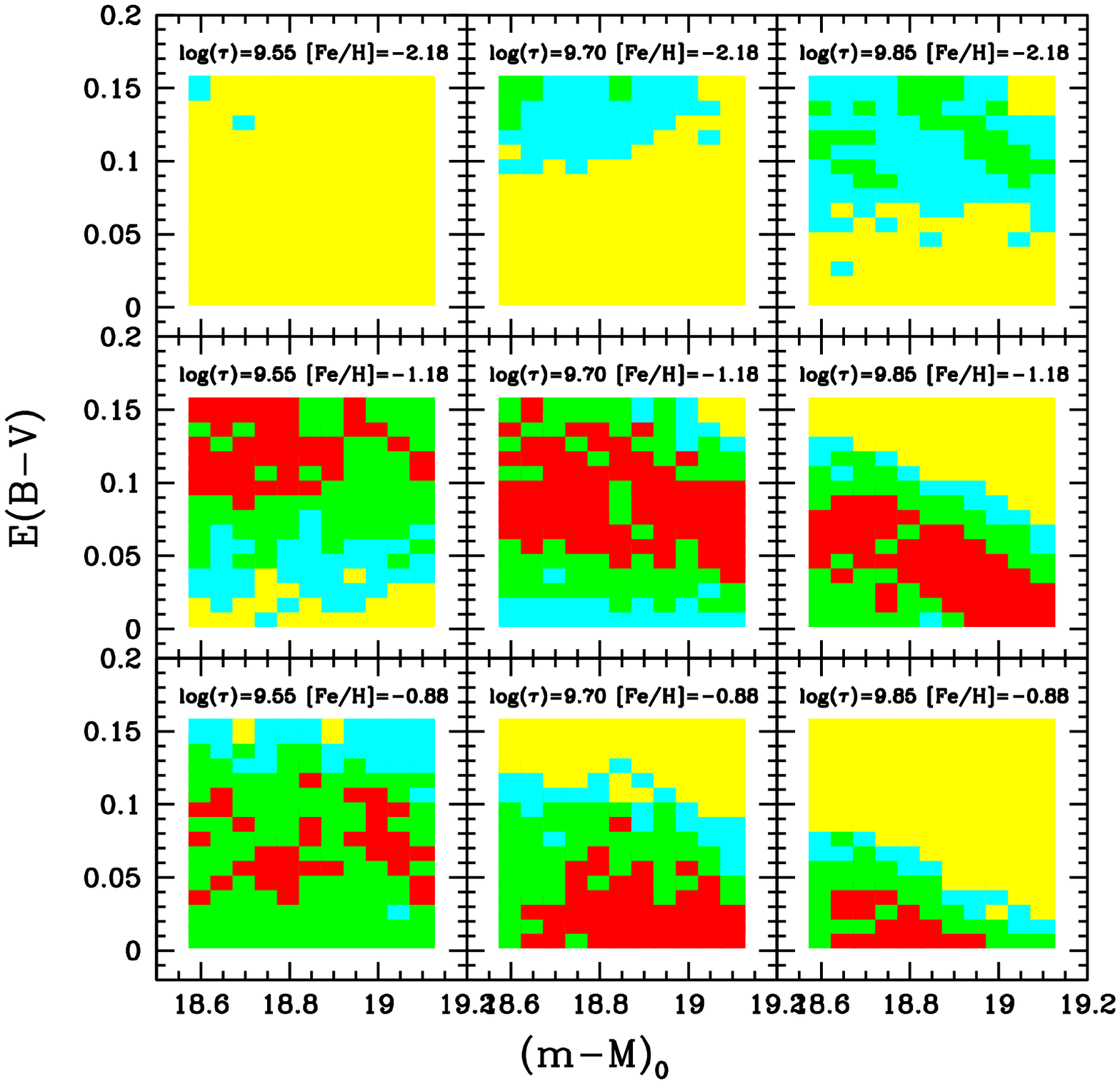}
   \caption{Same as Fig. \ref{likelihood-am3} for HW~1.}
   \label{likelihood-hw1}
   \end{figure}

\begin{figure}[!htb]
   \centering
   \includegraphics[height=0.3\textheight]{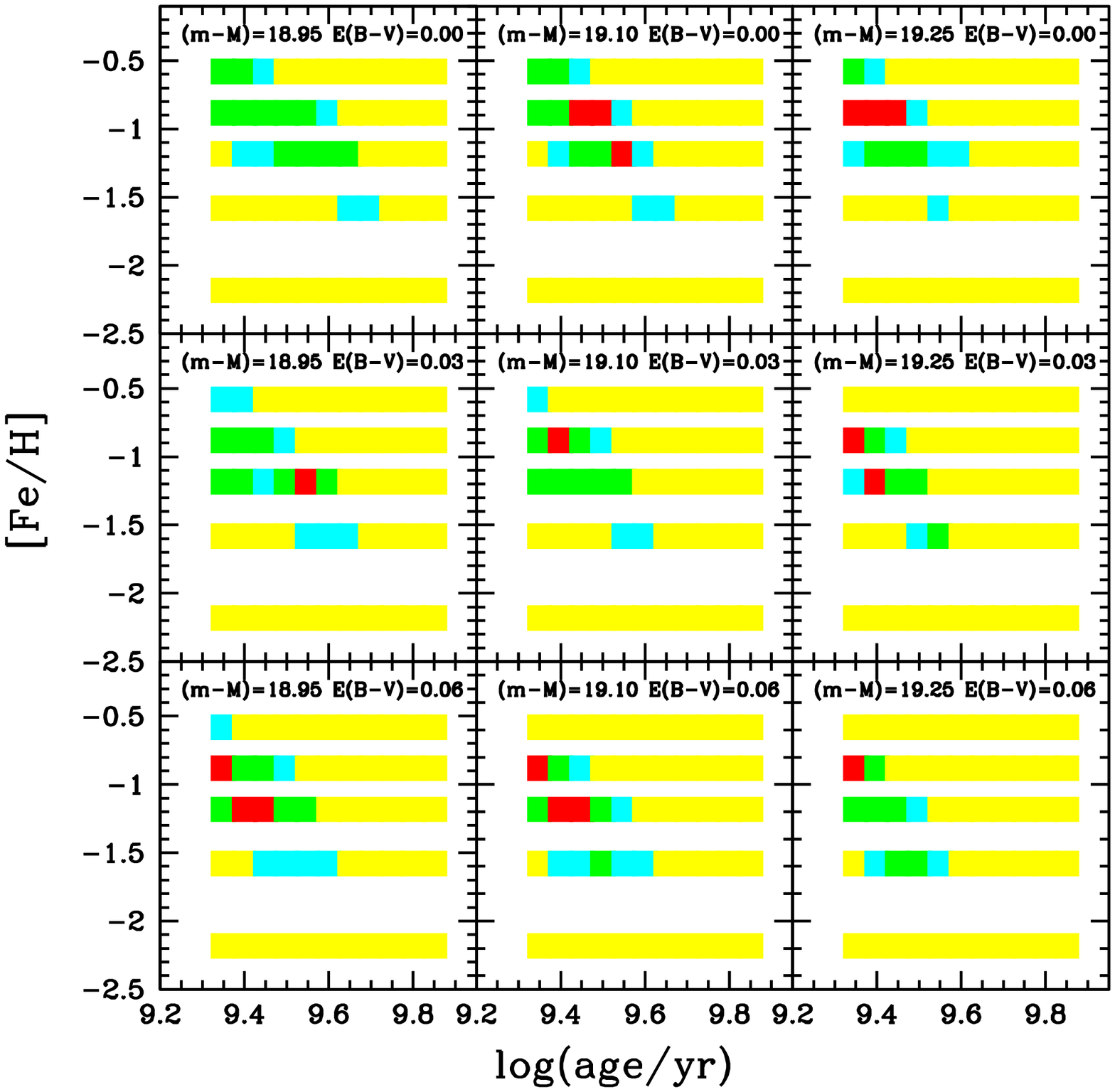}
   \includegraphics[width=0.3\textheight]{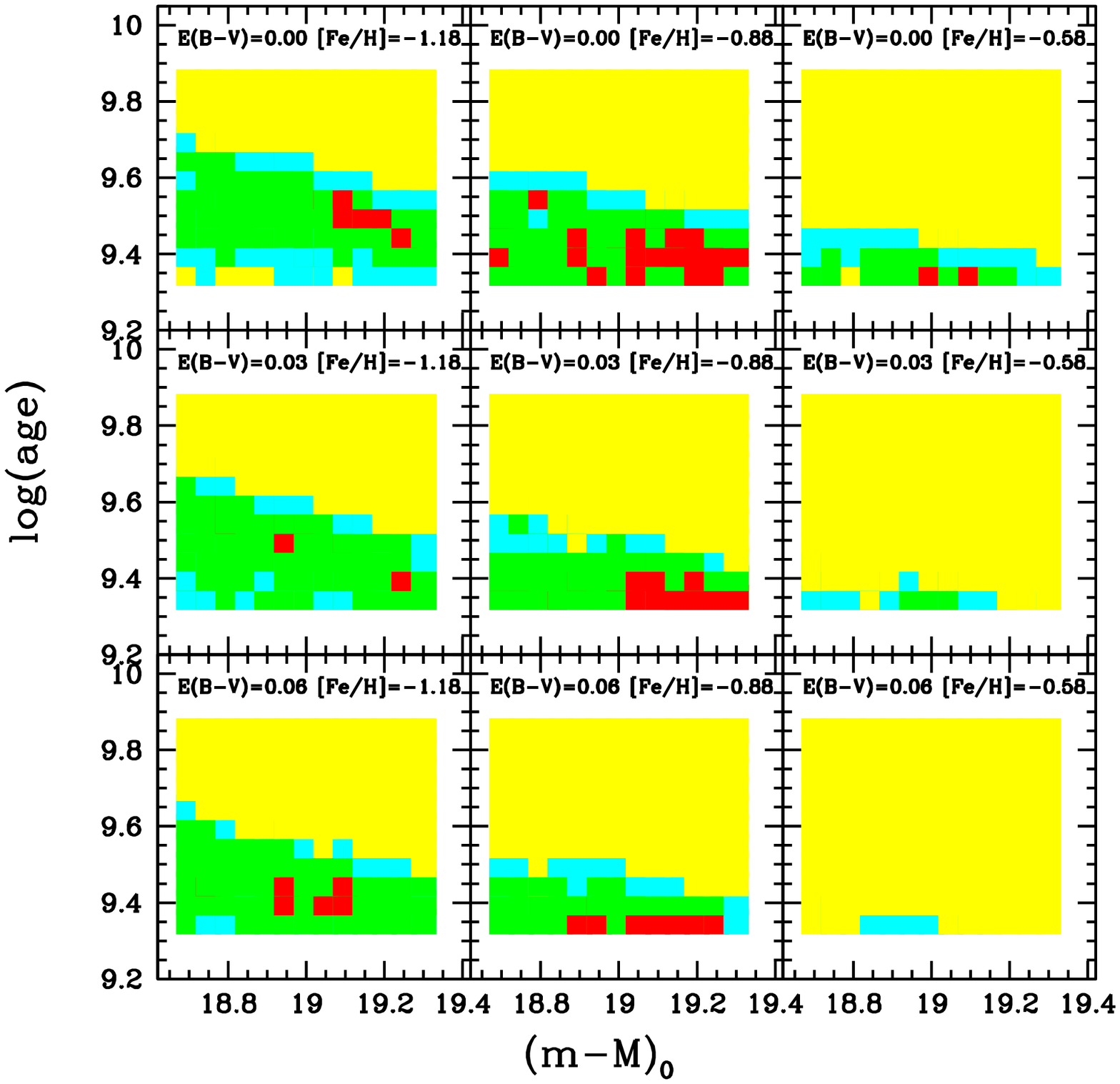}
   \includegraphics[width=0.3\textheight]{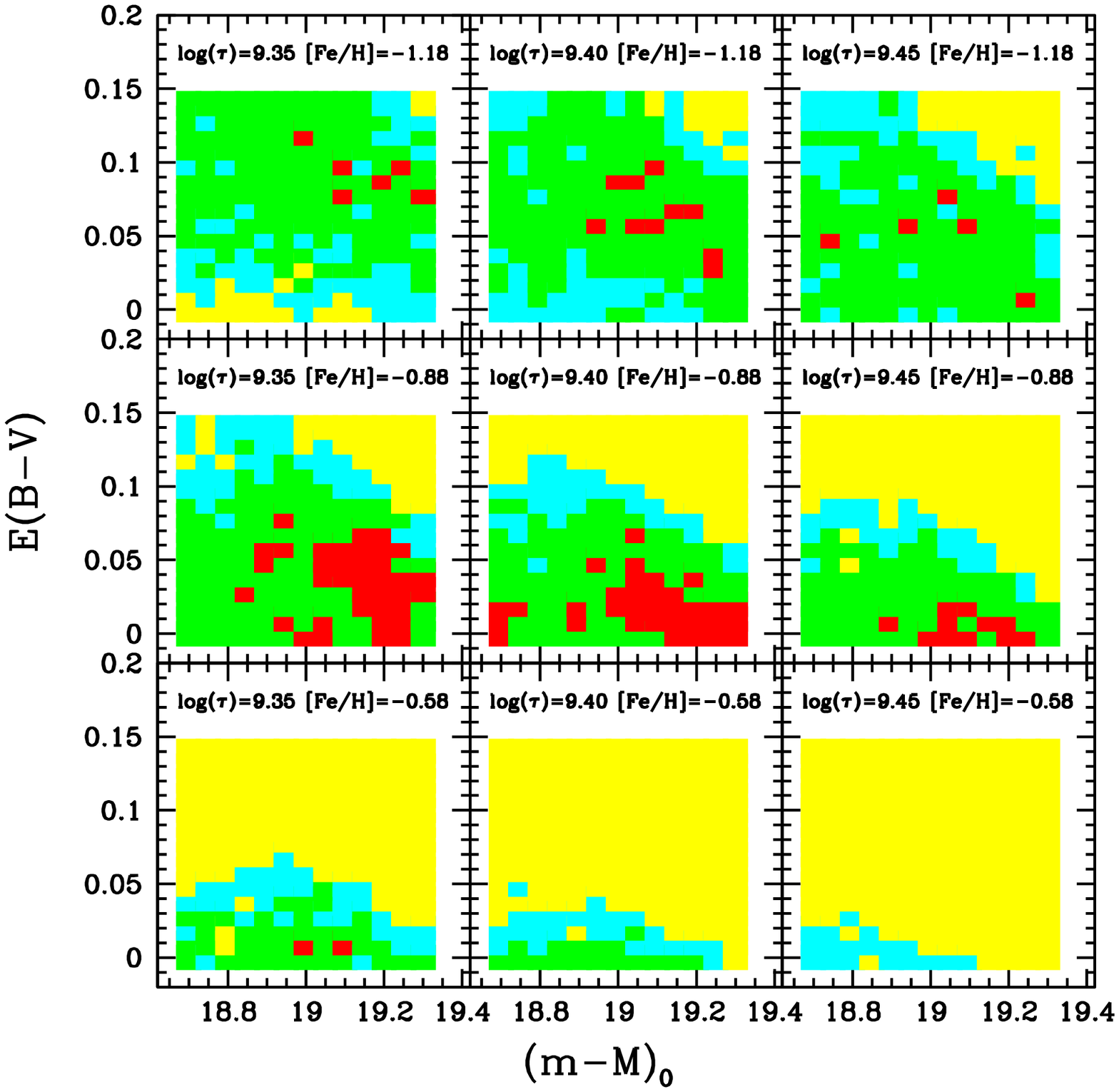}
   \caption{Same as Fig. \ref{likelihood-am3} for HW~40.}
   \label{likelihood-hw40}
   \end{figure}

\begin{figure}[!htb]
   \centering
   \includegraphics[height=0.3\textheight]{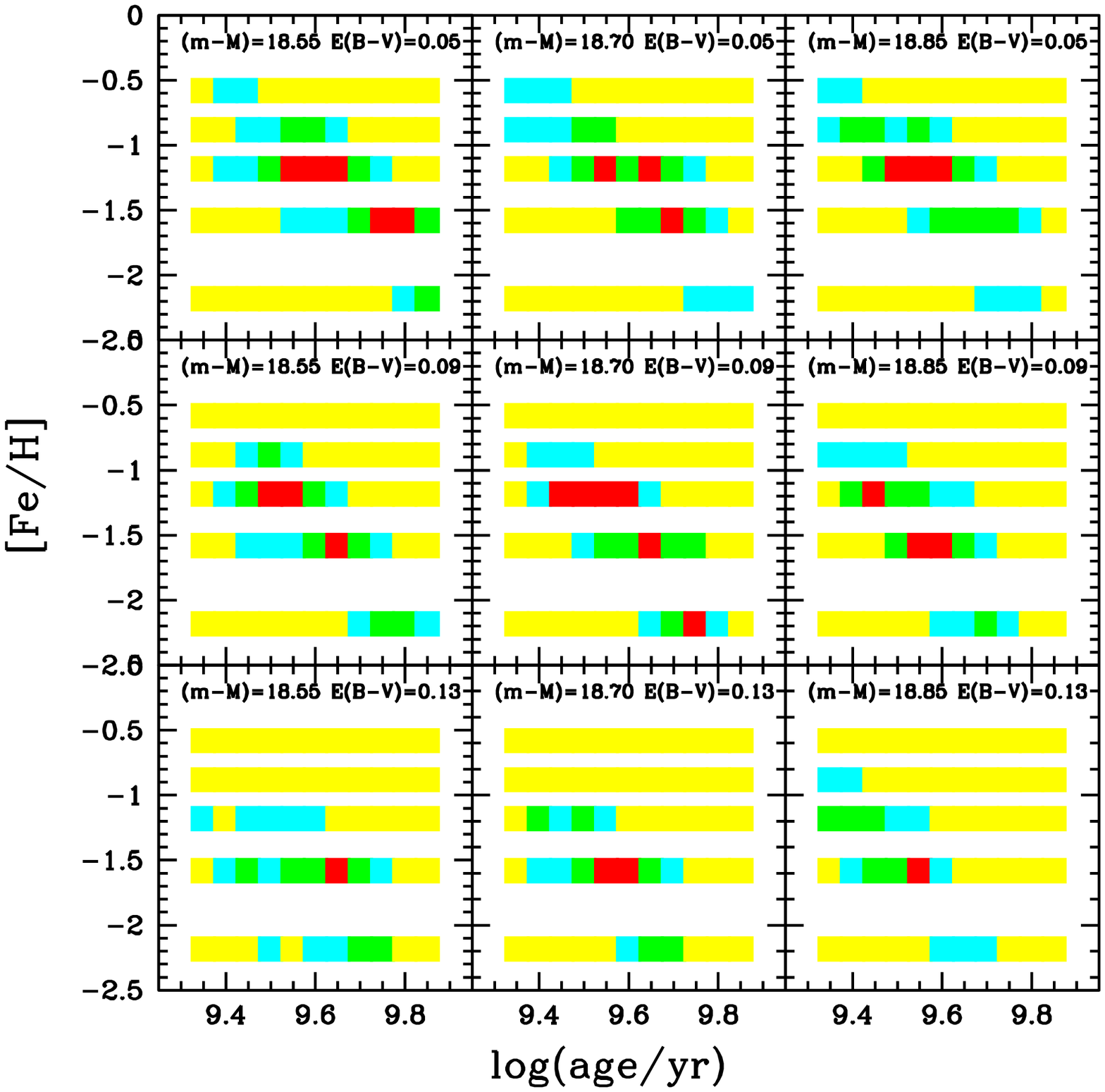}
   \includegraphics[width=0.3\textheight]{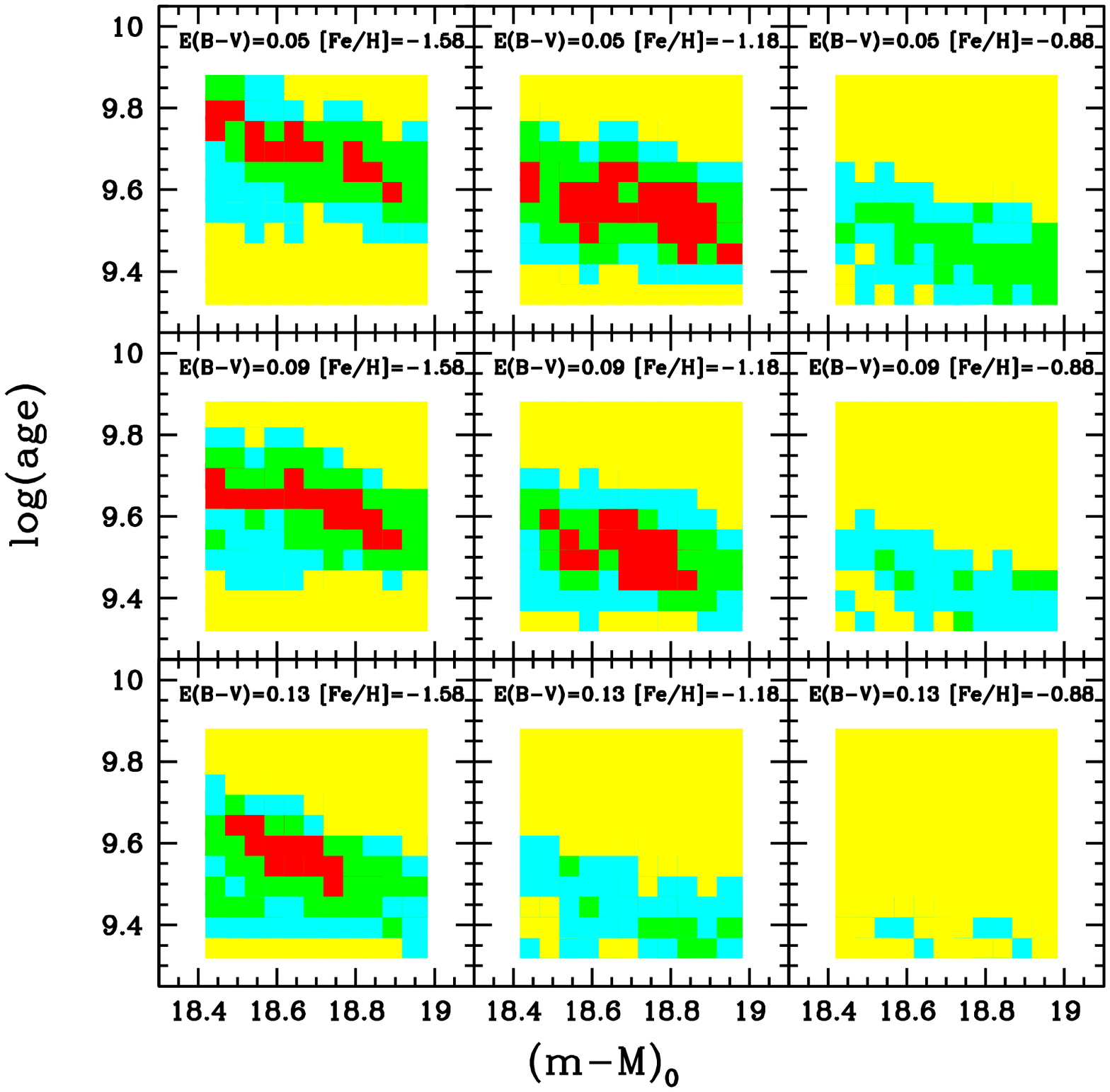}
   \includegraphics[width=0.3\textheight]{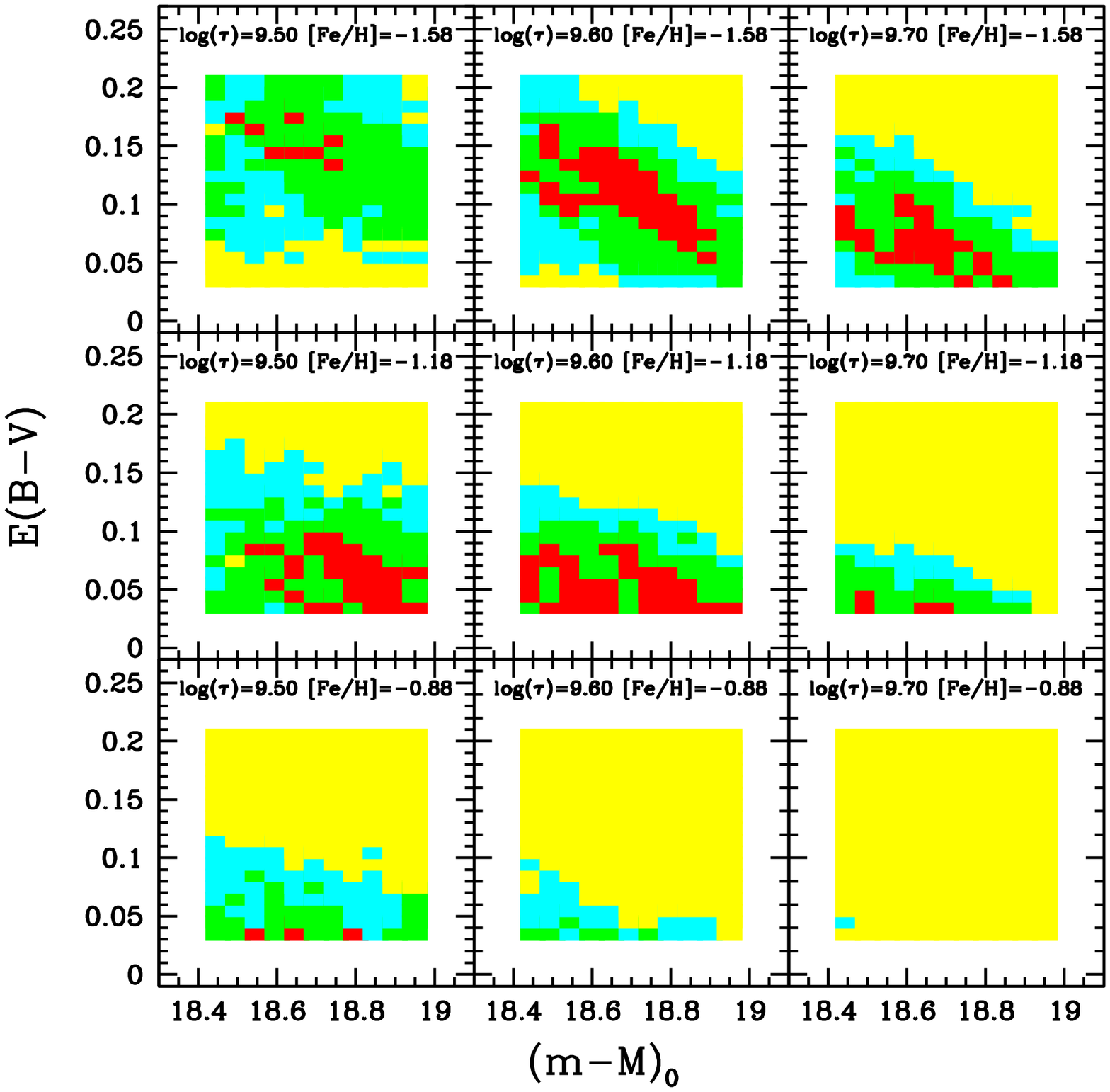}
   \caption{Same as Fig. \ref{likelihood-am3} for Lindsay~2.}
   \label{likelihood-l2}
   \end{figure}

\begin{figure}[!htb]
   \centering
   \includegraphics[height=0.3\textheight]{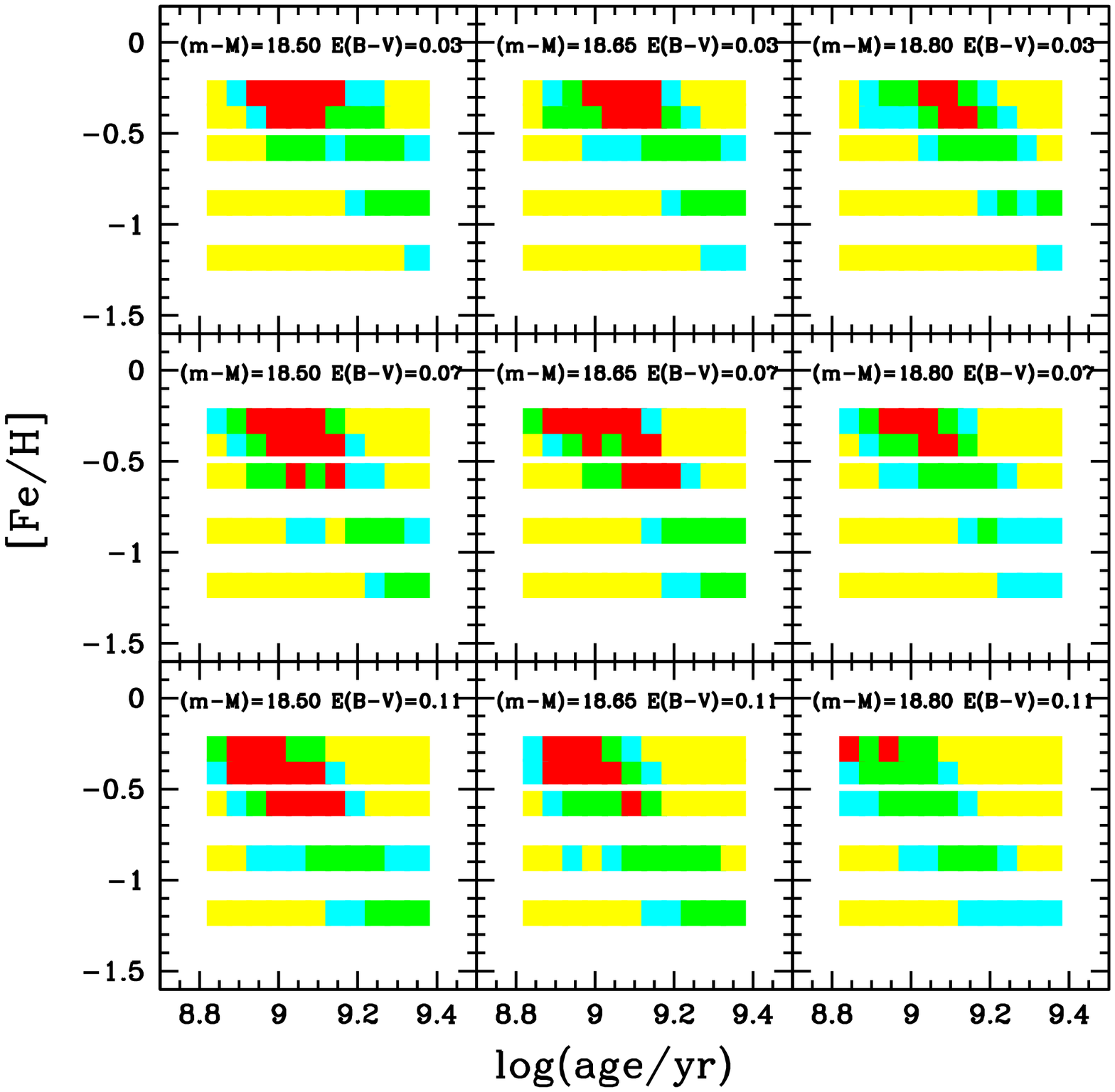}
   \includegraphics[width=0.3\textheight]{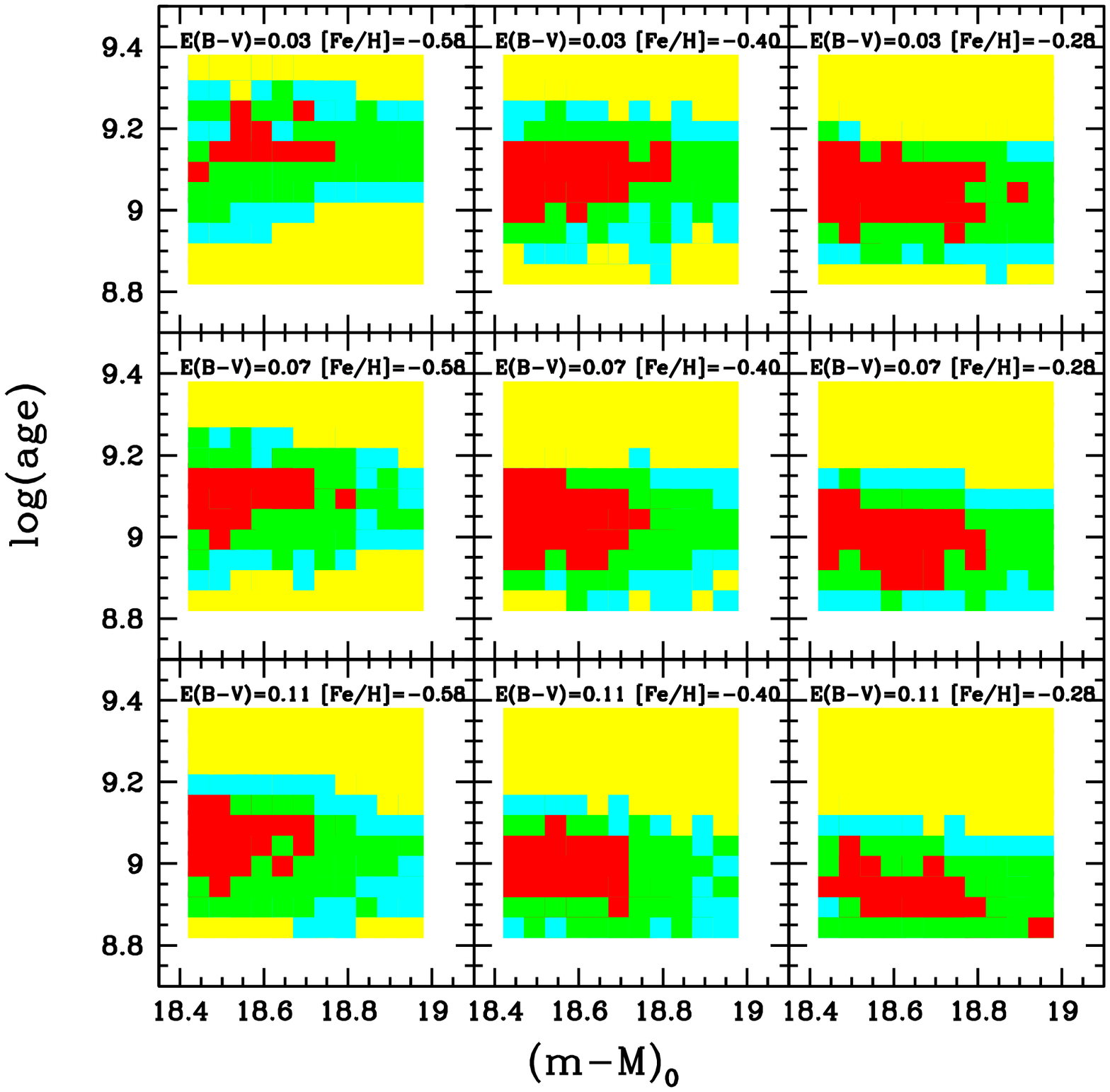}
   \includegraphics[width=0.3\textheight]{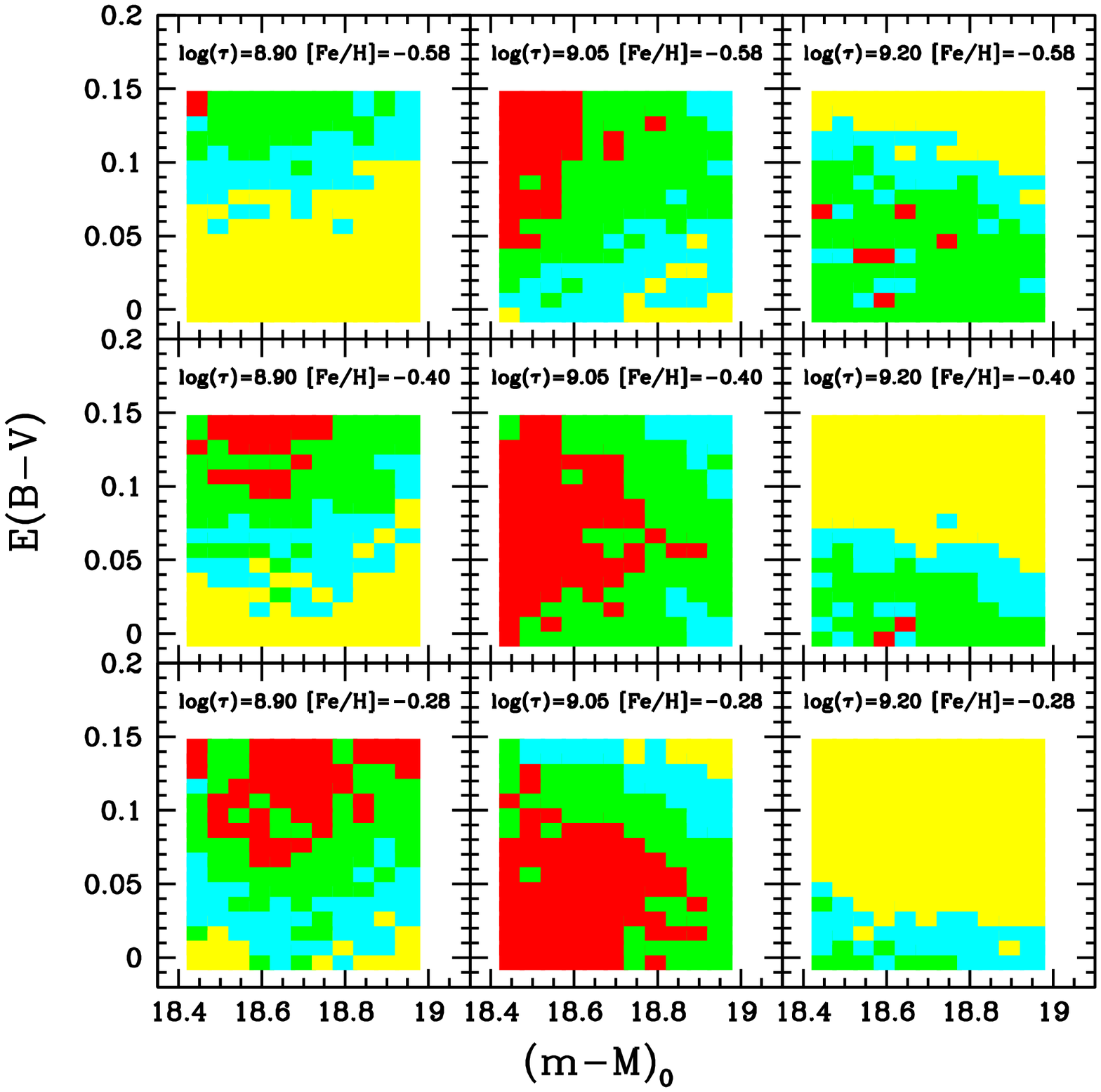}
   \caption{Same as Fig. \ref{likelihood-am3} for Lindsay~3.}
   \label{likelihood-l3}
   \end{figure}

The likelihood maps of these comparisons are displayed in
Figures \ref{likelihood-am3} to \ref{likelihood-l3}, in the space of
parameters of [Fe/H] vs. log(age), log(age) vs. (m-M), and E(B-V)
vs. (m-M), centred in the parameters of the best-fit. 
 The colour scale is red, green, cyan or yellow,
corresponding to a difference from the best fit
   solution up to 1-sigma, between 1- and 2-sigma,
    between 2- and 3-sigma, and larger than
   3-sigma of the best fit, respectively.
The central panel in each figure gives the
value obtained for the two fixed parameters, since in each
ma two parameters are varied and the two others are fixed.

These maps clearly reveal the expected anti-correlation beteween age
and metallicity, distance modulus and age, and between distance
modulus and reddening. These degeneracies are directly reflected in
the physical parameter uncertainties since they correspond to the
standard deviation in each parameter for the set of models with
likelihood values up to 1-sigma (red points) from the best-fit
solution. 

\end{appendix}

\end{document}